\documentclass[twocolumn,prd,floatfix,preprintnumbers,a4paper,nofootinbib,superscriptaddress,10pt]{revtex4-1}

\usepackage{epsfig}
\usepackage{graphics}
\usepackage{graphicx}
\usepackage{amsmath,amssymb}
\usepackage{amsfonts}
\usepackage{bm}
\usepackage[usenames,dvipsnames]{color}
\usepackage{amssymb}
\usepackage{psfrag}
\usepackage{times}
\usepackage[varg]{txfonts}
\usepackage{xspace}
\usepackage{placeins} 
\usepackage{capt-of} 
\usepackage{enumerate}

\usepackage[colorlinks, pdfborder={0 0 0}]{hyperref}
\definecolor{LinkColor}{rgb}{0.75, 0, 0}
\definecolor{CiteColor}{rgb}{0, 0.5, 0.5}
\definecolor{UrlColor}{rgb}{0, 0, 0.75}
\hypersetup{linkcolor=LinkColor}
\hypersetup{citecolor=CiteColor}
\hypersetup{urlcolor=UrlColor}
\let\oldtheequation\theequation
\makeatletter
\def\tagform@#1{\maketag@@@{\ignorespaces#1\unskip\@@italiccorr}}
\renewcommand{\theequation}{(\oldtheequation)}
\makeatother

\usepackage[utf8]{inputenc}
\usepackage{ulem}
\makeatletter
\g@addto@macro\normalsize{%
  \addtolength\abovedisplayskip{0.5\baselineskip}
  \addtolength\belowdisplayskip{0.5\baselineskip}
  \addtolength\abovedisplayshortskip{0.5\baselineskip}
  \addtolength\belowdisplayshortskip{0.5\baselineskip}
}
\makeatother

\newcommand{\avirgo}{Advanced Virgo\gdef\avirgo{Adv.~Virgo\xspace}\xspace}
\newcommand{\bh}[1][]{black hole#1 (BH#1)\renewcommand{\bh}[1][]{BH##1\xspace}\xspace}
\newcommand{\bbh}[1][]{binary black hole#1 (BBH#1)\renewcommand{\bbh}[1][]{BBH##1\xspace}\xspace}
\newcommand{\gw}[1][]{gravitational wave#1 (GW#1)\renewcommand{\gw}[1][]{GW##1\xspace}\xspace}
\newcommand{\nr}{numerical relativity (NR)\gdef\nr{NR\xspace}\xspace}

\newcommand{\gr}{General Relativity (GR)\gdef\gr{GR\xspace}\xspace}

\newcommand{\smbh}[1][]{super-massive black hole#1 (SMBH#1)\renewcommand{\smbh}[1][]{SMBH##1\xspace}\xspace}

\newcommand{\emri}[1][]{extreme-mass-ratio inspiral#1 (EMRI#1)\renewcommand{\emri}[1][]{EMRI##1\xspace}\xspace}

\newcommand{\SpeC}{Spectral Einstein Code~(SpEC)\gdef\SpeC{SpEC\xspace}\xspace}

\newcommand{\rmse}{root-mean-square error~(RMSE)\gdef\rmse{RMSE\xspace}\xspace}
\newcommand{\PN}{Post-Newtonian~(PN)\xspace\gdef\PN{PN\xspace}}
\newcommand{\ffi}{fixed-frequency-integration~(FFI)\gdef\ffi{FFI\xspace}\xspace}
\newcommand{\RWZ}{Regge-Wheeler-Zerilli~(RWZ)\gdef\RWZ{RWZ\xspace}\xspace}

\newcommand{\eob}{effective-one-body (EOB)\gdef\eob{EOB\xspace}\xspace}
\newcommand{\ADM}{Arnowitt Deser Misner (ADM)\gdef\ADM{ADM\xspace}\xspace}
\newcommand{\BSSN}{Baumgarte-Shapiro-Shibata-Nakamura (BSSN)\gdef\BSSN{BSSN\xspace}\xspace}

\newcommand{\nmax}{n_{\mathrm{max}}}
\definecolor{myblue}{rgb}{0.2,0.3,0.7}
\definecolor{darkgreen}{rgb}{0,0.3,0}
\definecolor{mygreen}{rgb}{0,0.5,0}
\definecolor{grey}{rgb}{0.5,0.5,0.5}
\definecolor{orange}{rgb}{1,0.5,0}
\definecolor{mypurple}{rgb}{0.5,0,0.4}
\newcommand{\revision}[1]{#1}
\begin{document}

\addtolength{\textheight}{-4\baselineskip}


\title{Quasinormal modes and their overtones at the common horizon in a binary black hole merger}
\newcommand{\aei}{Max Planck Institute for Gravitational Physics (Albert Einstein Institute), Callinstra\ss e 38, 30167, Hannover, Germany}
\newcommand{\luh}{Leibniz Universit\"at Hannover, 30167, Hannover, Germany}

\author{Pierre Mourier}
\email{pierre.mourier@aei.mpg.de}
\affiliation{Max-Planck-Institut f\"ur Gravitationsphysik (Albert
  Einstein Institute), Callinstr. 38, 30167 Hannover, Germany}
\affiliation{Leibniz Universit\"at Hannover, 30167 Hannover, Germany}

\author{Xisco Jim\'enez Forteza}\email{frjifo@aei.mpg.de}
\affiliation{Max-Planck-Institut f\"ur Gravitationsphysik (Albert
  Einstein Institute), Callinstr. 38, 30167 Hannover, Germany}
\affiliation{Leibniz Universit\"at Hannover, 30167 Hannover, Germany}

\author{Daniel Pook-Kolb}\email{daniel.pook.kolb@aei.mpg.de}
\affiliation{Max-Planck-Institut f\"ur Gravitationsphysik (Albert
  Einstein Institute), Callinstra\ss e 38, 30167 Hannover, Germany}
\affiliation{Leibniz Universit\"at Hannover, 30167 Hannover, Germany}

\author{Badri Krishnan}\email{badri.krishnan@aei.mpg.de}
\affiliation{Max-Planck-Institut f\"ur Gravitationsphysik (Albert
  Einstein Institute), Callinstr. 38, 30167 Hannover, Germany}
\affiliation{Leibniz Universit\"at Hannover, 30167 Hannover, Germany}

\author{Erik Schnetter}\email{eschnetter@perimeterinstitute.ca}
\affiliation{Perimeter Institute for Theoretical Physics, Waterloo, 
  ON N2L 2Y5, Canada}
\affiliation{Department of Physics \& Astronomy, University of Waterloo,
  Waterloo, ON N2L 3G1, Canada}
\affiliation{Center for Computation \& Technology, Louisiana State
  University, Baton Rouge, Louisana 70803, USA}

\date{2021-01-22}

\begin{abstract}

  It is expected that all astrophysical black holes in equilibrium are
  well described by the Kerr solution.  Moreover, any black hole far
  away from equilibrium, such as one initially formed in a compact
  binary merger or by the collapse of a massive star, will eventually
  reach a final equilibrium Kerr state.  At sufficiently late times in
  this process of reaching equilibrium, we expect that the black hole
  is modeled as a perturbation around the final state. The emitted
  gravitational waves will then be damped sinusoids with frequencies
  and damping times given by the quasinormal mode spectrum of the
  final Kerr black hole.  An observational test of this scenario,
  often referred to as black hole spectroscopy, is one of the major
  goals of gravitational wave astronomy.  It was recently suggested
  that the quasinormal mode description including the higher
  overtones might hold even right after the remnant black hole is
  first formed.  At these times, the black hole is expected to be
  highly dynamical and nonlinear effects are likely to be important.
  In this paper we investigate this remarkable scenario in terms of
  the horizon dynamics.  Working with high accuracy simulations of a
  simple configuration, namely the head-on collision of two
  nonspinning black holes with unequal masses, we study the dynamics
  of the final common horizon in terms of its shear and its multipole
  moments. We show that they are indeed well described by a
  superposition of ringdown modes as long as a sufficiently large
  number of higher overtones are included.  This description holds
  even for the highly dynamical final black hole shortly after its
  formation. We discuss the implications and caveats of this result
  for black hole spectroscopy and for our understanding of the
  approach to equilibrium.

\end{abstract}

\maketitle

\section{Introduction}
\label{sec:intro}

The process of binary black hole coalescence, the formation of a
remnant black hole and the associated emission of gravitational waves,
provides a rich arena for tests of general relativity (GR).  The
inspiral regime where we have two distinct black holes inspiralling
into each other is well described by the post-Newtonian
approximation. A useful framework for tests of general relativity in
this regime is provided by the parametrized post-Newtonian framework.
It can be argued however that it is the merger regime, which involves
the formation of the remnant black hole and its approach to
equilibrium, that is the most promising in the search for new physics.
It is during the merger that the nonlinear and nonperturbative
effects of general relativity are most prominent.  Moreover, the
approach of the remnant black hole to equilibrium is closely related
to one of the important predictions of general relativity, namely the
so-called black hole no-hair theorem (see
e.g. \cite{Chrusciel:2012jk,heusler_1996,Cardoso:2016ryw} for reviews
with diverse viewpoints).  The final state of the remnant black hole
in astrophysical situations is predicted to be a Kerr black hole
determined by just two parameters, namely the final mass and angular
momentum.  When the remnant black hole is initially formed, the
spacetime in the vicinity of the horizon is highly dynamical and
nonlinear, and it is responsible for the emitted gravitational
radiation.  In classical general relativity, the horizon itself cannot
emit any radiation.  Rather, it absorbs part of the emitted radiation
to reach equilibrium.  The gravitational wave emission at late times
during this approach to equilibrium is expected to be described by a
superposition of exponentially damped sinusoidal signals, with the
frequency and damping times determined just by the final black hole's
mass and angular momentum
\cite{Vishveshwara:1970zz,Chandrasekhar:1975zza,detweiler:1980gk} (we
can neglect the electric charge for astrophysical black holes).  It is
an important goal of gravitational wave astronomy to verify (or
disprove) this scenario observationally.

Towards this goal, the notion of ``black hole spectroscopy'' has been
proposed \cite{Dreyer:2003bv,Berti:2005ys,Berti:2016lat}.  The basic
idea is straightforward: Given that the ringdown frequencies and
damping times are determined by just two parameters, if we are able to
observe multiple ringdown modes, then the masses and spins inferred
from each mode must be consistent.  This is then potentially a
stringent test of the no-hair theorem; see e.g. \cite{Cardoso:2016ryw}
for a more detailed discussion.  Moreover, the test applies in
principle to any astrophysical process which leads to the formation of
a remnant black hole which approaches equilibrium.  A binary black
hole merger is the obvious target, but it also applies to binary
neutron star mergers or the gravitational collapse of sufficiently
massive stars.  In its original formulation, it was assumed that black
hole spectroscopy should only work once the black hole is sufficiently
close to equilibrium.  Consider for example the remnant black hole
formed from a binary black hole merger.  When the final black hole is
initially formed, it is highly distorted and dynamical, and far from
equilibrium.  There is thus no \emph{a priori} reason why the
perturbatively defined quasinormal mode frequencies should be
associated with the black hole at this point.  This issue of isolating
the perturbative regime where black hole spectroscopy can be applied
is considered \emph{e.g.}  in \cite{Thrane:2017lqn}.

An important recent development was the suggestion that it might in
fact be possible to associate the remnant black hole almost
immediately after merger with quasinormal modes
\cite{giesler2019,isi2019,Okounkova:2020vwu}; see also
\cite{forteza2020,Bhagwat:2019dtm,cook2020}. Given the considerations
mentioned at the end of the previous paragraph, this would seem to be
a very unlikely proposition.  However, as shown in these works, it is
clearly true that it is possible to model the gravitational waveform
immediately after the merger phase as a superposition of quasinormal
modes. For this, it is essential to include the higher ringdown
overtones which had, for the most part, not been included in previous
analyses.  If true, it could greatly improve the prospects of black
hole spectroscopy and would indicate a remarkable simplicity in black
hole mergers.  It is therefore necessary to investigate this scenario
from different perspectives, and one such perspective is in the strong
field region near the black hole horizons.  The goal of this paper is
to investigate whether the dynamics of the remnant black hole horizon
can be described by a superposition of quasinormal modes (including
the higher overtones).

Towards this end, in this paper we carry out a numerical study of the
remnant black hole formed by the head-on collision of two nonspinning
black holes with unequal masses.  This simple configuration, while not
of great astrophysical significance, allows one to obtain very
accurate numerical relativity simulations.  The manifest axisymmetry
of such systems also ensures that there is no ambiguity in the choice
of coordinate systems and that physical gauge invariant quantities can
be extracted in a straightforward manner.  The nonlinearities and
dynamics of general relativity are of course still present: a common
horizon is formed when the two individual black holes get sufficiently
close to each other; it settles to a final Schwarzschild black
hole, and gravitational radiation is emitted in the process.  This
provides us with a simple case where the physical question of interest
can be fruitfully explored without worrying about many of the
complications present in astrophysically realistic situations.  Two
geometrical quantities related to the final black hole are of interest
for our purposes: the angular modes $\sigma_l$ ($l = 2,3,\ldots$) of
the shear $\sigma$ of the outgoing light rays at the horizon, and the
nontrivial mass multipole moments $I_l$ ($l=2,3,\ldots$) of the horizon.
We calculate $\sigma_l$ and $I_l$ as functions of time and we attempt
to describe each of them by a superposition of quasinormal modes.  We
find that, indeed, including the higher overtones can allow for
obtaining excellent fits for $\sigma_l(t)$ and $I_l(t)$ starting
almost immediately after the merger.  The high precision of our
numerical simulation allows us to include angular modes with
$2\leq l \leq 12$, \emph{i.e.}, a total of 11 independent time series, and we
show that all of these modes are described by combinations of quas-normal modes
provided higher overtones are included.
Furthermore, while the multipole
moments $I_l$ are not fully independent of the shear as we shall see,
they do provide yet another 11 functions for testing the hypothesis.
Similar studies of the gravitational waves at infinity,
e.g. \cite{giesler2019}, typically consider only the dominant $l=|m|=2$ wave mode,
with a recent extension to a joint analysis of the $|m|=2$, $l=2,3,4$ wave modes
(which are coupled, due to spheroidal/spherical mode mixing
in the Kerr final state considered, unlike in our more symmetric case) \cite{cook2020}.
Thus, this work represents a significant additional evidence compared
to previous work in the literature.

The reader might legitimately ask: i) why should the behavior of
$\sigma_l$ and $I_l$ at the horizon have anything to do with the
actual observable quantity, namely the outgoing gravitational waves
which could be observed by gravitational wave detectors? Are the
horizons not causally disconnected from the outside observers and thus
observationally irrelevant?  ii) Even if one finds these calculations
to be of interest, and even though we are careful in extracting gauge
invariant quantities, is the apparent horizon not itself dependent on
which time slicing the numerical simulation uses?  How can we
guarantee that the results would not be entirely different with a
different choice of the time coordinate?  Let us address these in
turn.

For question i), we point out the remarkable correlations that exist
between the outgoing radiation seen by a far away observer, and the
in-falling radiation that could be seen by a hypothetical observer
living near the horizon.  Even though these two observers are not in
causal contact with each other, the gravitational radiation they would
see comes from the same source, namely the nonlinear and time
dependent gravitational field in the vicinity of the binary system
\cite{Jaramillo:2011rf,Jaramillo:2011re,Jaramillo:2012rr,Rezzolla:2010df,Gupta:2018znn}.
It is thus not a surprise that both observers will see qualitatively
similar features. In fact it was shown in \cite{Prasad:2020xgr} that
the two observations agree qualitatively.  The present study can be
viewed as further evidence of these correlations.  Thus, by studying
the behavior of the horizon, we can learn something about the outgoing
radiation (and vice versa).

Regarding ii), it is likely true that one could
have
chosen a particularly ``bad'' slicing and time coordinate which could
have obscured any of the correlations mentioned above.  First, we
could have made a different choice of spatial Cauchy surfaces for the
numerical evolution which would generically lead to different
dynamical horizons.  Even though there are known constraints on how
different the dynamical horizons can be \cite{Ashtekar:2005ez}, it is
possible in principle to choose spatial slices such that the horizons
could be extremely distorted \cite{eardley:1997hk,Bendov:2006vw}.
However, we are not aware of any numerical simulations which use, or
can practically use, such extreme choices.  Second, even within a
given choice of slicing, there is still the possibility of choosing a
different time parameter adapted to the slicing,
$t \mapsto t^\prime = F(t)$. This would change the functional
dependence of any relevant function of time $f(t)$ into
$f^\prime(t^\prime) = f(F^{-1}(t'))$.  We shall make no attempt to do
any such reparametrizations in this paper, and we shall simply work
with the slicing and time coordinate used in the simulation.

What is significant is that our results show that there is at least
one choice of slicing and of an adapted time coordinate, which happens
to be a widely used one, for which the correlations are manifestly
present. Specifically, we employ the
${1+\log}$ slicing, along with a $\Gamma$-driver shift condition
\cite{Alcubierre:2000xu,Alcubierre:2002kk}. These gauge conditions also
set the time parameter and spatial coordinates of the simulation. An important property of
these gauge conditions is that they are ``symmetry seeking'',
\emph{i.e.}, they attempt to find a timelike Killing vector if there is one,
and thus define reasonable local observers.  

The remainder of the paper is structured as follows.  A brief summary
of the basic quantities we calculate and study is provided in
Sec.~\ref{sec:basics}.  This section defines and identifies the
horizon shear and multipole moments as quantities of interest.
In the following sections we describe the methods
used and the results of attempts to fit these quantities
using the ringdown frequencies and damping times associated with the
final black hole.  The fitting procedure is described in
Sec.~\ref{sec:fitting} and applied to the shear and multipole moments
in Sec.~\ref{sec:results}.  Sec.~\ref{sec:overtones-discussion}
discusses the implications of these results and whether it is possible
to conclude, and in what sense, whether overtones are really
associated with the highly distorted remnant black hole immediately
after its formation.  Sec.~\ref{sec:conclusion}
concludes with a summary and suggestions for future work.

\section{Basic notions}
\label{sec:basics}

There are two main aspects relevant to our study: i) the quasinormal
modes (QNMs) of a black hole, which are usually defined within the
context of black hole perturbation theory, and ii) the
nonperturbative study of quasilocal black hole horizons.  This
section briefly summarizes the basic notions and results for both of
these aspects.

\subsection{Quasinormal modes}
\label{subsec:qnmbasics}

The metric perturbations of a Schwarzschild black hole (which is the final geometry
relevant for our study), for both polar
and axial perturbations, can be combined into scalar functions $\psi$
which satisfy equations of the form
\cite{Regge:1957td,Zerilli:1970se,chandrasekhar:1985kt}
\begin{equation}
  \frac{d^2\psi}{dr_\star^2} + \nu^2\psi = V_{\pm}\psi\,.
\end{equation}
Here, as usual, $r_\star = r + 2M\log(r/2M - 1)$ with $M$ being the
black hole mass, and $r$ is the usual Schwarzschild areal coordinate.
The potentials $V_\pm$ for the polar and axial perturbations are
functions of $r$ and they depend on $M$ and on the mode index $l$.
The potentials also differ depending on the nature of the
perturbation, and we shall here be concerned almost exclusively with
spin-2 fields (see Sec.~\ref{subsubsec:observables}).

Quasinormal modes are obtained by imposing outgoing boundary
conditions at both infinity, and at the horizon.  Only a discrete set
of (complex) values of the frequency $\nu$ allow for these dissipative
boundary conditions, and these are labeled by the integers $(l,m,n)$ where
$(l,m)$ are the usual angular mode indices in a decomposition into
spherical harmonics\footnote{%
  In general, \emph{i.e.}, for perturbations of a Kerr black hole, the
  quasinormal modes are obtained from a decomposition into spheroidal
  harmonics. The latter equivalently reduce to spherical harmonics for
  the Schwarzschild case considered here due to its spherical
  symmetry.%
}, and $n=0,1,2,\ldots$ is the overtone index.  See
\cite{leaver:1985ax} for an analytic method for calculating this
spectrum and \cite{berti-webpage,berti:2009kk} for a compilation of the values in
different situations. We also show a sample of $m=0$ quasinormal mode frequencies
(real and imaginary parts) for a Schwarzschild black hole of unit mass
below in Table~\ref{tab:QNMmodes}. Finally, there are several interesting
mathematical and numerical issues related, in particular, to the
non-self-adjoint nature of the problem.  For example, of great
potential interest is the recent suggestion that the higher overtones
might in fact be unstable \cite{Jaramillo:2020tuu}.  Similarly, the
issue of the completeness of the quasinormal modes is also of great
interest; see
e.g. \cite{Ansorg:2016ztf,PanossoMacedo:2018hab,Beyer:1998nu}.

\subsection{Nonperturbative framework for studying quasilocal horizons}
\label{sec:quasilocal-horizons}

\subsubsection{Horizon definition}

The study of horizons here is based on the notions of marginally
trapped surfaces and dynamical horizons (see
e.g. \cite{ashtekar:2004cn,Booth:2005qc,krishnan:2013saa,Gourgoulhon:2005ng,Faraoni:2015pmn,Hayward:2000ca}
for reviews).  Here we shall only briefly summarize the basic notions
required for our purposes.  The first is that of a marginally outer
trapped surface (MOTS).  This is a closed spacelike $2$-surface
$\mathcal{S}$ whose outer null-normal $\ell^a$ has vanishing
expansion:
\begin{equation}
  \Theta_{(\ell)} := q^{ab}\nabla_a\ell_b = 0\,.
\end{equation}
Here $q^{ab}$ is the intrinsic metric on $\mathcal{S}$.  MOTSs are
closely related to trapped surfaces with negative expansions for
both the outgoing and ingoing null normals, and the significance of these
notions goes back to the singularity theorems
\cite{Penrose:1964wq,Hawking:1969sw}. Their presence implies the
existence of a spacetime singularity to its future, and thus indicates
the presence of a black hole.  Well developed methods exist to locate
MOTSs in numerical relativity (NR) simulations
\cite{Thornburg:2006zb}.  Here we shall employ the method developed in
\cite{pook-kolb:2018igu,PhysRevD.100.084044} and available from
\cite{pook_kolb_daniel_2020_3885191},
which in turn uses libraries described in
\cite{erik_schnetter_2019_3258858,
      mike_boyle_2018_1221354,
      2020scipy-nmeth,
      van_der_walt_numpy,
      mpmath,
      meurer2017sympy,
      hunter:2007,
      michael_droettboom_2018_1202077}.

As a MOTS evolves in time, it traces out a $2+1$-dimensional world-tube
$\mathcal{H}$ which we shall refer to as a dynamical horizon.  Several
mathematical and physical properties of $\mathcal{H}$ are known and
summarized in the review articles referred to above.  The behavior of
dynamical horizons in black hole mergers has been studied in detail
recently
\cite{PhysRevLett.123.171102,PhysRevD.100.084044,pook-kolb2020I,pook-kolb2020II}.

\subsubsection{Setup and numerical simulation employed}
\label{subsec:numerical-setup}

The configuration we consider here is the head-on merger of two
nonspinning black holes initially at rest.  The initial data is the
time symmetric Brill-Lindquist puncture data \cite{Brill:1963yv}.
This data describes a spatial slice $\Sigma$ with vanishing
extrinsic curvature $K_{ab}=0$, and conformally flat 3-metric
$h_{ab} = \Phi^4\delta_{ab}$.  The conformal factor $\Phi$ is a
harmonic function on three-dimensional Euclidean space with two points
removed (the punctures).  At a point $\mathbf{x}$,
\begin{equation}
  \Phi(\mathbf{x}) = 1 + \frac{m_1}{2r_1} + \frac{m_2}{2r_2}\,,
\end{equation}
where $r_{1,2}$ are the respective distances from $\mathbf{x}$ to each of the two
punctures, and $m_{1,2}$ are known as the bare masses of the two
punctures.  We will note the total \ADM mass as $M = m_1 + m_2$. We study here a particular configuration with
$m_2/m_1= 1.6$. The ADM mass has a value of $1.3$ in the code units used,
but we instead set it as the mass unit here, \emph{i.e.}, $M=1$ in this work.

The simulations are carried out based on the \BSSN formulation of the
Einstein equations using the {\scshape Einstein Toolkit}
\cite{Loffler:2011ay,EinsteinToolkit:web}, with the initial data being
generated by {\scshape TwoPunctures} \cite{Ansorg:2004ds}.  We evolve the
spacetime using an axisymmetric version of {\scshape McLachlan}
\cite{Brown:2008sb}, which uses {\scshape Kranc}
\cite{Husa:2004ip,Kranc:web} to generate efficient C++ code.  As
mentioned earlier, our gauge conditions use a $1+\log$ slicing and a
$\Gamma$-driver shift condition
\cite{Alcubierre:2000xu,Alcubierre:2002kk}.  Further details of our
simulation method are described in \cite{PhysRevD.100.084044}.  The
results presented in the present paper use data obtained from a
simulation with a spatial grid resolution of ${\rm res} = 240$.
Additional simulations with resolutions of ${\rm res} = 60$,
$120, 180$, and restricted simulations with higher resolutions of
${\rm res} = 480$, $960$, have been used to ensure convergence of our
results\footnote{%
  The convergence of the results is already achieved at
  ${\rm res} = 240$~\cite{PhysRevD.100.084044}. The two extra datasets
  with ${\rm res} = 480$, $960$ have been produced to further test the
  dependence of the numerical error with the discretization scheme,
  which is relevant for our definition of the NR error in
  Sec.~\ref{sec:results}. Due to the high computational cost involved,
  their total simulation times have been reduced to $t_f^{(480)} \simeq 15M$ and
  $t_f^{(960)} \simeq 5 M$ respectively, thus being too short for producing
  accurate fits.%
}.
We do not use mesh refinement and instead choose our numerical domain
large enough to ensure that boundary effects do not reach the horizons
up to the final time of $t_f \simeq 38.5\,M$ of the
simulations.

In the resulting spacetime, we initially have two disjoint MOTSs
$\mathcal{S}_{1,2}$.
As the time evolution proceeds, $\mathcal{S}_1$
and $\mathcal{S}_2$ approach each other, touch at a particular time
labeled $t_{\rm touch}$,
and then go through each other after that.
Sometime before $t_{\rm touch}$, at a time labeled
$t_{\mathrm{bifurcate}} \simeq 1.06 M$, a common horizon forms
  and immediately bifurcates into two MOTSs representing an
outer and an inner branch $\mathcal{S}_{out}$ and $\mathcal{S}_{in}$
respectively.  $\mathcal{S}_{in}$ moves inwards, becomes increasingly
distorted and eventually merges with
$\mathcal{S}_1\bigcup\mathcal{S}_2$ at $t_{\rm touch}$, and then
develops self-intersections.  The focus of this paper is not any of
these phenomena, but rather the behavior of $\mathcal{S}_{out}$ which
moves outwards and loses its distortions as it approaches its final
state as that of a spherically symmetric Schwarzschild black hole.  We
shall in particular look at two particular quantities on
$\mathcal{S}_{out}$ as functions of time, namely the shear $\sigma$ of
the outward null normal $\ell^a$ and the mass multipoles of
$\mathcal{S}_{out}$.  In the remainder of this section, we shall
define these quantities and explain why they are of interest.

\subsubsection{Observables on the outer common horizon}
\label{subsubsec:observables}

We begin with the definition of the shear.  Here it will be convenient
to introduce a complex basis for tangent vectors on a MOTS:
$m^a$ and $\bar{m}^a$, that satisfy $m\cdot \bar{m} = 1$ and
$m\cdot m = 0$.  Then, the shear of the outgoing null normal is
defined as
\begin{equation}
  \sigma = m^am^b\nabla_a\ell_b\,.  
\end{equation}  
Such a complex basis is determined up to a spin rotation freedom
$m\rightarrow e^{\iota \psi}m$.  Under this transformation, the shear
transforms as $\sigma \rightarrow e^{2 \iota \psi}\sigma$, thus $\sigma$ is
said to have spin weight $+2$.  This means that $\sigma$ can be expanded
in angular modes using spin-weighted spherical harmonics ${}_{2 \!}Y_{lm}(\theta,\phi)$ of spin weight $+2$.

There still remains
the question of whether there is a preferred choice of angular
coordinates $(\theta,\phi)$; we will end up with different mode
decompositions for different choices.  The general solution to this is
given in \cite{Ashtekar:2013qta}. (In the present case, since we have
manifest axial symmetry, a simpler approach suffices.)
On a surface
$\mathcal{S}$ of spherical topology equipped with an axial symmetry
$\varphi^a$, we can introduce preferred angular coordinates
$(\theta,\phi)$.  First, we assume that $\varphi^a$ vanishes at only
two points, which are taken to be the poles.
On the integral curves
of $\varphi^a$, take $\phi$ to be the affine parameter along
$\varphi^a$, normalized to lie in the range $0\leq \phi < 2\pi$.
One of the meridians, \emph{i.e.}, the lines joining both poles
and everywhere orthogonal to $\varphi^a$, can be arbitrarily selected.
The intersection of this meridian with each integral curve of $\varphi^a$
then defines the point on that curve where $\phi$ is set to zero.
The other coordinate $\theta$, is defined via $\zeta = \cos\theta$
according to
\begin{equation}
  D_a\zeta = \frac{4\pi}{A_{\mathcal{S}}}\widetilde{\epsilon}_{ba}\varphi^b\,,\quad \oint_{\mathcal{S}}\zeta\, \mathrm{d}A = 0\,.
\end{equation}
Here $A_{\mathcal{S}}$ is the area of $\mathcal{S}$,
$\widetilde{\epsilon}_{ab}$ is the volume $2$-form, and $D_a$ is the
covariant derivative compatible with $q_{ab}$. The first equation
ensures that $\varphi^aD_a\zeta = 0$. Hence, $\zeta$ is constant on
each integral curve of $\varphi^a$, and the meridians are integral
curves of $D^a \zeta$.  The second equation fixes the freedom to add
an additive constant to $\zeta$ in the first equation.  With these
choices, it is shown in \cite{Ashtekar:2004gp} that the metric
$q_{ab}$ is written as
\begin{equation}
\label{eq:canonical-metric}
q_{ab} = R_{\mathcal{S}}^2\left(\frac{\partial_a\zeta \partial_b\zeta}{F(\zeta)} + F(\zeta) \, \partial_a\phi\partial_b\phi \right) \,,
\end{equation}
where $R_{\mathcal{S}} := \sqrt{A_{\mathcal{S}}/4\pi}$ and
\begin{equation}
  F(\zeta) = \frac{4\pi\varphi_a\varphi^a}{A_{\mathcal{S}}}\,.
\end{equation}
It can be shown that $-1<\zeta<+1$, and it goes from $+1$ to $-1$ as
we go from one pole to the other.  Therefore we can set
$\cos\theta=\zeta$ with $0 < \theta < \pi$ (and extend it to
$\theta = 0$ or $\pi$ at the poles).

We have thus specified $(\theta,\phi)$ on $\mathcal{S}$ (up to a rigid
rotation by adding a constant to $\phi$ corresponding to choosing the
$\phi=0$ meridian).  A suitable choice for $m^a$ is given by the
following form for its dual $1$-form $\underline{m}$:
\begin{equation}
  \label{eq:m}
  \underline{m} = \frac{R_{\mathcal{S}}}{\sqrt{2}}\left(\frac{\mathrm{d}\zeta}{\sqrt{F}} + \iota \sqrt{F} \, \mathrm{d}\phi\right) \,.
\end{equation}
We can now expand $\sigma$ as
\begin{equation}
  \sigma = \sum_l  {}_{2 \!}Y_{l 0}(\theta,\phi)\, \sigma_l \, .
\end{equation}
We take only the $(l,0)$ modes because of the manifest
axisymmetry.
\revision{This symmetry and our specific choice for $m^a$ also imply here
that $\sigma$ and the $\sigma_l$ are real.}
Under time evolution, the mode amplitudes $\sigma_l$
will \revision{then be real-valued}
functions of time that we aim to model with a combination of damped
sinusoids.

The importance of $\sigma$ lies in the fact that the shear, or more
precisely $|\sigma|^2$, yields the dominant part of the energy flux
in-falling into the black hole \cite{Ashtekar:2002ag,Ashtekar:2003hk}.
Based on the discussion in the introduction, we expect the energy
fluxes across the horizon to be highly correlated with the outgoing
radiation which is determined by the $|\mathcal{N}|^2$ with
$\mathcal{N}$ being the News function \cite{bondi:1962px}.  Thus one
would expect $\sigma$ to be closely correlated with
$\mathcal{N}$. This has been shown to be indeed the case for the
inspiral regime \cite{Prasad:2020xgr}.  Here our focus is on the
postmerger regime.  The outgoing radiation is represented by the two
polarizations $h_{+,\times}$, or equivalently by a complex combination
$h = h_+ + \iota h_\times$. The News function is given by
$\mathcal{N} = \dot{h}$. Thus, when $h$ is a combination of damped
sinusoids then so is $\mathcal{N}$ and thus, if the proposed
correlations mentioned above do exist, the same should be true for
$\sigma$.  Thus, if the higher overtones appear in $h$, then they
should also appear in the shear $\sigma$, and vice versa.

Let us now turn to the multipole moments.  As for any mass or charge
distribution, it is possible to define suitable mass and current
multipoles for black hole horizons \cite{Ashtekar:2004gp}.  For
nonspinning configurations where the individual black holes are
nonspinning and the orbital angular momentum is also vanishing, as in
our case, we only need to consider the mass multipoles. These are
moments $I_l$ of the intrinsic scalar curvature $\mathcal{R}$ of
$\mathcal{S}$ calculated from Eq.~\eqref{eq:canonical-metric}:
\begin{equation}
\label{eq:mass-moment-l}
  I_l = \frac{1}{4}\oint_{\mathcal{S}}\mathcal{R}\, Y_{l,0}(\zeta)\, \mathrm{d}A\,.
\end{equation}
Just as for the shear, we calculate $I_l$ as functions of time and
look for the presence of ringdown modes therein. We will only consider
$l \geq 2$ since $I_0$ is constant as a topological invariant (here
$I_0 = \sqrt{\pi}$) and $I_1$ vanishes at all times due to the
symmetries of the angular coordinates used
\cite{Ashtekar:2004gp,Ashtekar:2013qta}.

\begin{figure*}
  \centering    
  \includegraphics[width=\linewidth]{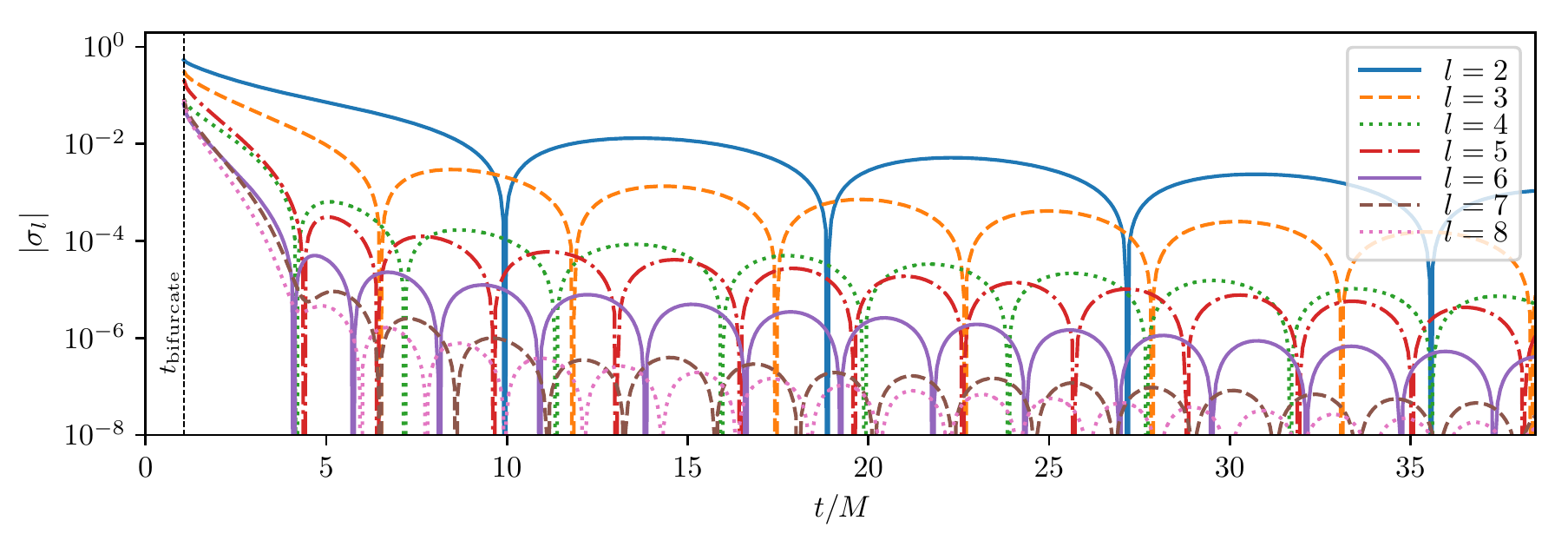}
  \includegraphics[width=\linewidth]{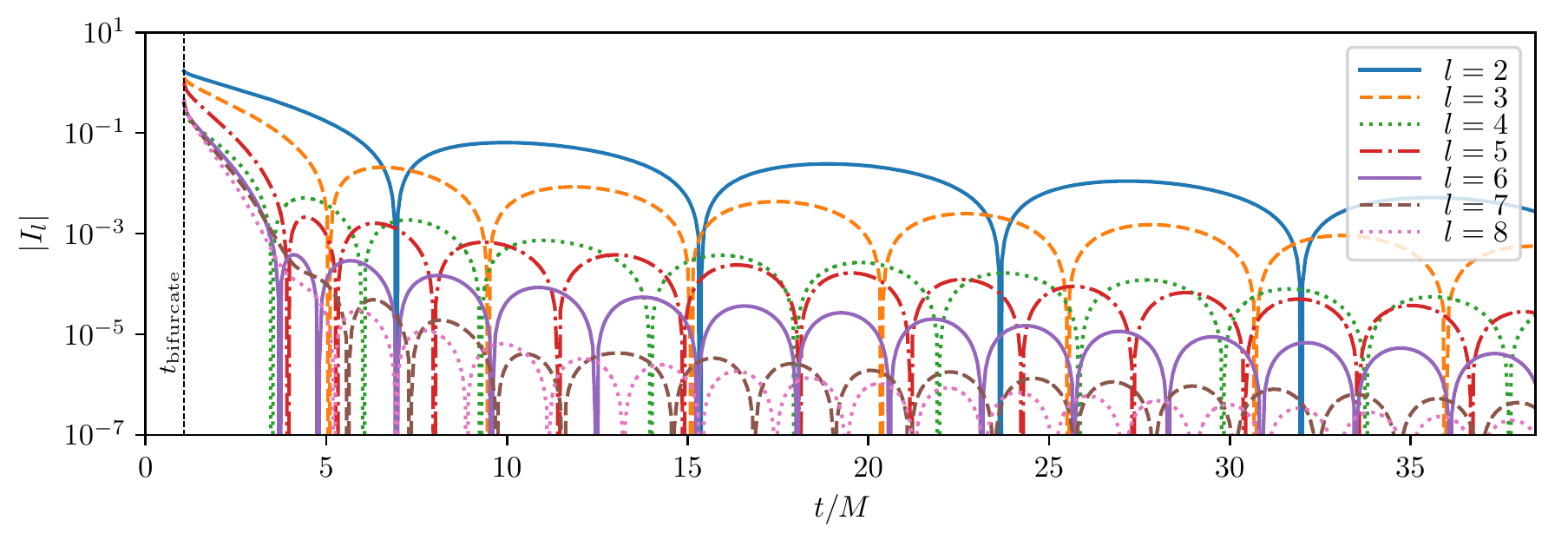}
  \caption{Shear modes (top) and mass multipoles (bottom) for
    $2 \leq l \leq 8$ for $\mathcal{S}_{out}$ as a function of the
    simulation time.  The multipoles $I_l$ for $l=0,1$ are constant
    and not shown. See Sec.~\ref{subsubsec:observables} for further discussion.}
  \label{fig:shear-multipole-common}
\end{figure*}

\begin{figure*}
  \centering
  \includegraphics[width=\linewidth]{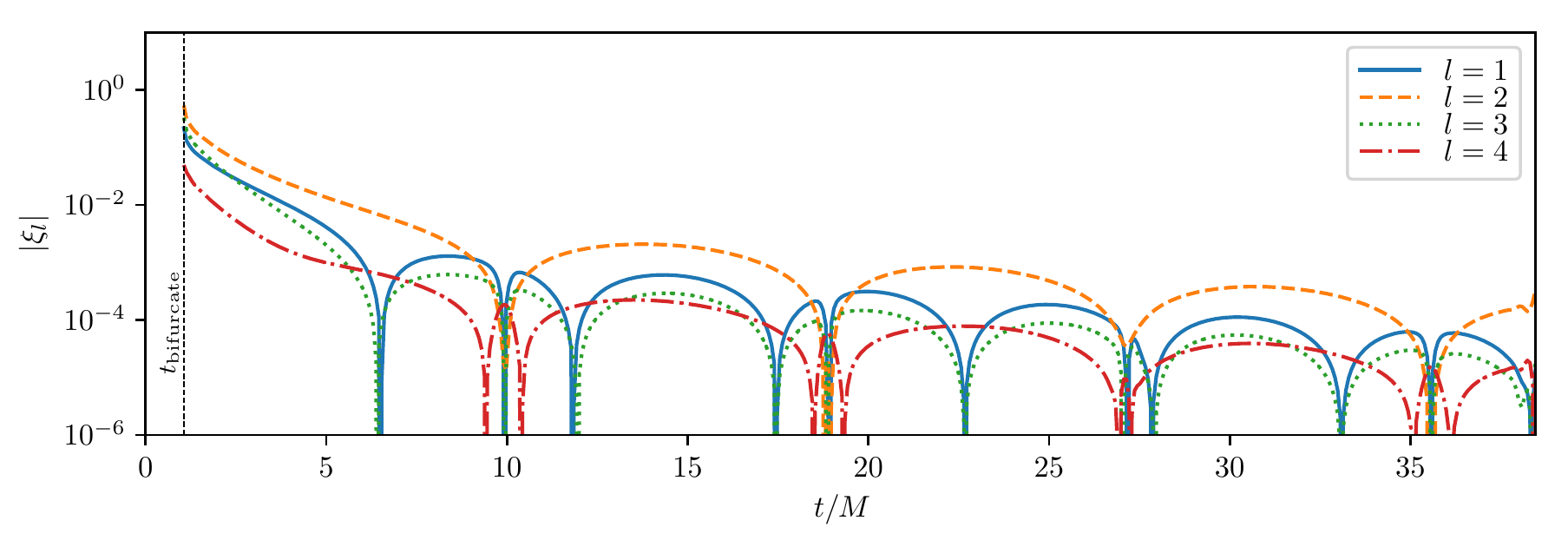}
  \caption{Vector modes $\xi_l$ on the outer common horizon for
    $1 \leq l \leq 4$, as a function of the simulation time $t$.}
  \label{fig:xi-modes}
\end{figure*}

Preliminary investigations of $\sigma_l(t)$ and $I_l(t)$ are given in
\cite{pook-kolb2020II}.  Given that we will analyze essentially the
same dataset as in \cite{pook-kolb2020II} (here obtained from
performing the same simulation with a higher resolution), it will be
useful to summarize the results.  We begin with plots of $\sigma_l$
and $I_l$ as functions of time, shown in
Fig.~\ref{fig:shear-multipole-common}.  The behavior of the modes
$\sigma_l(t)$ and $I_l(t)$ all have similar qualitative behaviors: a
rapid initial decay followed by a slower decay with oscillations. The
higher the $l$, the more rapid the initial decay.  At late times on
the other hand, the damping rates of different modes seem very
similar, but the higher modes have higher oscillation frequencies.
While we shall not discuss it in this paper, we mention in passing
that the in-falling energy flux also has a contribution from a vector
field $\xi^a$ \cite{Ashtekar:2002ag,Ashtekar:2003hk} (denoted
$\zeta^a$ in these references).  As for the shear, we can perform a
mode decomposition for the vector field as well, but using
spin-weight-1 spherical harmonics. The time dependence of these modes
$\xi_l$, $l \geq 1$, is shown in Fig.~\ref{fig:xi-modes}. For
$l \neq 2$, it is evidently more complex than the shear.  This vector
contribution is however subdominant, and we shall study this in
detail elsewhere.

\begin{figure}    
  \includegraphics[width=\columnwidth]{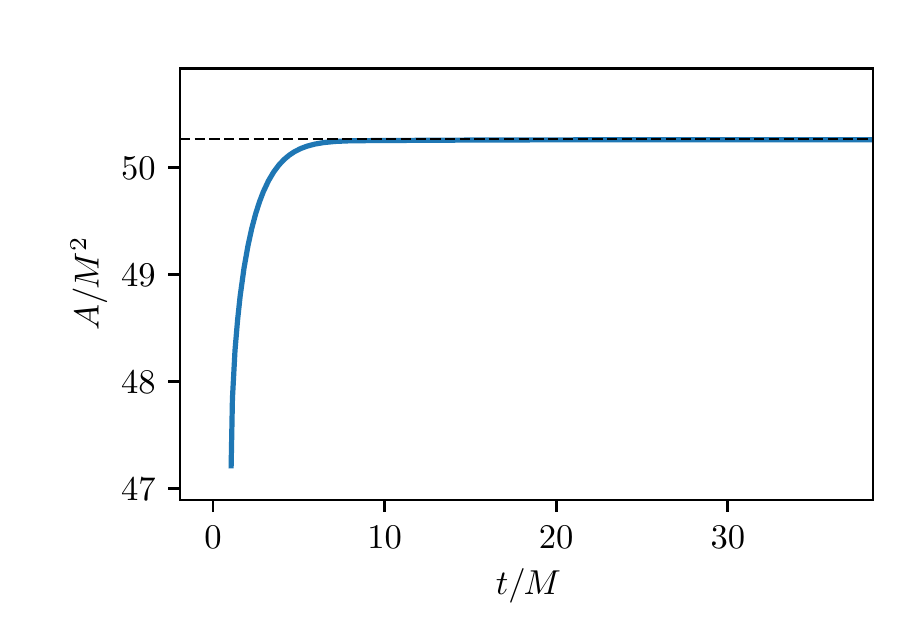} 
  \caption{Area of the outer common horizon as a function of simulation
    time. The final area shown here corresponds to a horizon mass of
    $\sim (1 - 7\cdot 10^{-5})\,M$.
  }
  \label{fig:bl5-area}
\end{figure}

It is also useful to note that the final black hole horizon is
not in equilibrium at early times just after it is formed.
An easy
way to see this is by looking at the area growth of the final black
hole.  Fig.~\ref{fig:bl5-area} shows the area of the final black hole
as a function of time starting from when it is initially formed.  We
see a rapid initial increase showing unambiguously the dynamical
nature of the black hole in this regime.  The analysis of
\cite{pook-kolb2020II} shows, using many different criteria all of
which give approximately the same answer, that the black hole can be
considered close to equilibrium after $\sim 10M$ after its formation.

It was shown in \cite{pook-kolb2020II} that the late-time behavior, on
the other hand, is consistent with the principal (fundamental)
quasinormal mode.  It was also shown there that at early times after
the merger, the observed high decay rates were at first glance not
consistent with any of the higher overtones considered separately.
However, this early-time postmerger behavior analysis was rather
simplistic.  Here we perform a more sophisticated analysis by
considering the entire time series of the shear and multipoles
(instead of breaking it up into early and late portions), and model it
with a superposition of quasinormal modes including the higher
overtones.

We conclude this section with a brief discussion of the relation between
the multipoles and the shear.  The shear is a spin-weight-2 field,
hence it is expanded in spin-weighted spherical harmonics, and it is
natural to expect its decay rates to follow the spin-2 quasinormal
modes.  The scalar 2-curvature of the horizon $\mathcal{R}$,
on the other
hand, is a spin-weight-0 field which is why Eq.~\eqref{eq:mass-moment-l}
uses, in effect, the spin-weight-0 spherical harmonic in defining its
moments. Should we then expect the decay rates of $\mathcal{R}$ to
follow the spin-0 quasinormal modes?  If not, then what should we
expect?  To answer this question, using the Gauss-Codazzi relations
applied to 2-surfaces, we can relate $\mathcal{R}$ to the spacetime
Riemann curvature:
\begin{equation}
  \mathcal{R} = 4 \mathrm{Re}[\Psi_2] - q^{ac}q^{bd}\sigma^{(\ell)}_{ab}\sigma^{(n)}_{cd}\,.
\end{equation}
Specifically in this paragraph we denote the shear of $\ell^a$
(elsewhere simply denoted $\sigma$) with a
superscript ${}^{(\ell)}$ in order to distinguish it from the shear $\sigma^{(n)}$
of the ingoing null normal.  $\Psi_2$ is a component of the Weyl
tensor and $\mathrm{Re}[\Psi_2]$ is its real part.  We see that $\mathcal{R}$
depends linearly on $\sigma^{(\ell)}$.  The shear $\sigma^{(n)}$ is
not directly associated with the in-falling radiation, and $\Psi_2$ is
also not associated with the radiative part of the gravitational
field.  Thus, $\sigma^{(\ell)}$ controls the time dependence of
$\mathcal{R}$, and it is reasonable to expect the decay rates of
$\mathcal{R}$ to follow the spin-2 quasinormal ringdown modes.  It is
also useful to note that for a black hole in equilibrium when there is
no in-falling radiation (formally modeled as an isolated horizon
\cite{Ashtekar:1998sp,Ashtekar:1999yj,Ashtekar:2001is,Ashtekar:2001jb,Lewandowski:1999zs,Lewandowski:2018khe,Korzynski:2004gr,Ashtekar:2000hw,Krishnan:2012bt,Booth:2001gx,Booth:2012xm}),
the shear $\sigma^{(\ell)}$ vanishes, $\mathcal{R}$ is
time-independent and $\mathcal{R} = 4\mathrm{Re}[\Psi_2]$.

\section{Overtone models and fitting procedure}
\label{sec:fitting}

In this section we introduce the basic concepts and framework commonly
used for the modeling of ringdown-type waveforms (including the outer
horizon shear modes and multipoles in our case) with overtones and we
the discuss the statistical tools used to fit such models to NR data.

\subsection{The overtone model}
\label{subsec:fitting}


At late times, we can decompose a spin-weight-$s$ field ${}_{s}X$
propagating in a Schwarzschild (or more generally Kerr) background
as a sum of damped sinusoids, namely,
\begin{equation}
\label{eq:rdmodel}
\revision{   _sX = \sum_{l \, \geq \, |s|, \, m, \, n} {\mathcal{A}}^{\pm}_{lmn} \, \exp \left[- \iota \omega_{lmn}^\pm(t-t_r) -\frac{t-t_r}{\tau_{lmn}^{\pm}} \right] \, _{s \!}\tilde{Y}_{lm}\, . }
\end{equation}
Here, the $(l,m)$ indices describe the angular decomposition of the
modes \revision{(with $m = -l, \dots, l$)}, and \revision{${}_{s\!}\tilde{Y}_{lm}$} are the spin-weighted spheroidal harmonics\footnote{%
\revision{In particular, the shear as defined above, following the usual convention, has spin weight +2 and is thus decomposed in spin-weight +2 harmonics.  One could as well have worked with a spin-weight -2 field by using the complex conjugate
of the shear instead, which would then have been expanded in spin-weight -2 harmonics.  Both types of fields have the same QNM spectrum in a Kerr (or Schwarzschild) background.  (See, \emph{e.g.}, Eqs.~(3.29) and (3.30) in \cite{teukolsky3} relating the Weyl tensor
components $\Psi_4$ and $\Psi_0$, which respectively have spin
weight -2 and  +2, showing that both variables are isospectral.)}
};
for a perturbed Schwarzschild black hole as in our case, they reduce to the spin-weighted
\revision{\emph{spherical} harmonics ${}_{s\!} Y_{lm}$.  $n = 0, 1, 2, \dots$} accounts for the $n$-tone excitations of a
given $(l, m)$ mode, with $n = 0$ being the fundamental ``tone'' and
$n = 1, 2, \dots$ corresponding to overtones, and $t_r$ is a suitable reference
time where the linear perturbation theory is expected to describe the
dynamics
accurately~\cite{bhagwat:2017tkm,Bhagwat:2019dtm,forteza2020,Ota:2019bzl}. 

If linear perturbation theory applies, the quasinormal mode
frequencies and damping times\footnote{%
  Alternatively, one can rewrite the exponential factors
  \revision{$\exp[- \iota \omega_{lmn}^{\pm} (t-t_r) -(t-t_r)/\tau_{lmn}^{\pm})]$} in
  Eq.~\eqref{eq:rdmodel} for each $(l,m,n)$ component as an
  \revision{$\exp[ - \iota \nu_{lmn}^{\pm}$ $(t-t_r)]$} with a complex frequency
  \revision{$\nu_{lmn}^{\pm}$} of which \revision{$\omega_{lmn}^{\pm}$} is the real part and the damping
  rate \revision{$1/\tau_{lmn}^{\pm}$} is (up to a sign change) the imaginary part.
} \revision{$\omega^{\pm}_{lmn}$ and $\tau^{\pm}_{lmn}$} are solely determined by the black
hole's final mass and angular momentum.
\revision{For a given choice of the $(l,m,n)$ indices,
one finds two families of solutions, those with $\omega_{lmn}^{+}>0$
and those with $\omega_{lmn}^{-} < 0$, corresponding to the co-rotating
and counter-rotating modes respectively,
with their associated damping times $\tau_{lmn}^{\pm}$ and complex amplitudes
$\mathcal{A}^{\pm}_{lmn}$~\cite{Berti:2005ys,leaver:1985ax,berti:2009kk,kokkotas:1999bd,ferrari:1984zz,cook2020,forteza2020}.
Note that for the  Schwarzschild case,
${\omega_{lmn}^{+}=-\omega_{lmn}^{-}}$
and ${\tau_{lmn}^{+}=\tau_{lmn}^{-}}$ for any $m$;
this also holds in the general Kerr case if $m=0$.
Hence, in these cases, the (absolute) values
of the frequencies and damping times
are independent of the family (co- or counter-rotating) of modes considered.
}

Throughout this work, we will
set $t_r$ to the time value used for the late-time fits
in~\cite{pook-kolb2020II}, that is in the units used in the present
work, $t_r / M = 20/1.3 \simeq 15.4$. The amplitudes
\revision{at $t_r$, ${\cal A}_{lmn}^{\pm}$,}
are unknown
complex numbers that only depend on the perturbation conditions set up
during the inspiral-plunge-merger phase of the binary black hole
evolution. \revision{Hence, they are fully determined by the initial parameters
of the binary prior to the merger --- in our head-on, nonspinning case,
the mass ratio of the two colliding black holes and their relative boost
at a given separation.}

To allow for deviations on the complex frequencies from the QNM
values, Eq.~\eqref{eq:rdmodel} may be replaced by
\revision{
\begin{align}
\label{eq:rdmodel_ab}
   {}_sX = \sum_{l \, \geq \, |s|, \, m, \, n} {\cal A}_{lmn}^{\pm} \,
   \exp \left[ - \iota \omega_{lmn}^{\pm}(1+\alpha_{lmn}^{\pm}) (t-t_r) \right] \times \nonumber \\
   \times \exp \left[- \frac{t-t_r}{\tau_{lmn}^{\pm}(1+\beta_{lmn}^{\pm})} \right] \, _{s \!} \tilde{Y}_{lm}\,,
\end{align}
}
where \revision{$\alpha_{lmn}^{\pm}$ and $\beta_{lmn}^{\pm}$} are two sets of perturbation
parameters \revision{for each co- or counter-rotating mode}.  These will measure the deviations to the QNM spectrum (as
predicted by perturbation theory within GR), while the latter spectrum
is recovered for \revision{${\alpha_{lmn}^{\pm} =\beta_{lmn}^{\pm}=0}$}. To perform black hole
spectroscopy, one shall require a) that the posterior distributions of
\revision{$\alpha_{lmn}^{\pm}$ and $\beta_{lmn}^{\pm}$} are consistent with zero and b) that
the frequency values can be resolved to a given ${n\sigma}$ credible
value~\cite{Bhagwat:2019dtm}. The latter is technically difficult due
to the low sparsity of the QNM real frequencies.  For instance, for
the $(l=2,m=0)$ QNM of a ${s=2}$ field in a Schwarzschild spacetime,
the real frequencies of the fundamental mode and first overtone
only differ by \revision{${1-\omega_{201}^{\pm}/\omega_{200}^{\pm} \simeq 7\%}$}
(see the corresponding frequency values in Table~\ref{tab:QNMmodes}, left panel),
making the separate resolution of the two tone frequencies
a challenging task~\cite{forteza2020,Bhagwat:2019dtm}.
An attempt to estimate the overtone frequencies by means of the Bayesian framework
on GW150914 data was partially tackled in~\cite{isi2019}
by performing a parameter estimation on a reduced parameter space.
Other recent studies have provided estimates on the QNM parameters by performing
fits to NR data~\cite{giesler2019,forteza2020,Bhagwat:2019dtm,cook2020,pook-kolb2020II}.
Fitting the data circumvents the extensive exploration of the parameter space
by estimating the physical parameters from maximum likelihood estimation algorithms. 

The  fields originated from head-on collisions of nonspinning black holes
---as in our case--- are fully described by the $m=0$ modes
due to the rotational symmetry of such collisions. In this scenario,
all angular \revision{${m \neq 0}$} modes vanish, \emph{i.e.},
\revision{${{}_sX_{l, \, m \neq 0, \, n}}=0$.
We will only allow for deviations from the QNM spectrum
respecting its symmetries, $\omega^{+}_{l0n} = - \omega^{-}_{l0n}$
and $\tau^{+}_{l0n} = \tau^{-}_{l0n}$: \emph{i.e.}, we set
$\alpha_{l0n}^{+} = \alpha^{-}_{l0n}$ and $\beta_{l0n}^{+} = \beta^{-}_{l0n}$.
Moreover, the fields we are considering, \emph{i.e.},
the multipole moments $I_l$, and the shear modes $\sigma_l$
as defined above in Sec.~\ref{subsubsec:observables},
are real-valued functions.
For such variables, the two families (co- and counter-rotating)
of modes combine, with their complex amplitudes at $t_r$ satisfying
$\mathcal{A}_{l0n}^{-} = ( \mathcal{A}_{l0n}^{+} )^{*}$, so that
\begin{align*}
 & {\cal A}_{l0n}^{+} \,
   \exp \left[ - \iota \omega_{l0n}^{+}(1+\alpha_{l0n}^{+}) (t-t_r) \right] \\
  &  \quad + {\cal A}_{l0n}^{-} \,
   \exp \left[ - \iota \omega_{l0n}^{-}(1+\alpha_{l0n}^{-}) (t-t_r) \right] \\
   & = {\cal A}_{l0n}^{+} \,
   \exp \left[ - \iota \omega_{l0n}^{+}(1+\alpha_{l0n}^{+}) (t-t_r) \right] \\
   & \quad \; + \left( {\cal A}_{l0n}^{+} \right)^{*} \,
   \exp \left[ + \iota \omega_{l0n}^{+}(1+\alpha_{l0n}^{+}) (t-t_r) \right] \\
 & {} = A_{l0n} \, \cos\left[{\omega_{l0n}^{+} (1+\alpha_{l0n}^{+}) \, (t-t_r) + \phi_{l0n}} \right] \, ,
\end{align*}
with $\mathcal{A}_{l0n}^{+} \equiv (1/2) \, A_{l0n} \exp \left[ - \iota \phi_{l0n} \right]$
and for a real amplitude $A_{l0n}$ and phase $\phi_{l0n}$.
With the above remarks,
from now onwards we can drop the ${}^{\pm}$ superscripts on all parameters
and simplify} the ansatz of Eq.~\eqref{eq:rdmodel_ab} \revision{into,
  \begin{align}
\label{eq:rdmodel_ansatz}
 {}_s X_{l0}(t) =  \sum_{n=0}^{\nmax}
   & \; A_{l0n} \,
   \exp \left[ - \frac{t-t_r}{\tau_{l0n}(1+\beta_{l0n})} \right] \times \nonumber \\
   & \quad \times \cos\left[{\omega_{l0n}(1+\alpha_{l0n}) (t-t_r) + \phi_{l0n}} \right] \, ,
\end{align}
with ${}_s X_{l0} = \mathrm{Re}({}_s X_{l0})$.
As a sign flip on the real frequencies would still be possible in principle,
we specify that the co-rotating (positive real part) choice is implied
for the complex frequencies $\nu_{l0n}$ and their real part $\omega_{l0n}$.
We will further}
drop the fixed $s=2$ and $m=0$ subscripts on $X_l \equiv {}_s X_{l0}$
in the following.

The parameters $\alpha_{l0n}$ and $\beta_{l0n}$ help
one test the effects of eventual systematic errors sourced by
a) including an insufficient number of tones $\nmax$
when modeling the data with Eq.~\eqref{eq:rdmodel_ansatz}
or b) the presence of non-negligible nonlinearities
in the data~\cite{bhagwat:2017tkm,Bhagwat:2019dtm, forteza2020}. We note in
passing that introducing the parameters $\alpha_{l0n}$ and
$\beta_{l0n}$ may also be used in a more general context
to parametrize deviations from general relativity.

In this work we model the data for the shear modes $\sigma_l$ and
multipoles $I_l$ with $2 \leq l \leq 12$, using multiple values of
$\nmax$, up to $\nmax=15$. In general, we fit for the amplitude
$A_{l0n}$ and for the phase $\phi_{l0n}$, and we either set
the frequency deviation parameters $\beta_{l0n}$ and $\alpha_{l0n}$ to zero
or additionally fit for them.

We compute the QNM spectrum values $\{\omega_{l0n}, \tau_{l0n} \}$ of
the final black hole using the \textsc{qnm} Python script
\cite{Stein:2019mop}, which combines a Leaver solver with the
Cook-Zalutskiy spectral approach to the angular
sector~\cite{leaver:1985ax, Cook:2014cta}.  
Our final black hole has no spin
and its mass is slightly lower
than $M$ due to the gravitational radiation. This relative mass
decrease with respect to $M$ can be estimated at about
$7 \cdot 10^{-5}$ from the outer horizon area at late times.  We
simply approximate the final mass as $M$ when computing the QNM
spectrum, implying a similar relative error on $\tau_{l0n}$ and
$\omega_{l0n}$ which are proportional and inversely proportional to
the final mass, respectively.
In Table~\ref{tab:QNMmodes}, we show as an example,
a sample of the resulting QNM frequencies $\omega_{l0n}$ and damping rates
 $1/\tau_{l0n}$, for $l=2,\dots,12$ and
$n=0,\dots,5$.
\begin{table*}[]{
\setlength{\tabcolsep}{1.4ex}
\begin{tabular}{c||c c c c c c } 
\hline
 l &n=0 & n=1   & n=2         & n=3       & n=4 & n=5   \\ \hline
\hline
 2 & 0.3737 & 0.3467 & 0.3011 & 0.2515 & 0.2075 & 0.1693 \\
 3 & 0.5994 & 0.5826 & 0.5517 & 0.5120 & 0.4702 & 0.4314 \\
 4 & 0.8092 & 0.7966 & 0.7727 & 0.7398 & 0.7015 & 0.6616 \\
 5 & 1.0123 & 1.0022 & 0.9827 & 0.9550 & 0.9211 & 0.8833 \\
 6 & 1.2120 & 1.2036 & 1.1871 & 1.1633 & 1.1333 & 1.0988 \\
 7 & 1.4097 & 1.4025 & 1.3882 & 1.3674 & 1.3407 & 1.3093 \\
 8 & 1.6062 & 1.5998 & 1.5872 & 1.5687 & 1.5449 & 1.5163 \\
 9 & 1.8018 & 1.7961 & 1.7848 & 1.7682 & 1.7466 & 1.7205 \\
 10 & 1.9968 & 1.9916 & 1.9815 & 1.9664 & 1.9467 & 1.9227 \\
 11 & 2.1913 & 2.1866 & 2.1773 & 2.1635 & 2.1455 & 2.1234 \\
 12 & 2.3855 & 2.3812 & 2.3727 & 2.3600 & 2.3433 & 2.3228 \\
\hline
\end{tabular}
\hspace{6ex}
\begin{tabular}{c||c c c c c c } 
\hline
 l &n=0 & n=1   & n=2         & n=3       & n=4 & n=5   \\ \hline
\hline
2 & 0.0890 & 0.2739 & 0.4783 & 0.7051 & 0.9468 & 1.1956 \\
3 & 0.0927 & 0.2813 & 0.4791 & 0.6903 & 0.9156 & 1.1522 \\
4 & 0.0942 & 0.2843 & 0.4799 & 0.6839 & 0.8982 & 1.1230 \\
5 & 0.0949 & 0.2858 & 0.4803 & 0.6806 & 0.8882 & 1.1042 \\
6 & 0.0953 & 0.2866 & 0.4806 & 0.6786 & 0.8821 & 1.0921 \\
7 & 0.0955 & 0.2872 & 0.4807 & 0.6773 & 0.8782 & 1.0841 \\
8 & 0.0957 & 0.2875 & 0.4808 & 0.6765 & 0.8755 & 1.0786 \\
9 & 0.0958 & 0.2877 & 0.4809 & 0.6759 & 0.8736 & 1.0747 \\
10 & 0.0959 & 0.2879 & 0.4809 & 0.6755 & 0.8723 & 1.0718 \\
11 & 0.0959 & 0.2880 & 0.4810 & 0.6752 & 0.8712 & 1.0696 \\
12 & 0.0960 & 0.2881 & 0.4810 & 0.6749 & 0.8704 & 1.0679 \\
\hline
\end{tabular}
%
}
\caption{QNM real frequencies \revision{$\omega_{l0n} \equiv \left| \omega_{l0n}^{\pm} \right| = \left| \mathrm{Re}(\nu_{l0n}^{\pm}) \right|$} (left)
and damping rates \revision{$1/\tau_{l0n} \equiv 1/\tau_{l0n}^{\pm} = - \mathrm{Im}(\nu_{l0n}^{\pm})$} (right)
for a Schwarzschild black hole with unit mass, for $l=2, \dots, 12$ and $n = 0, \dots, 5$,
computed with the \textsc{qnm} Python script \cite{Stein:2019mop}.}
\label{tab:QNMmodes}
\end{table*}

The fitting algorithm is explained in Sec.~\ref{subsec:fit-alg}. In
Sec.~\ref{subsec:singlemode} we explore the fit results for a
single-tone ($\nmax=0$) analysis. We observe that the single-tone
model is not sufficient to fully describe even the late-time data. In
Sec.~\ref{subsec:fit-overtones} we extend the results to the
multiple-tone ($\nmax>0$) analysis and to the whole dataset, with all
the $\alpha_{l0n}$ and $\beta_{l0n}$ parameters set to zero. In this
case and for large enough $\nmax$, we find that the model is
sufficient to describe the data even including the early times where
the horizon is not in equilibrium. In
Sec.~\ref{sec:overtones-discussion} we discuss this, and investigate
whether one can infer from it an actual presence and predominance of
overtones over nonlinear contributions right from shortly after the
horizon is formed.

\subsection{The fitting algorithm}
\label{subsec:fit-alg}
We use a maximum likelihood estimation algorithm to obtain the
best-fit parameters $\lambda_i$. Those correspond to the
parameter values that minimize the $\chi^2$, namely,
\begin{equation}
  \label{eq:chi2}
  \chi^2=\sum_{k} \left| h_x \left[\vec{\lambda} \right] (t_k) - h_{\mathrm{NR}} (t_k) \right|^2,
\end{equation} 
where $h_x \left[ \vec\lambda \right]$ stands for the model given by
Eq.~\eqref{eq:rdmodel_ansatz} and evaluated at the parameters
$\vec{\lambda} =\left\lbrace A_{l0n},\phi_{l0n} \right\rbrace$ or
$\vec\lambda = \left\lbrace
  A_{l0n},\phi_{l0n},\alpha_{l0n},\beta_{l0n}\right\rbrace$, and
$h_{\rm NR}=\left\lbrace \sigma_l, I_l \right\rbrace$ stands for the
numerical data for the shear modes or multipoles, respectively.
We sum over the data points
$k$ at all times $t=t_k \in [t_0, t_f]$, for a certain fit starting
time $t_0$ which may be picked at any value
$t_{\mathrm{bifurcate}} \leq t_0 \leq t_f$, and where, as above, $t_f \simeq 38.5 M$
is the end time of the simulation. Minimization of
\eqref{eq:chi2} is performed running the Levenberg-Marquardt algorithm
for nonlinear fitting\footnote{%
Note that in the case where the free parameters are only the amplitude and phase
of each mode, $\vec{\lambda} =\left\lbrace A_{l0n},\phi_{l0n} \right\rbrace$,
\emph{i.e.} their complex amplitude, and writing the expansion under its complex form
as in Eq.~\eqref{eq:rdmodel}, the fitting problem is linear
and a dedicated scheme could have been used instead~\cite{cook2020}.
In this work, we however use the same (nonlinear) algorithm
for either choice of the set of free parameters to fit for.
This allows for a consistent approach throughout our investigation
and for direct comparisons between models
where some of the frequencies are left as free parameters
and models with all frequencies set to the QNM values
(as in Sec.~\ref{subsec:discussion-equal-number-of-DoF}).%
}
as implemented in {\scshape Mathematica}~\cite{mathematica}.

To assess the fit goodness we use the mismatch
$\mathcal{M}$ as
in~\cite{giesler2019,Bhagwat:2019dtm,forteza2020,cook2020}, which is defined as
\begin{equation}
    \mathcal{M} = 1 - \frac{\langle h_{\rm NR}|h_x\rangle}{\sqrt{\langle h_{\rm NR}|h_{\rm NR}\rangle \langle h_x|h_x\rangle}}\,
    \label{eq:mismatch}
\end{equation}
with
\begin{equation}
    \langle f|g\rangle = \int_{t_0}^{t_f} f(t) g(t) \, \mathrm{d}t\,.
\end{equation}
The standard errors $\delta \lambda_i$ on the parameters are computed from the diagonal terms of the covariance matrix as~\cite{mathematica},
\begin{equation}
\label{eq:fiterrors}
\delta \lambda_i=\sqrt{\frac{2\, \text{RSS} \times [H^{-1}]_{ii}}{N-p}}\,,
\end{equation} 
where $[H^{-1}]_{ii}$ stands for the diagonal terms of the $[H^{-1}]_{ij}$ matrix (without implicit summation on the $i$ indices);
\begin{equation}
\text{RSS}=\sum_k \big(h_{NR}(t_k) -h_{x}(t_k) \big)^2
\end{equation} is the residual sum of squares; $N$ is the number of data points in the time range considered, \emph{i.e.} in $[ t_0,t_f ]$; and $p$ is the number of parameters one wants to fit for. $H_{ij}$ is the Hessian matrix defined as
\begin{equation}
H_{ij}=\frac{\partial^2 (\text{RSS})}{\partial \lambda_i \, \partial \lambda_j}\Bigm|_{\vec{\lambda}} \;\; ,
\end{equation}
which is evaluated at the best fit parameters $\vec{\lambda}$.

\subsection{Exponential rescaling procedure and numerical errors}
\label{subsec:rescaling-and-errors}

The NR data appears to be exponentially damped at late times, and so
are the damped-sinusoidal models of the class
\eqref{eq:rdmodel_ansatz} that we fit to this data. Due to this
damping, the fitting procedure based on the residual sum of squares
---rather than relative differences--- will capture better the
behavior of the NR data towards the beginning of the time interval
considered (close to $t_0$) than towards the latest times (close to
$t_r$). A reliable estimate of the oscillation frequencies, damping
rates, and amplitudes of each tone in the model would rather require
an accurate match of the \emph{relative} amplitudes and positions of
the successive extrema (or zeros), and thus a small relative
deviation to the NR data, on the entire time interval considered.

To this aim, when looking specifically for the best-fit frequencies, damping rates and/or amplitudes\footnote{%
The rescaling procedure presented here is dedicated to improving the accuracy of the determination of these parameters. To ease the interpretation, we will \emph{not} use this rescaling when we rather simply wish to evaluate the quality of the fit of a model to the NR shear modes and multipoles (either by computing the corresponding mismatch or \emph{via} direct visualization). In this case we shall simply compare the model $h_x(t)$ to the data $h_{\mathrm{NR}}(t)$ directly.
}, we will first apply a time-dependent rescaling of the NR data and of the model, determined by the damping rate of the fundamental QNM. That is, we will fit a rescaled dataset with the similarly rescaled model according to
\begin{align}
h_{\mathrm{NR}}(t) & \mapsto \tilde h_{\mathrm{NR}}(t)  = \exp \left[ + \frac{t-t_r}{\tau_{l00}} \right] \, h_{\mathrm{NR}}(t) \; ; \nonumber \\
h_x(t) & \mapsto \tilde h_x(t) = \exp \left[ + \frac{t-t_r}{\tau_{l00}} \right] \, h_x(t) \; ,
\label{eq:rescaling}
\end{align}
where the unrescaled model $h_x(t)$ is given by
Eq.~\eqref{eq:rdmodel_ansatz} for a certain $\nmax$, and with the
parameters $\alpha_{l0n}$ and $\beta_{l0n}$ either set to
zero, or left as free parameters.  In this way, provided the late-time
decay rate of the NR data is comparable to the fundamental QNM value
on the time range considered, the rescaled dataset will have an
approximately constant (rather than decaying) amplitude towards late
times. This then allows for a more accurate retrieval of the complex
frequency parameters $\alpha_{l0n}$ and $\beta_{l0n}$, if left free,
and of the amplitudes $A_{l0n}$.

\begin{figure*}
  \centering    
  \includegraphics[width=\columnwidth]{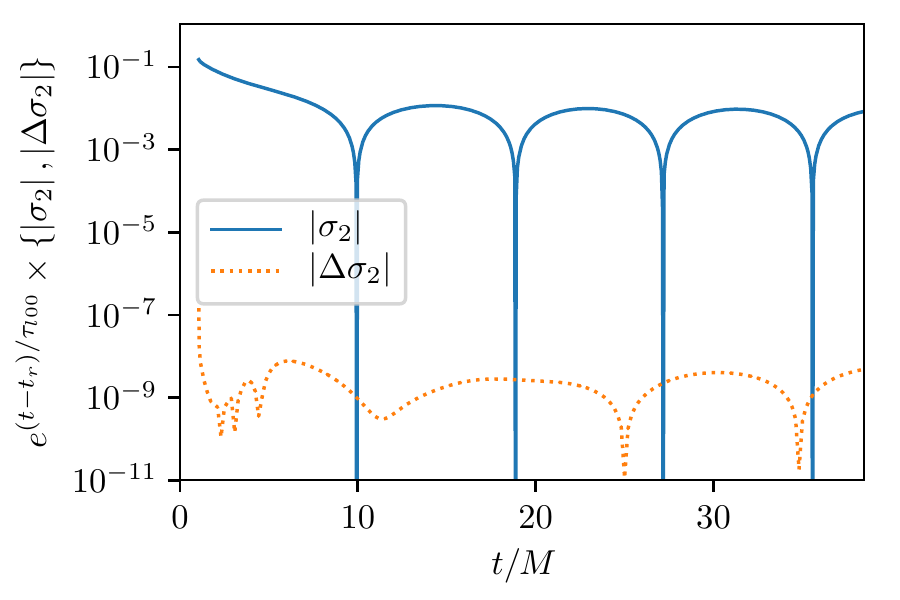}
  \includegraphics[width=\columnwidth]{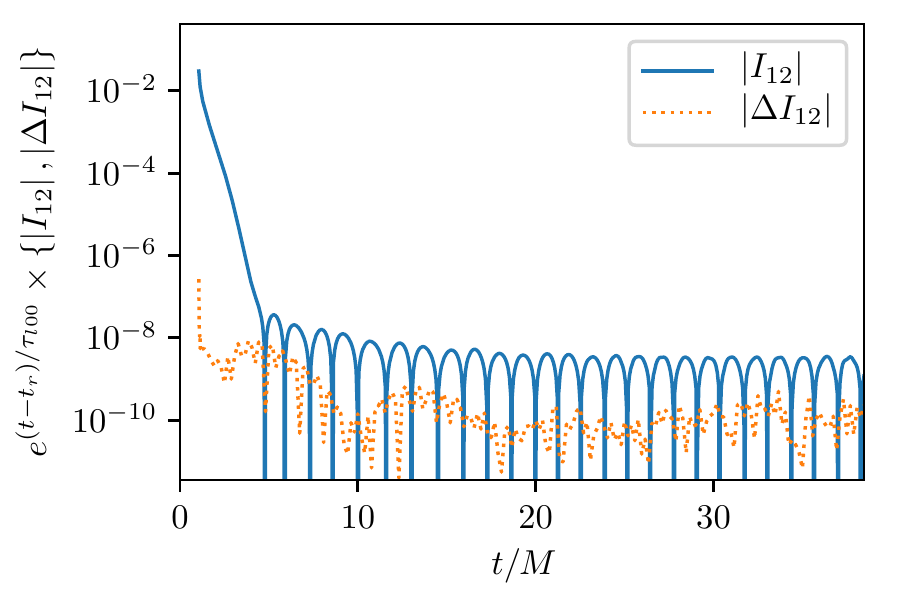}
  \caption{Left: rescaled highest-resolution NR results for the shear $l=2$ mode (continuous line) and numerical error on this rescaled dataset evaluated as the difference with the next-highest resolution available (dashed line; see Sec.~\ref{subsec:rescaling-and-errors}), as a function of simulation time. Right: same results for the $l=12$ multipole. These two cases represent respectively the smallest and the largest relative numerical errors among the modes considered in this analysis.}
  \label{fig:numerical-errors}
\end{figure*}

Such a rescaling will of course also scale up the numerical errors at late times. However, the high resolution used here means that the relative error on the computed shear modes and multipoles remains rather low for all modes $l \leq 12$.  We conservatively estimate the error on the NR results ---assuming it to be dominated by discretization error--- as the difference between the values obtained at the highest resolution, $\mathrm{res} = 240$ ---used throughout this paper--- and the next highest resolution available\footnote{%
\label{fn:NRerror}
To test the validity of this measure of the error (let us denote it as $\epsilon_{180-240}$),
we have computed the differences $\epsilon_{240-480}$
between the $\mathrm{res} = 240$ datasets
and higher resolved datasets with $\mathrm{res} = 480$
but with a shorter simulation time $t_f^{(480)} \simeq 15M$.
We have checked that the error $\epsilon_{180-240}$ is larger than  $\epsilon_{240-480}$,
which is expected given that the discretization scheme
of the datasets is globally fifth-order accurate
and that all datasets are shown to be in the convergent regime~\cite{PhysRevD.100.084044}.
This trend has been confirmed for the examples of the ${l=2,4,8,12}$ modes
for both the shear and the multipoles data.%
},  $\mathrm{res} =  180$. The relative error on the shear modes and multipoles computed in this way increases with $l$ as the amplitude of the modes decreases. Away from the zeros of the modes, it ranges from about $10^{-6}$ at $l=2$ to about $1 \%$ at $l=12$ for the shear modes, and up to $1$ order of magnitude larger for the multipoles. We accordingly consider $l=12$ as a threshold value for sufficiently small numerical uncertainties and we shall not consider the higher-$l$ modes in the present work. Fig.~\ref{fig:numerical-errors} shows the rescaled NR data $\tilde h_{\mathrm{NR}}$ and the numerical error on this rescaled data as a function of time for the two extreme cases of the smallest ($\sigma_2$, left panel) and the largest ($I_{12}$, right panel) relative error among the modes we consider.

\section{Fit results}
\label{sec:results}

\subsection{Single-mode fits with variable frequency and damping rate}
\label{subsec:singlemode}

Before considering higher overtones, we first focus on the $s=2$ fundamental QNMs of the remnant black hole. We aim at checking whether we recover the good agreement found in~\cite{pook-kolb2020II} (with a different method as used here) between the damped oscillating behavior of the shear modes and multipoles at late times on the one hand, and the complex frequencies of these fundamental modes on the other hand.

To this end, we consider a single-tone model where the frequency and damping rate are free parameters, \emph{i.e.}, the ringdown model \eqref{eq:rdmodel_ansatz} restricted to the $n=0$ fundamental tone:
\begin{equation}
X_l (t) = A_{0} \exp \left[ - \frac{t-t_r}{\tau_{l\,00} (1+\beta_{0})} \right] \cos \left[(1+\alpha_{0}) \, \omega_{l00} (t-t_r) + \phi_{0} \right] .
\label{eq:onetonemodel}
\end{equation}
We will then check to what extent the best-fit frequencies and damping times
in such a model match the fundamental QNM values,
corresponding to $\alpha_{0} = 0$ and $\beta_{0} = 0$,
for the late-time shear modes and multipoles.
Thanks to the high numerical resolution, we can perform this analysis
up to the $l=12$ mode of the shear and multipoles, so that
we can also consider whether the conclusions of~\cite{pook-kolb2020II}
on the late-time behavior of these variables do extend beyond the $l=7$ mode.

Note that for convenience, we here drop the $l$ and the $m=0$ indices
on the free parameters, that is on $A_0 \equiv A_{l00}$,
$\phi_0 \equiv \phi_{l00}$, $\alpha_0 \equiv \alpha_{l00}$
and $\beta_0 \equiv \beta_{l00}$.
It should however be understood that their values will, of course,
depend on the variable $X_l$ considered, \emph{i.e.},
on the observable $X$ (shear or multipoles) and on the mode $l$.

\subsubsection{Late-time best-fit frequencies and comparison to the fundamental QNM values}
\label{subsubsec:singlemode_latetimes}

We first turn our attention to the late-time data
as selected in the same way as in~\cite{pook-kolb2020II},
\emph{i.e.}, we consider $t \in [ t_0, t_f ]$ for $t_0/M = 20/1.3 \simeq 15.4$.
(In this case, the fit starting time $t_0$
then coincides with the constant value we have set for $t_r$.)

\begin{figure*}
  \centering    
  \includegraphics[width=\columnwidth]{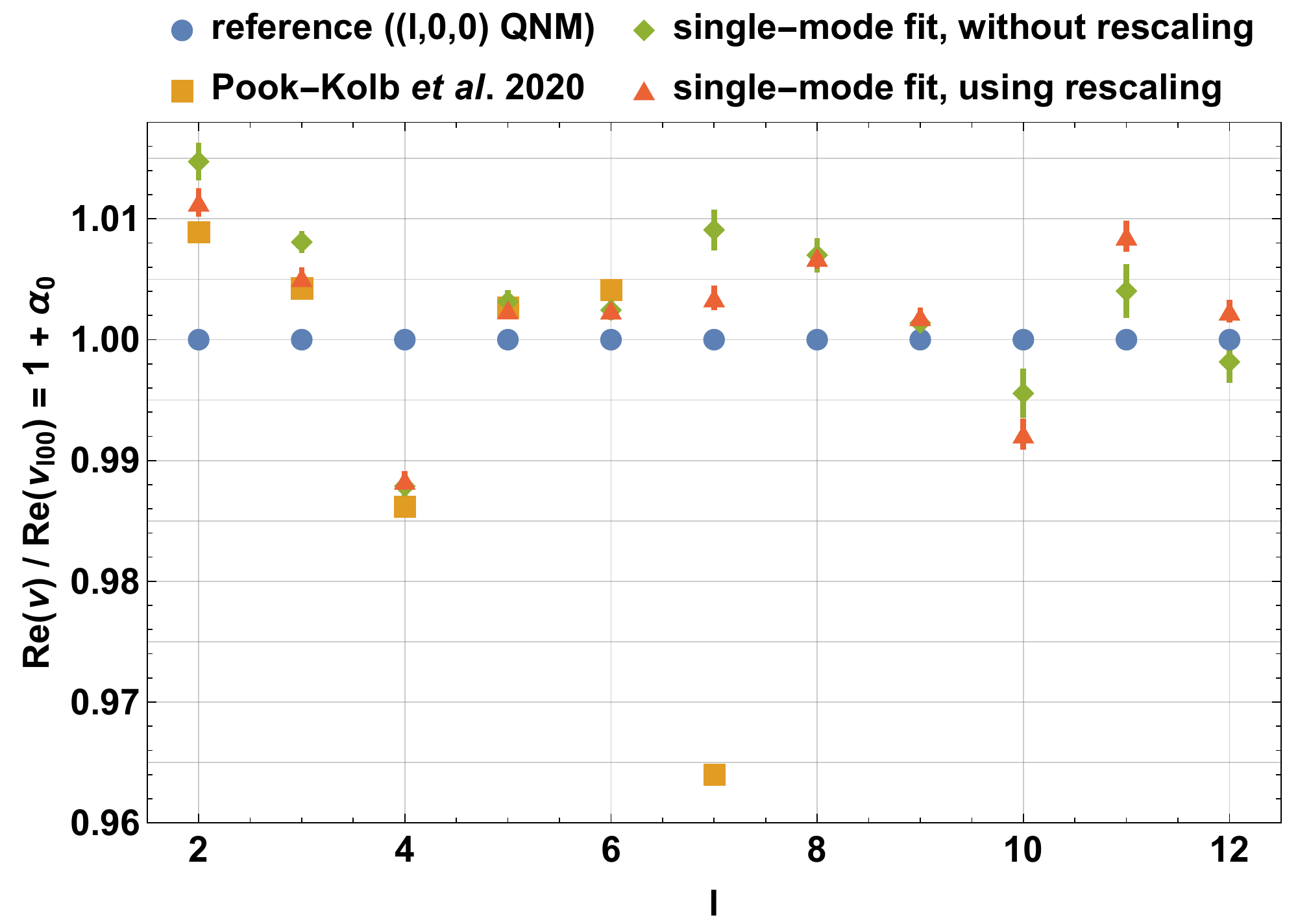}
  \includegraphics[width=\columnwidth]{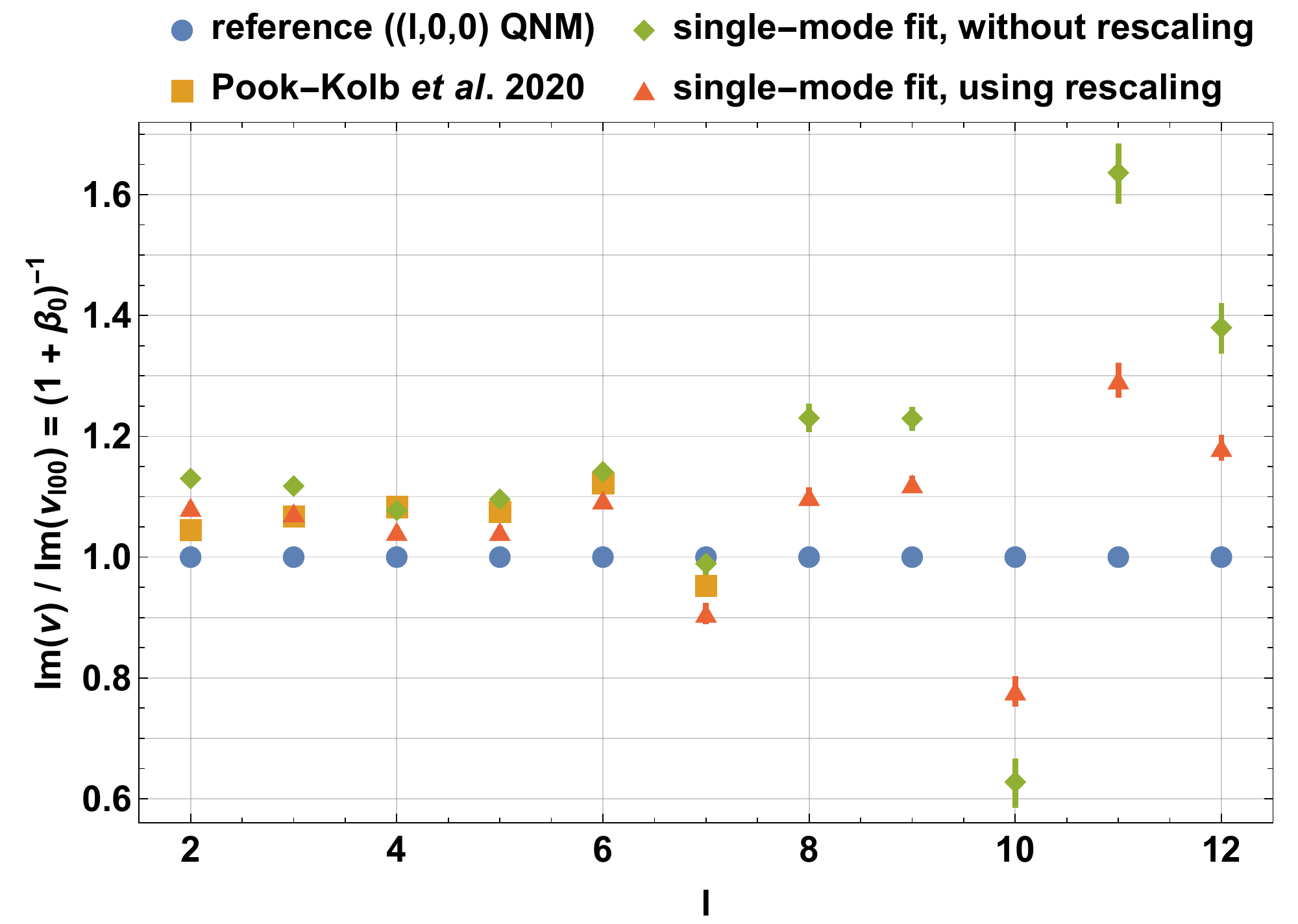}
  \caption{Best-fit real (left) and imaginary (right) frequencies for the shear $l=2$ to $l=12$ modes obtained for $t_0 = (20/1.3) \, M$ using the single-mode model of Eq.~\eqref{eq:onetonemodel} and the rescaling procedure given by Eq.~\eqref{eq:rescaling}, shown as red triangles with $1 \sigma$ error bars (computed from Eq.~\eqref{eq:fiterrors}). The values are normalized to the fundamental QNM values and the reference unity value for this ratio is marked as blue disks. The best-fit values we obtain without using the rescaling procedure are also shown for reference (green diamonds) with their associated $1 \sigma$ uncertainties, and the values quoted from~\cite{pook-kolb2020II} for $l \leq 7$ are also given for comparison (orange squares). For the latter, we do not show error bars on the figure as precise uncertainties were not given for every $l$. These uncertainties were estimated to be of the order of $\delta(\mathrm{Re}(\nu)) / \mathrm{Re}(\nu) \simeq \pm 1 \%$ for the real part and $\delta(\mathrm{Im}(\nu)) / \mathrm{Im}(\nu) \simeq \pm 10 \%$ for the imaginary part for each $l \leq 6$, and larger for $l=7$.}
  \label{fig:single-mode-comparison-shear}
\end{figure*}

We begin by considering the shear modes $\sigma_l$ for $2 \leq l \leq 12$.
Fig.~\ref{fig:single-mode-comparison-shear} shows the best-fit values for each mode $l$
for the real (left panel) and imaginary (right panel) parts
of the model's single frequency
$\nu = (1+\alpha_0) \, \omega_{l00} - \iota \, \tau_{l00}^{-1} (1+\beta_0)^{-1}$,
using the rescaling procedure \eqref{eq:rescaling}
to obtain a better accuracy in the recovery of $\nu$.
The results for $\mathrm{Re}(\nu)$ and $\mathrm{Im}(\nu)$
are normalized to the corresponding fundamental QNM values,
and we also include as error bars on these results, the $1 \sigma$ (standard) deviations
on the best-fit estimates as computed from the covariance matrix
using Eq.~\eqref{eq:fiterrors} (see Sec.~\ref{subsec:fit-alg}).
The results from~\cite{pook-kolb2020II} are also indicated for comparison when available,
\emph{i.e.}, for $l \leq 7$.

For the real part, we find that the relative deviations to the fundamental QNM values, $\mathrm{Re}(\nu) / \mathrm{Re}(\nu_{l00}) - 1 = \alpha_0$, stay within $\pm 1.5 \%$ for all modes $l \leq 6$, in agreement with~\cite{pook-kolb2020II}. We extend this conclusion to $7 \leq l \leq 12$ with deviations ${| \alpha_{0}|}$ below $\sim 1 \%$ in these cases (while a deviation of about ${\alpha_{0} \simeq -3.6 }\%$ was found in~\cite{pook-kolb2020II} for $l=7$, with a larger uncertainty). We also obtain deviations $\left| \, \mathrm{Im}(\nu) / \mathrm{Im}(\nu_{l00}) -1 \right|$ within $\sim 10 \%$ to the fundamental QNM values for the imaginary part for $l \geq 7$, in consistency with~ \cite{pook-kolb2020II}.

The disagreement to the QNM values for the imaginary part
does however increase for larger values of $l$, up to the order of
$\left| \, \mathrm{Im}(\nu) / \mathrm{Im}(\nu_{l00}) -1 \right| \simeq 20 \%$ to $\sim 30\%$
for $l=10$, $11$ and $12$. For these large-$l$ modes
the quasinormal fundamental mode
is thus no longer an accurate model of the shear modes for the range $t \gtrsim 15.4 M$,
at least regarding their damping rates.
The $\left| \, \mathrm{Im}(\nu) / \mathrm{Im}(\nu_{l00}) -1 \right| \simeq 10 \%$ deviations
at several of the smaller-$l$ values despite uncertainties much smaller than this number
suggest that this may even be the case for most of the modes.
This may be a consequence of the presence of residual nonlinear deviations to equilibrium
at these times, or of the residual presence of higher overtones for these modes.
The latter hypothesis would explain the larger decay rates found for most modes
and the very small deviations on the real parts of the frequencies:
the respective real frequencies $\omega_{l01}$ and $\omega_{l00}$
of the QNM first overtone and fundamental mode differ by
${1-\omega_{201}/\omega_{200}\simeq 7 }\%$ for $l=2$
and by even smaller amounts for all higher-$l$ modes,
while the decay rates of the QNM first overtones
are typically $3$ times larger than those of the fundamental modes
(see Table~\ref{tab:QNMmodes}).
The decay rates smaller than the fundamental QNM value found for $l=7$ and $l=10$
would be harder to explain in this scenario, but could be caused by a modulation
induced by a higher overtone if this latter is nearly in antiphase with the fundamental mode.
This would cause a decrease in the overall amplitude
in the early part of the time range considered (\emph{i.e.}, for $t$ close to $t_0$)
before the overtone fully decays away.

Note that the systematics due to the errors in the numerical data are not included in the error bars shown. We can estimate these errors by comparing the best-fit frequencies to those found by instead fitting the (rescaled) next-highest-resolution ($\mathrm{res} = 180$) NR data. The relative deviations $\Delta(\mathrm{Re}(\nu))/\mathrm{Re}(\nu)$, $\Delta(\mathrm{Im}(\nu))/\mathrm{Im}(\nu)$ obtained in this way on the best-fit real and imaginary frequencies are very small for the low values of $l$. They remain under $10^{-4}$ for all modes for the real part, and under $10^{-3}$ for all modes for the imaginary part.
There are further systematic errors if, \emph{e.g.}, nonlinearities or higher overtones are present since they are not accounted for in the model.

For comparison, we also include in Fig.~\ref{fig:single-mode-comparison-shear} the results obtained from directly fitting the model \eqref{eq:onetonemodel} to the NR data on the same time range, this time without applying the rescaling \eqref{eq:rescaling}. For the real frequency $\mathrm{Re}(\nu)$, these results remain within $\pm 1.5 \%$ of the QNM values $\omega_{l00}$ for all modes. They however display substantial systematic errors with much larger deviations to the QNM values $\mathrm{Im}(\nu_{l00})$ for the imaginary parts $\mathrm{Im}(\nu)$ for almost all modes, due to an inaccurate fitting of the decaying amplitude over time.

\begin{figure*}
  \centering    
  \includegraphics[width=\columnwidth]{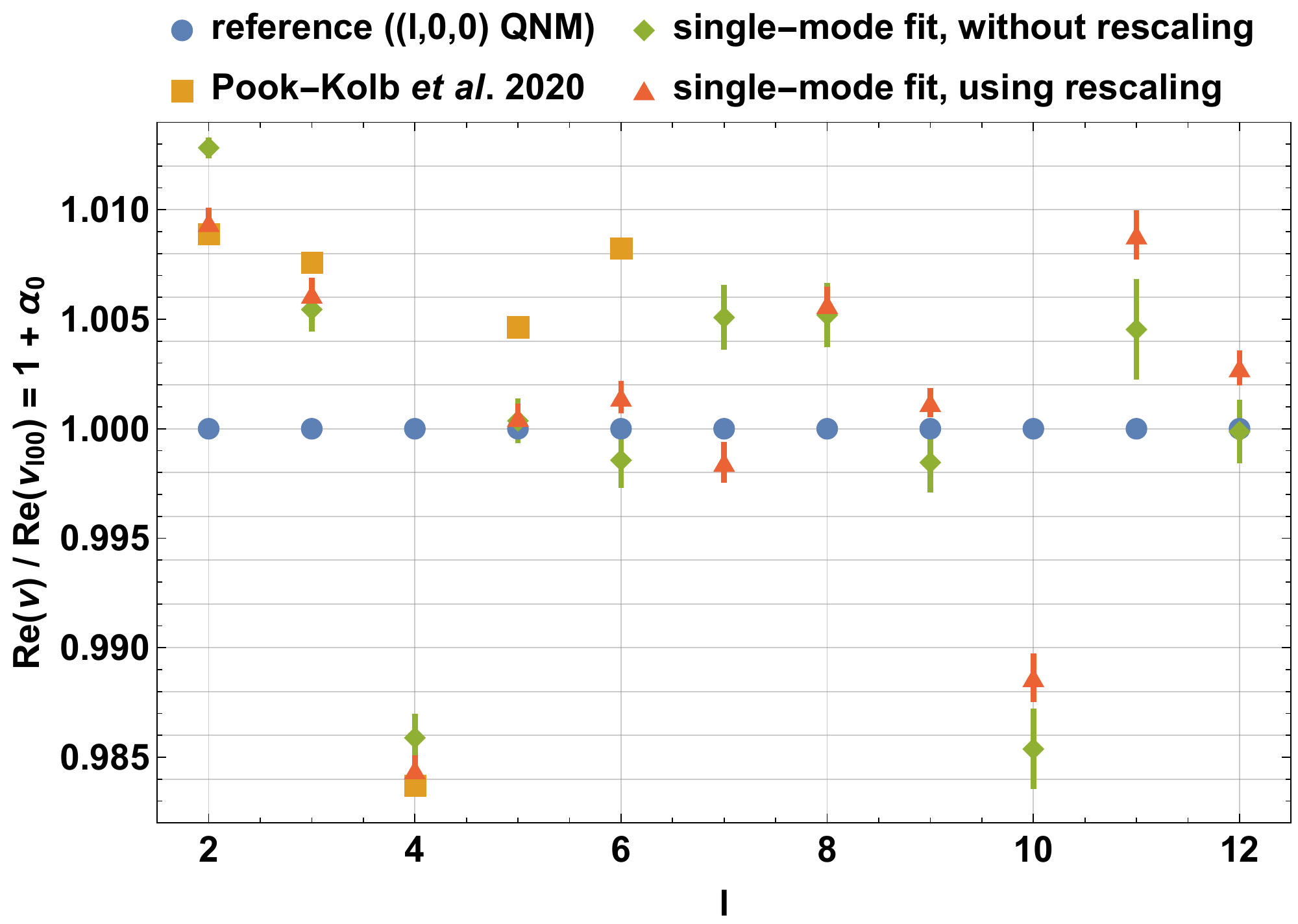}
  \includegraphics[width=\columnwidth]{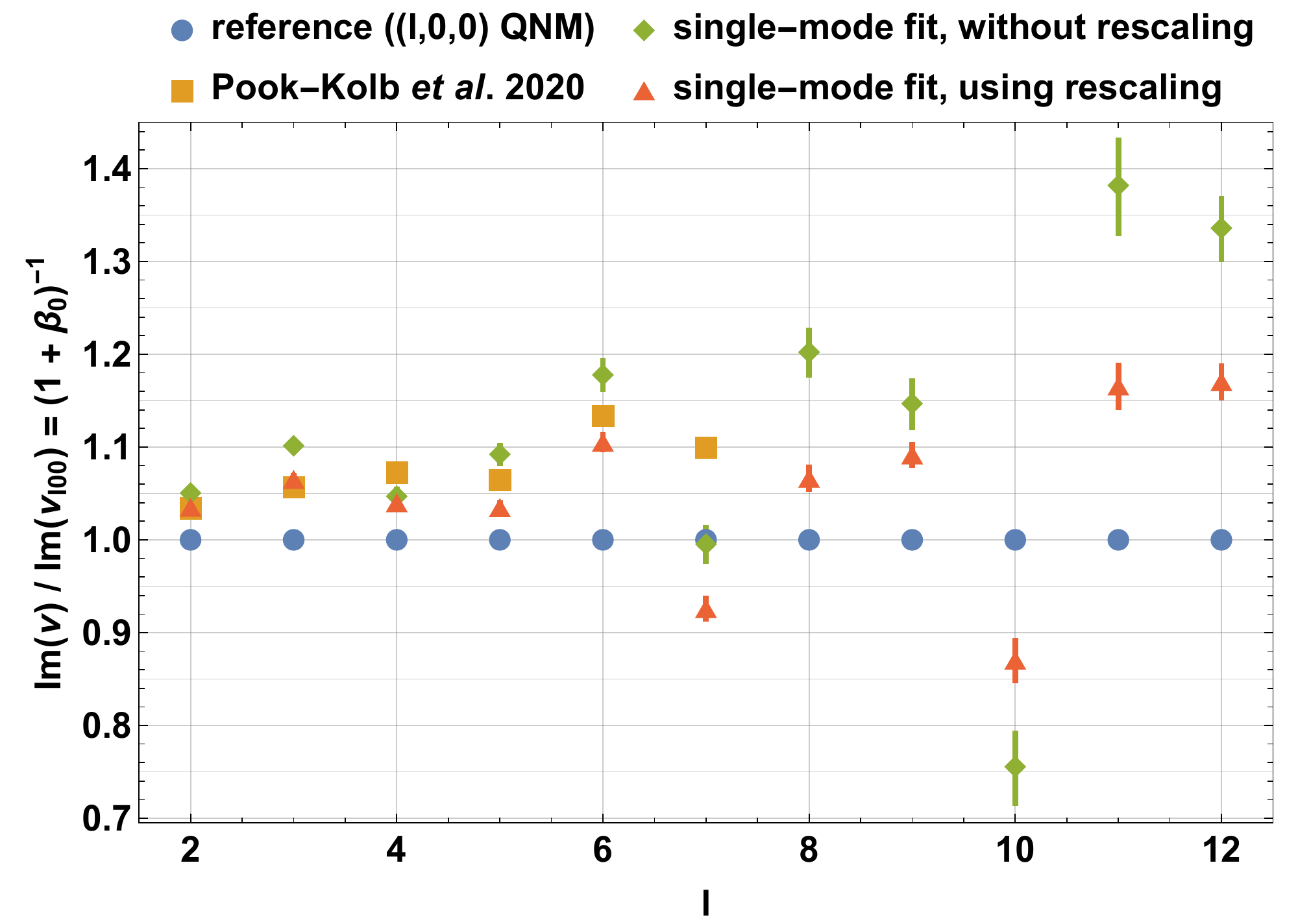}
  \caption{Same as Fig.~\ref{fig:single-mode-comparison-shear}, for the successive multipoles $I_2$ to $I_{12}$.}
  \label{fig:single-mode-comparison-multipole}
\end{figure*}

We then repeat this analysis for the multipole moments $I_l$ for $2 \leq l \leq 12$. The results are shown in Fig.~\ref{fig:single-mode-comparison-multipole} where the available ($l \leq 7$) results from~\cite{pook-kolb2020II} for the multipoles are again also given for comparison. As also found in the latter reference, the results we obtain for the multipoles are qualitatively very similar to those obtained for the shear modes. We here again focus on the results obtained after applying the rescaling procedure given by Eq.~\eqref{eq:rescaling}, which is expected to improve the determination of the frequencies.

The real frequencies $\mathrm{Re}(\nu)$ again remain within $\pm \sim 1.5\%$ (although no longer within $ \pm \sim 1\%$ at large $l$ values) of the fundamental QNM real frequencies $\mathrm{Re}(\nu_{l00}) = \omega_{l00}$. The imaginary frequencies $\mathrm{Im}(\nu)$ still feature relative deviations to the QNM values $\mathrm{Im}(\nu_{l00}) = \tau_{l00}^{-1}$ of up to $\pm \sim10\%$ for $l \leq 9$ and larger deviations for $l \geq 10$, although they do remain below $\sim 20\%$  in magnitude for all modes. While these magnitudes do change, we note that the signs of the deviations $\mathrm{Im}(\nu) / \mathrm{Im}(\nu_{l00})-1$ are the same as those found for the shear for every mode $l$.

The numerical relative errors on the best-fit frequency values, estimated in the same way as for the shear above, are again very small at small $l$. They reach slightly larger values than for the shear, from $\Delta(\mathrm{Re}(\nu))/\mathrm{Re}(\nu) \sim 1 \cdot 10^{-3}$ to just above $2 \cdot 10^{-3}$, for the real frequencies and $l =10$ to $12$.  For the imaginary part, these relative error estimates stay below $\Delta(\mathrm{Im}(\nu))/\mathrm{Im}(\nu) \sim 10^{-3}$, as for the shear, for $l \leq 9$; but they reach about $4$ to $9 \cdot 10^{-3}$ for the last three modes, implying a small but non-negligible possible systematic error on the values of $\mathrm{Im}(\nu) / \mathrm{Im}(\nu_{l00})$ shown in these cases.

Note that the QNM complex frequencies used here are still those of a
field of spin-weight $2$, as for the shear. The late-time oscillations
of the multipoles match well the corresponding real frequencies (as
well as the damping rates to a lesser extent), even though ---unlike the
spin-weight-$2$ shear scalar--- the multipoles are scalar fields of
spin weight $0$. For instance, the fundamental $l=2$ QNM real
frequency for a spin-0 perturbation is nearly $30 \%$ larger than the
corresponding frequency for spin-2 perturbations; a difference which
would be easily noticeable. Hence, the dynamics of the geometry of the outer
common horizon as measured by the mass multipoles may be determined
by the shear flux at the horizon, at least at the late times considered so far.
This is entirely consistent with
the discussion at the end of Sec.~\ref{subsubsec:observables}.

\subsubsection{Dependence on the fit starting time}
\label{subsubsec:singlemode_varyingt0}

We can also let the fit starting time $t_0$ vary and span the available interval
$[ t_{\mathrm{bifurcate}}, t_f]$. One can expect the behavior of the shear scalar
and multipoles to be fully described by the fundamental QNM $(n=0)$ at large $t$,
\emph{i.e.}, in the near-equilibrium regime and after the higher overtones have decayed away.
If this is the case, the best-fit complex frequencies to the shear modes
and to the multipoles should converge towards the fundamental QNM values at late times.

\begin{figure}
  \centering    
  \includegraphics[width=\columnwidth]{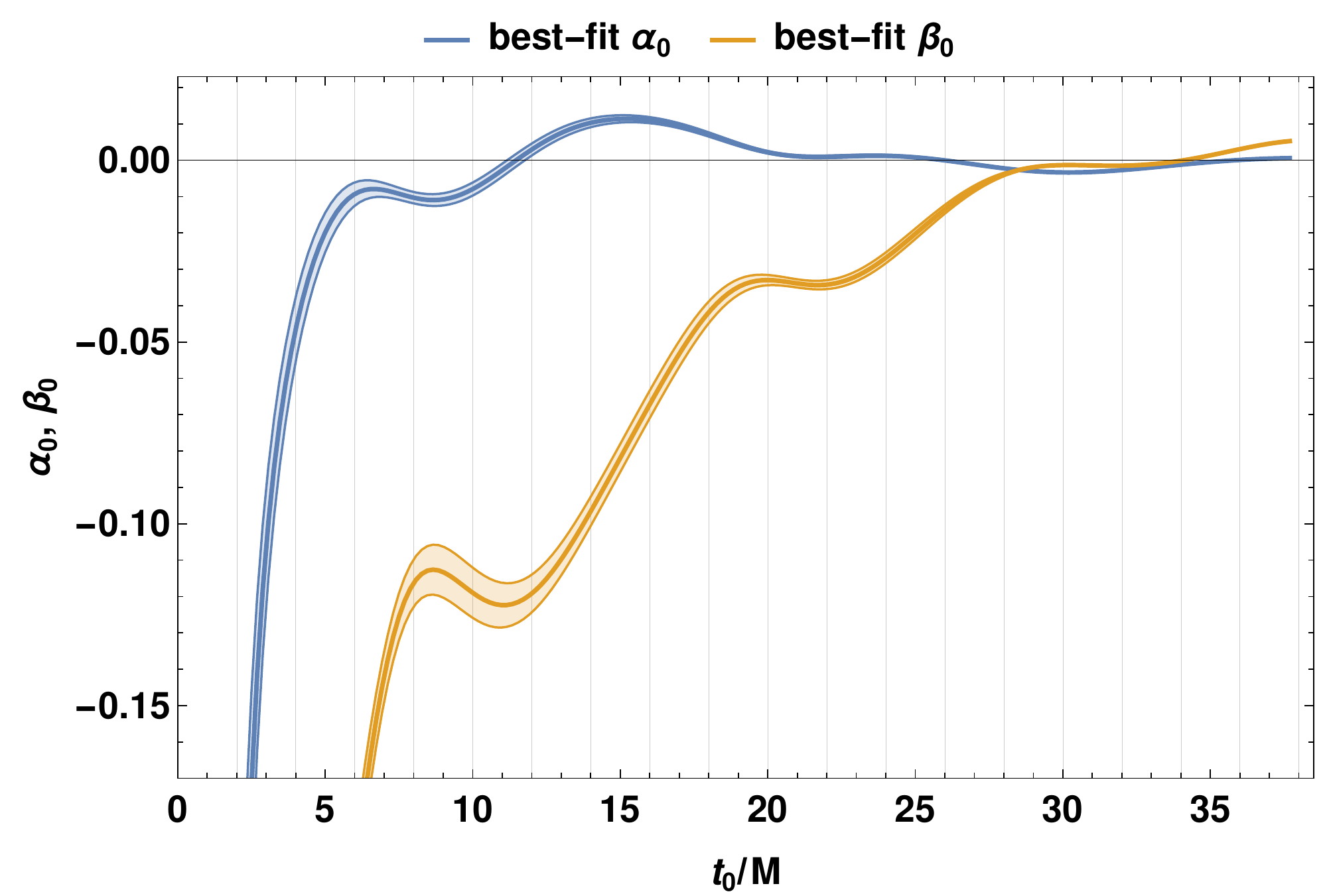}
  \caption{Best-fit parameters $\alpha_0$ (blue) and $\beta_0$ (orange)
  as functions of the fit starting time $t_0$, for the shear $\sigma_2$ mode
  and for the single-tone model \eqref{eq:onetonemodel}.
  The rescaling procedure \eqref{eq:rescaling} has been applied prior to fitting.
  $1 \sigma$ deviations (Eq.~\eqref{eq:fiterrors}) around the best-fit values are included.}
  \label{fig:single-mode-alpha0-beta0-shear-l2}
\end{figure}

\begin{figure}
  \centering    
  \includegraphics[width=\columnwidth]{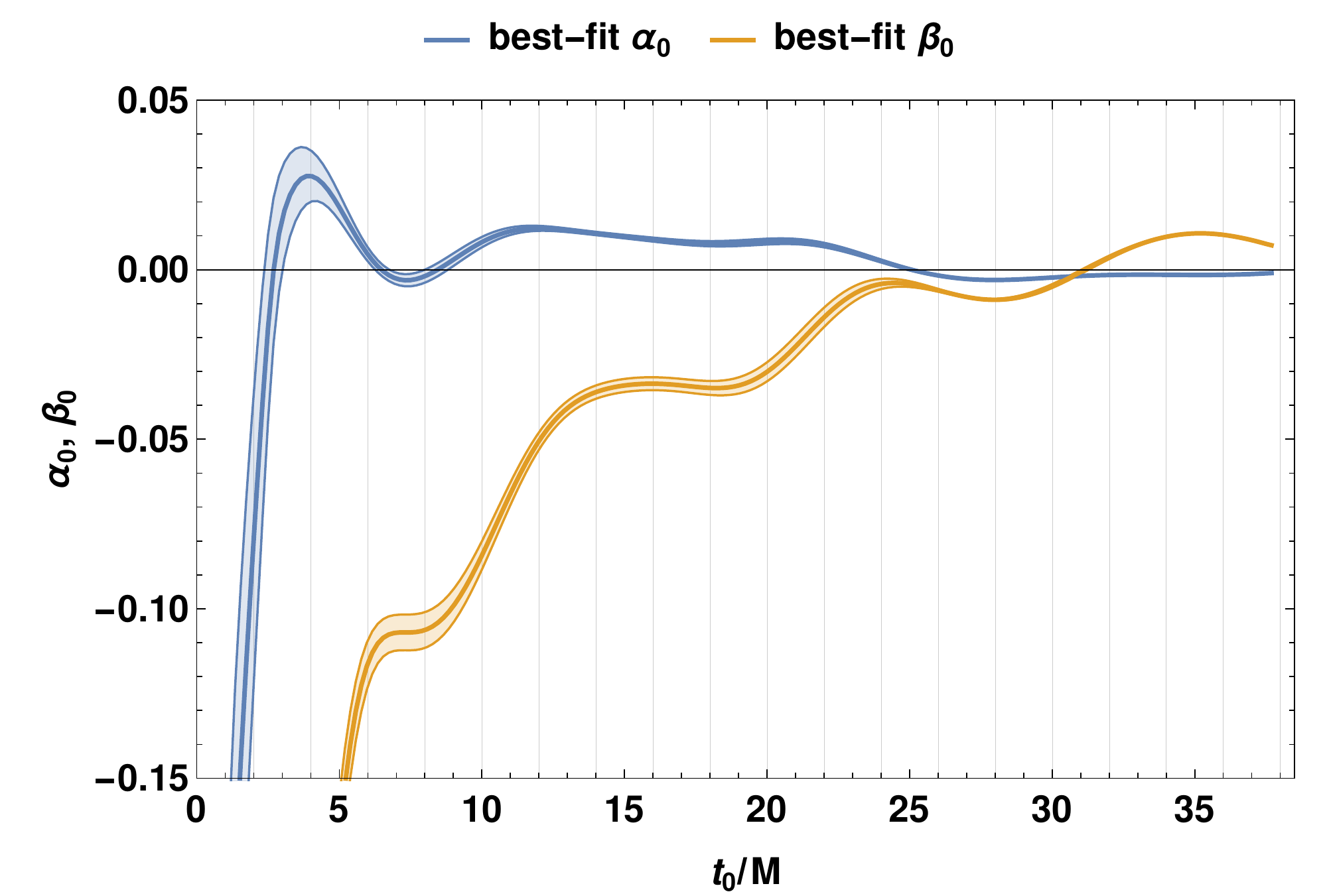}
  \caption{Same as Fig.~\ref{fig:single-mode-alpha0-beta0-shear-l2}, for the $I_2$ multipole.}
  \label{fig:single-mode-alpha0-beta0-multipole-l2}
\end{figure}

Fig.~\ref{fig:single-mode-alpha0-beta0-shear-l2}
shows the best-fit complex frequency deviation parameters $\alpha_0$ and $\beta_0$
for the time range $[t_0, t_f]$ as a function of $t_0$, for the shear $l=2$ mode as an example,
along with $1 \sigma$ uncertainties on these parameters as given by Eq.~\eqref{eq:fiterrors}.
The rescaling given by Eq.~\eqref{eq:rescaling} has again been applied to the model and to the NR data
for a more accurate retrieval of the complex frequency.
The results similarly obtained for the $l=2$ multipole
are shown in Fig.~\ref{fig:single-mode-alpha0-beta0-multipole-l2}
---still using the $s=2$ fundamental QNM as the reference complex frequency value---
with very similar behaviors. In both cases, both parameters $\alpha_0$ and $\beta_0$
do appear to converge towards zero,
or at least to a value of modulus $|\beta_0|<10^{-2}$ in the case of $\beta_0$,
as $t_0$ approaches $t_f$.

However, $\alpha_0$ and ---more prominently--- $\beta_0$
clearly deviate from zero for earlier fit starting times.
In particular, $\beta_0$ reaches increasingly negative values,
corresponding to larger and larger damping rates, as $t_0$ is decreased.
Such deviations are of course expected at early times
when nonlinear deviations to equilibrium should still be present.
The observed behavior of $\alpha_0$ and $\beta_0$
is however also compatible with the presence of QNM overtones,
which are damped faster than the fundamental mode.
Their presence may indeed be expected
at least at the intermediate times $t / M \simeq 10$ to $20$,
if the nonlinear dynamics have become sufficiently negligible,
and before these overtones have decayed much below the fundamental mode.

The best-fit $\alpha_0$ and $\beta_0$ values for the higher shear modes
and multipoles typically display similar behaviors as those observed here for $l=2$,
although the residual magnitudes of these quantities towards $t_f$ can be a little larger
than in the $l=2$ case, with $|\alpha_0| \lesssim 1 \%$ and $|\beta_0| \lesssim 5 \%$,
and with typically $\alpha_0 < 0$ and $\beta_0 > 0$. For both the shear and multipoles,
the $l=7$ and $l=10$ modes (already singled out in
Figs.~\ref{fig:single-mode-comparison-shear}
and~\ref{fig:single-mode-comparison-multipole} for their atypical best-fit damping rates)
are exceptions regarding the best-fit $\beta_0$, which takes again substantial positive values
(ranging from $\beta_0 \simeq 0.18$ to $\beta_0 \simeq 0.37$) for very late starting times
$30 M \lesssim t_0 < t_f$ ---where few data points remain--- after being first damped
with increasing $t_0$ for $t_0 \lesssim 30 M$.

\subsection{QNM models including overtones}
\label{subsec:fit-overtones}

In the previous subsection, we noted that the late-time oscillations
of each of the shear modes and multipoles for $2 \leq l \leq 12$ were well modeled
by the real frequency $\omega_{l00}$ of the corresponding
$s=2$ fundamental QNM of the remnant black hole. On the other hand, the damping rates
of these oscillations showed some deviation
to the fundamental QNM imaginary frequencies $\tau_{l00}^{-1}$, especially at large $l$.
Moreover, the deviations on both the real and the imaginary frequencies
generally appeared to increase as earlier times were taken into account,
and to nearly vanish if only the very end of the dataset was considered.
We have accordingly suggested that the shear modes and multipoles
are fully described asymptotically by the fundamental QNMs while the behavior
at more intermediate times (say, around $t = 15 M$)
may correspond to the additional residual presence of higher overtones.

Moreover, as already pointed out in~\cite{pook-kolb2020II}, each of the shear modes
and multipoles clearly displays a steep nonoscillating decay at early times
(at $t/M \lesssim 4$), with a substantially larger damping rate
than the late-time damped-oscillatory regime. Accordingly, none of the modes
can be correctly described by a single damped sinusoid
over the whole time range $[t_{\mathrm{bifurcate}}, t_f]$.
It was noted in~\cite{pook-kolb2020II} that the larger decay rate observed at early times,
while reminiscent of the large values of the imaginary frequencies
of the QNM overtones, generally did not appear to quantitatively match
the imaginary frequency of any particular overtone. It was left as a possibility
that this early damping regime may nevertheless correspond to combined contributions
of multiple overtones.

Accordingly, we will now examine the hypothesis
that the behavior of each of the outer common horizon shear modes and multipoles
is consistently described by a combination of QNMs, including overtones,
over the whole available time range from the very formation of this horizon
or shortly afterwards. To this end, we consider
the multiple-tone model given by Eq.~\eqref{eq:rdmodel_ansatz}
with all complex frequencies set to the QNM values, \emph{i.e.},
$\alpha_{l0n} = \beta_{l0n} =0 \; \forall l,n$:
\begin{equation}
 X_l = \sum_{n=0}^{\nmax} A_n \, \exp \left[ -\frac{t-t_r}{\tau_{l0n}} \right] \,
 \cos \left[ \omega_{l0n} (t-t_r) + \phi_n \right] \; ,
 \label{eq:QNM-multitone-model}
\end{equation}
where we have again dropped the $m=0$ index and the implicit $l$ dependence
on the free parameters: $A_n \equiv A_{l0n}$ and $\phi_n \equiv \phi_{l0n}$. 
We then check for the agreement of such a combination of QNMs
to the numerically computed shear modes and multipoles
as the total number $\nmax$ of overtones is varied.

We here aim at directly comparing the above class of models to the NR results,
rather than at accurately estimating best-fit frequencies or amplitudes.
Accordingly, for a more straightforward comparison and interpretation,
in this subsection we will directly use the models and NR datasets
without applying a rescaling procedure such as that of Eq.~\eqref{eq:rescaling}.

\subsubsection{Shear modes}
\label{subsubsec:shear_overtones}

\begin{figure*}
  \centering    
  \includegraphics[width=\columnwidth]{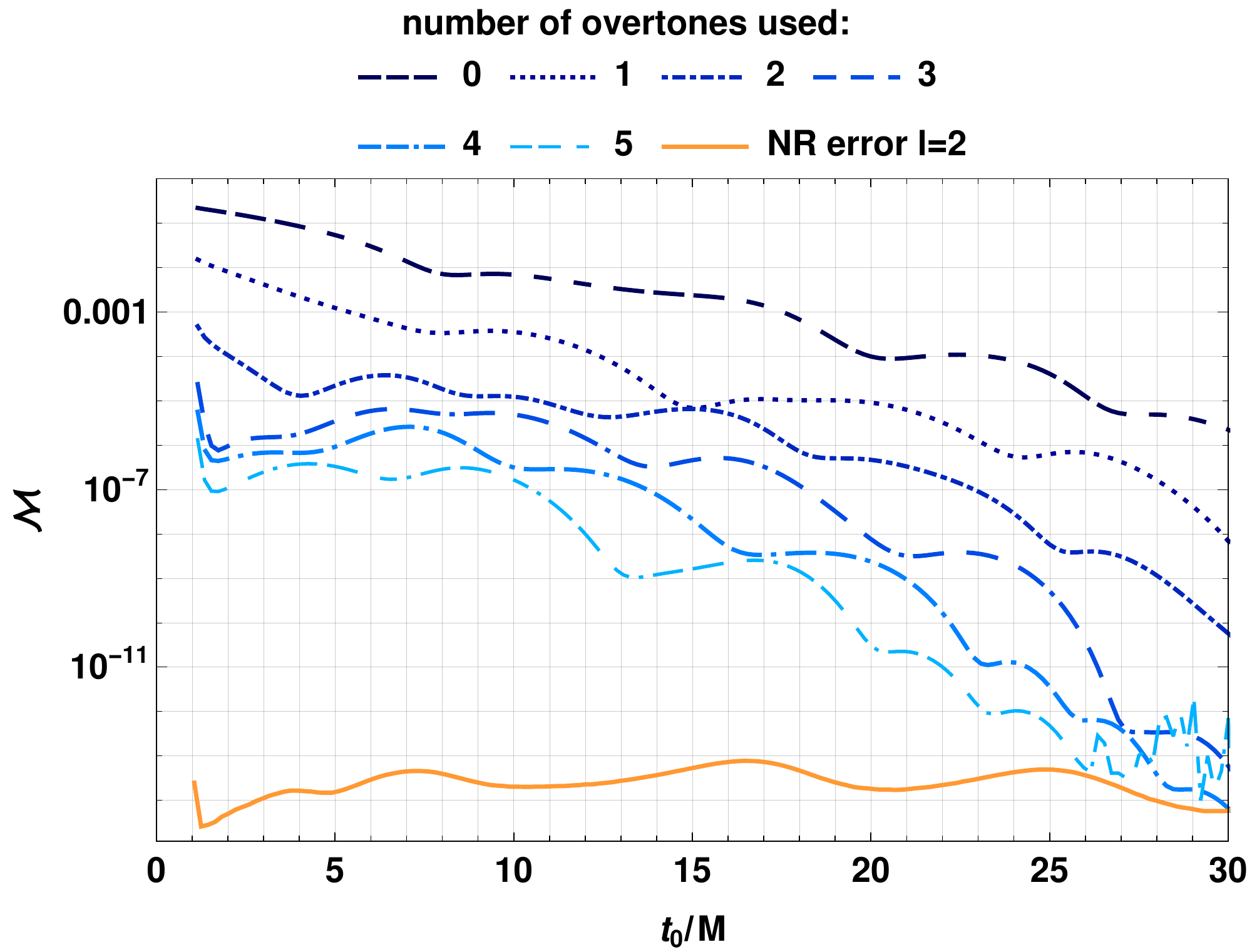}
  \includegraphics[width=\columnwidth]{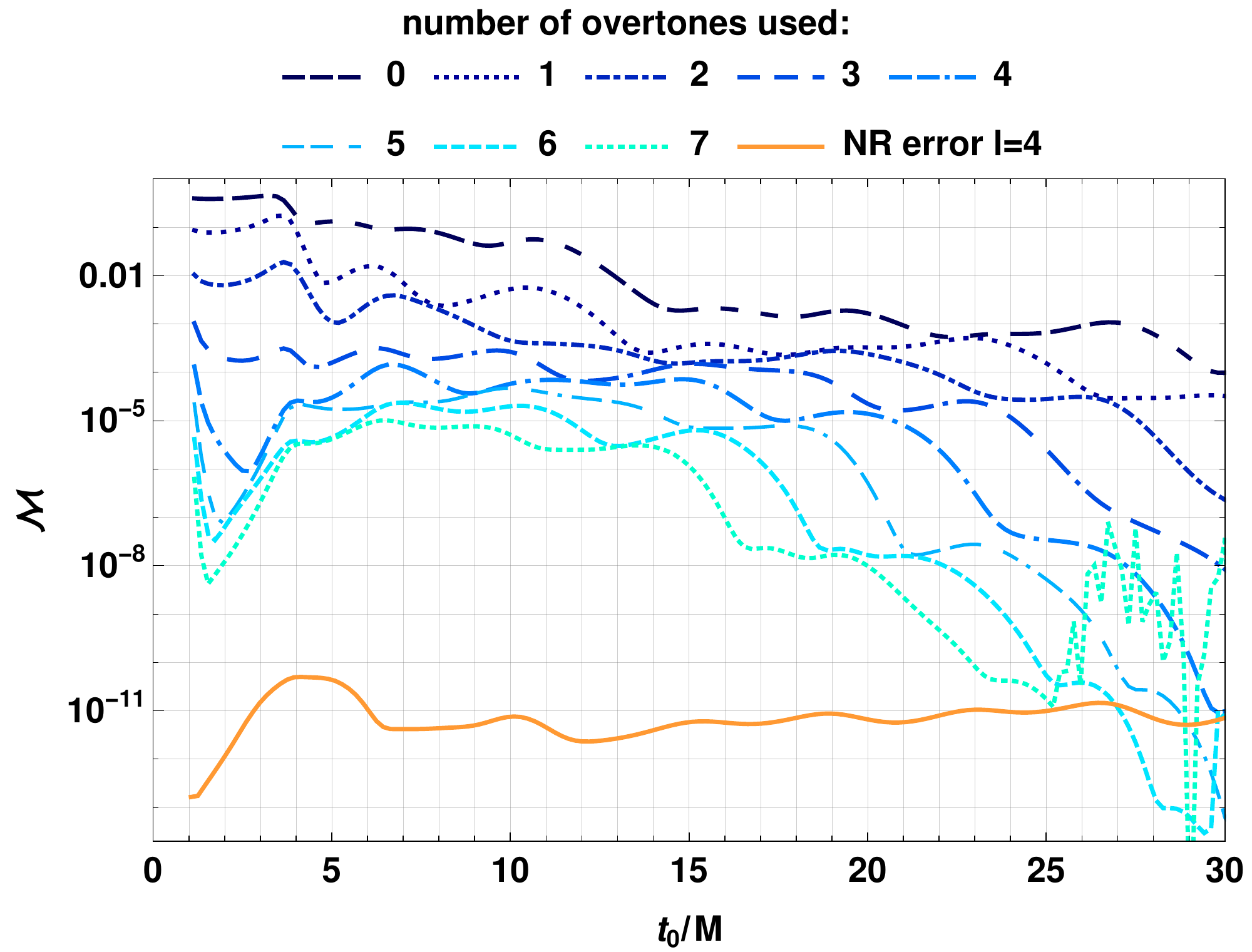}
    \includegraphics[width=\columnwidth]{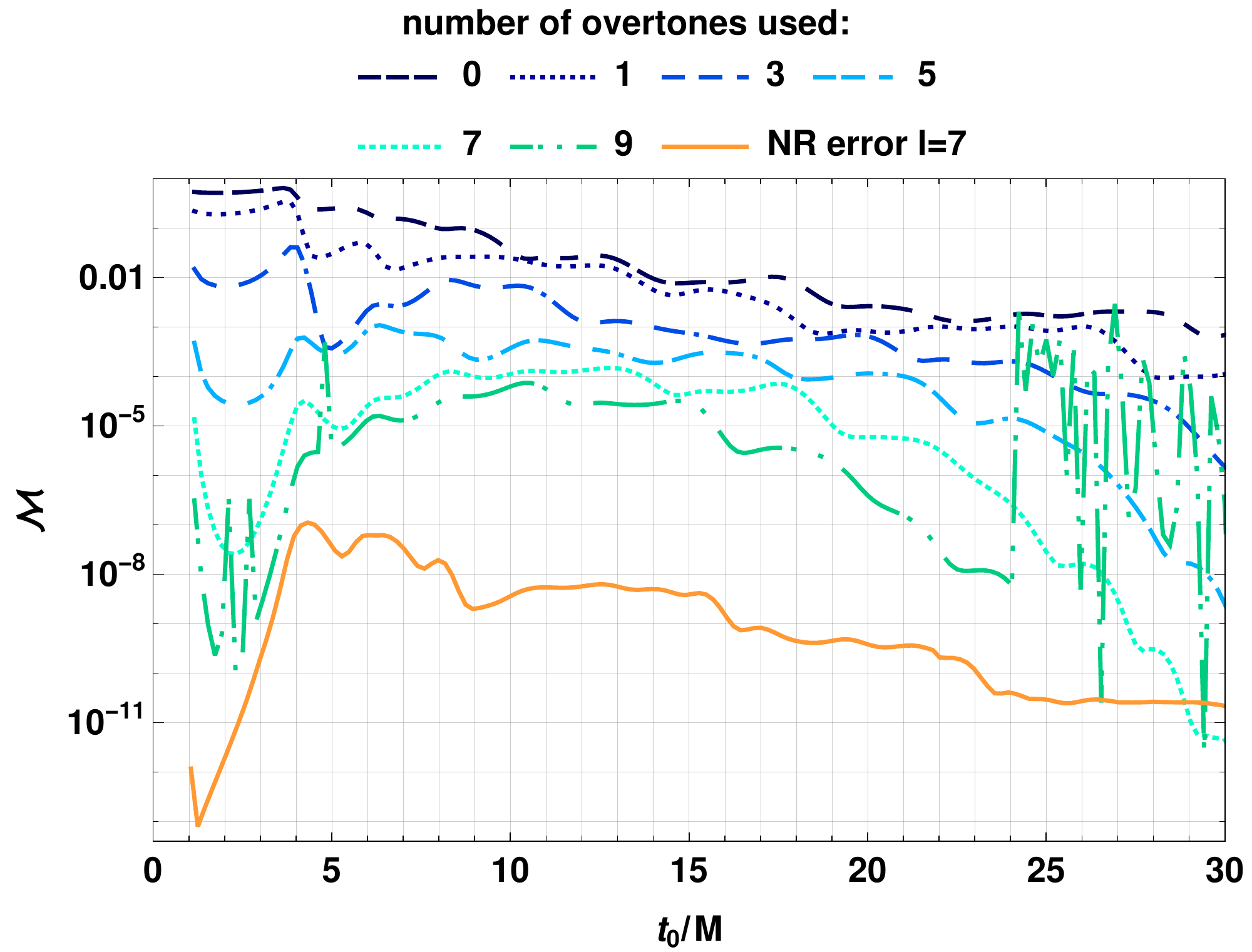}
    \includegraphics[width=\columnwidth]{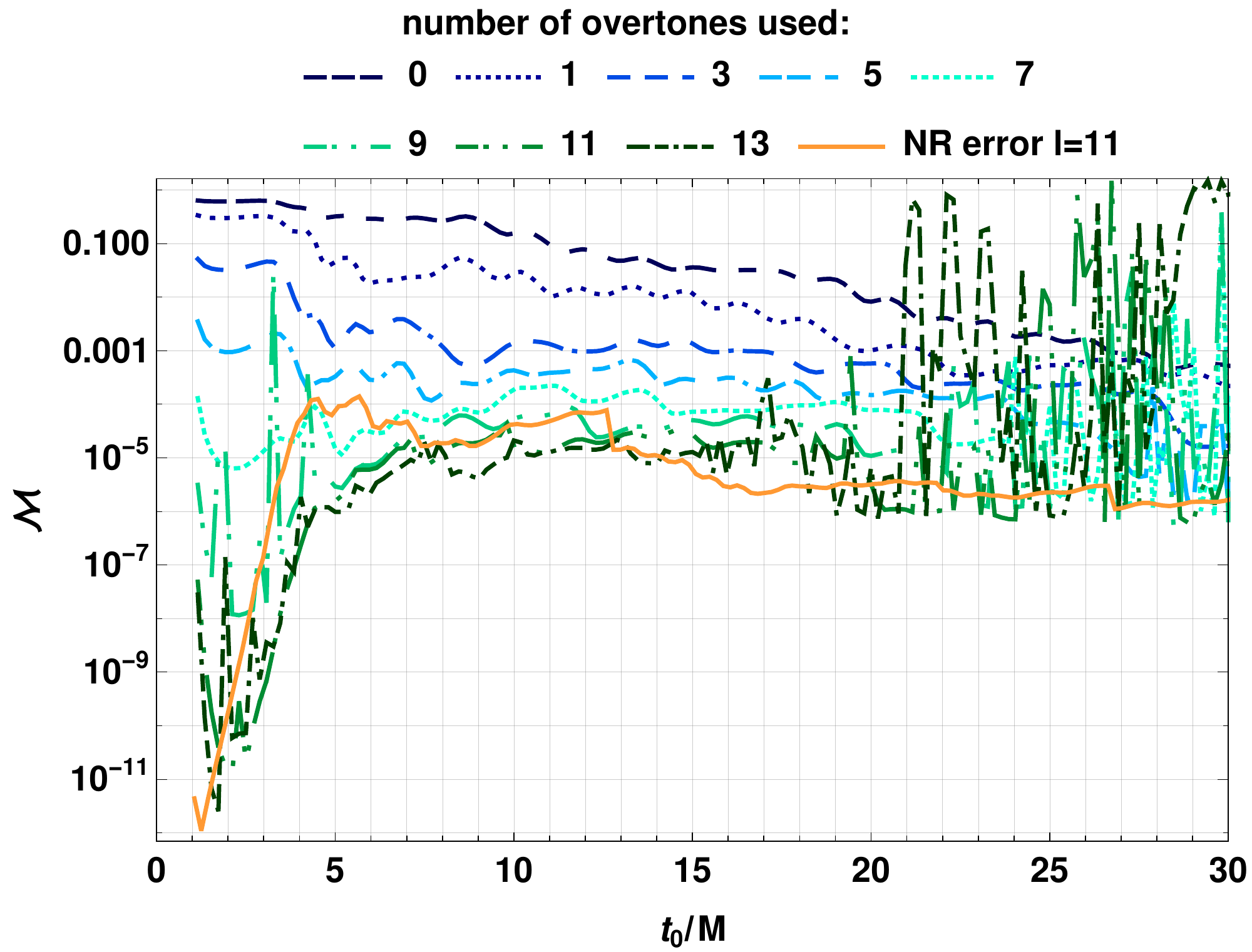}
  \caption{Mismatch between a sample of the NR outer horizon shear modes
  (upper left panel: $l=2$; upper right: $l=4$; lower left: $l=7$; lower right: $l=11$)
  and the sum-of-QNM-tones model of Eq.~\eqref{eq:QNM-multitone-model}
  at the best-fit parameter values, for a variable number of overtones $\nmax$
  and as a function of the fit starting time $t_0$. An estimate of the numerical uncertainty
  as a function of $t_0$ is also provided (orange continuous line;
  see Sec.~\ref{subsubsec:shear_overtones}).
  For the lower two panels,  for the sake of readability we have only included the models
  with the fundamental mode only ($\nmax=0$) or with odd numbers of overtones.
  Mismatch curves with even numbers of overtones
  remain consistent with the overall trend of decreasing $\mathcal{M}$
  with increasing $\nmax$ and typically lie in between the curves
  corresponding to the adjacent odd numbers.}
  \label{fig:mismatch-shear}
\end{figure*}

We first consider the shear modes $\sigma_l$, with, as in the previous subsection, $2 \leq l \leq 12$.
We begin by considering, for each $l$, how the mismatch $\mathcal{M}$
between the best-fit model on $[t_0,t_f]$ and the NR data, improves
as more and more overtones are included in the model,
depending on the fit starting time $t_0$.
The results are shown in Fig.~\ref{fig:mismatch-shear} for a sample of $l$ values
($l=2$, $l=4$, $l=7$ and $l=11$), with $\mathcal{M}$ as a function of $t_0$
and for multiple values of the total number of overtones $\nmax$.
We do not go beyond $t_0/M = 30$ here as the number of data points
in the remaining time interval would become too low for the fitting algorithm
to always converge, especially for large $\nmax$.
The mismatch between the highest-resolution NR data ($\mathrm{res} = 240$)
and the NR results at the next-highest resolution ($\mathrm{res} = 180$)
on the time range $[t_0,t_f]$
is also shown as a function of $t_0$, as an estimate of the numerical error, for comparison.
We have also checked the validity of this mismatch-based estimate of the NR error
in a similar way as explained for the local error $\epsilon_{180-240}$
in footnote~\ref{fn:NRerror}. The other values of $l$ not shown here
typically display characteristics intermediate between those presented below.

For $\nmax=0$ we find again
that the fundamental QNM alone matches the shear modes
better and better towards later times, but also that for a given $t_0$,
the quality of the fit provided by this fundamental mode alone
overall degrades for increasing $l$. Both of these trends still hold
for models additionally including a small number (\emph{e.g.}, $\nmax=1$ or $2$)
of QNM overtones.

Furthermore, we observe that the quality of the fit clearly improves
for almost any $t_0$ as the number of overtones is increased.
For $l=2$ and $l=4$ here, one can notice a relatively sharp decrease of the mismatch
at small $t_0$ to small values $\mathcal{M} < 10^{-5}$
for a certain number $\nmax$ of overtones,
beyond which adding more overtones decreases the mismatch less significantly.
This occurs at $\nmax=3$ for $l=2$ and around $\nmax=4$ for $l=4$.
This suggests that these numbers of QNM overtones
provide a good modeling of the entire dataset for these modes,
at least beyond the first $0.5 M$ or so after $t_{\mathrm{bifurcate}}$.
Such a trend does not appear as clearly for the higher $l$ values shown here,
but the mismatch still gets down to small values at early $t_0$
for a sufficiently large number of overtones.
The number of overtones needed for the mismatch to stay below a certain threshold
appears to increase with $l$. For large $l$,
the numerical error becomes too large for reliable constraints on large numbers of overtones,
and for estimating the number of overtones needed
for the mismatch to stay below a too small threshold. For $l=11$ for instance,
the mismatch with the $\mathrm{res} = 180$ results
reaches $\mathcal{M} \simeq 10^{-4}$ for certain values of $t_0$,
and it appears that any improvement in the mismatch
by increasing the number of overtones beyond $\nmax=6$
lies within this numerical uncertainty.

While a useful synthetic quantitative tool,
the mismatch is based on absolute deviations between the data and the model,
and for this reason it does not necessarily clearly reflect
how well the damped behavior of the modes is represented by the model at all times.
For such damped data, it will more accurately measure the relative deviations to the model
towards the early parts of the time interval considered.

\begin{figure*}
  \centering    
  \includegraphics[width=\columnwidth]{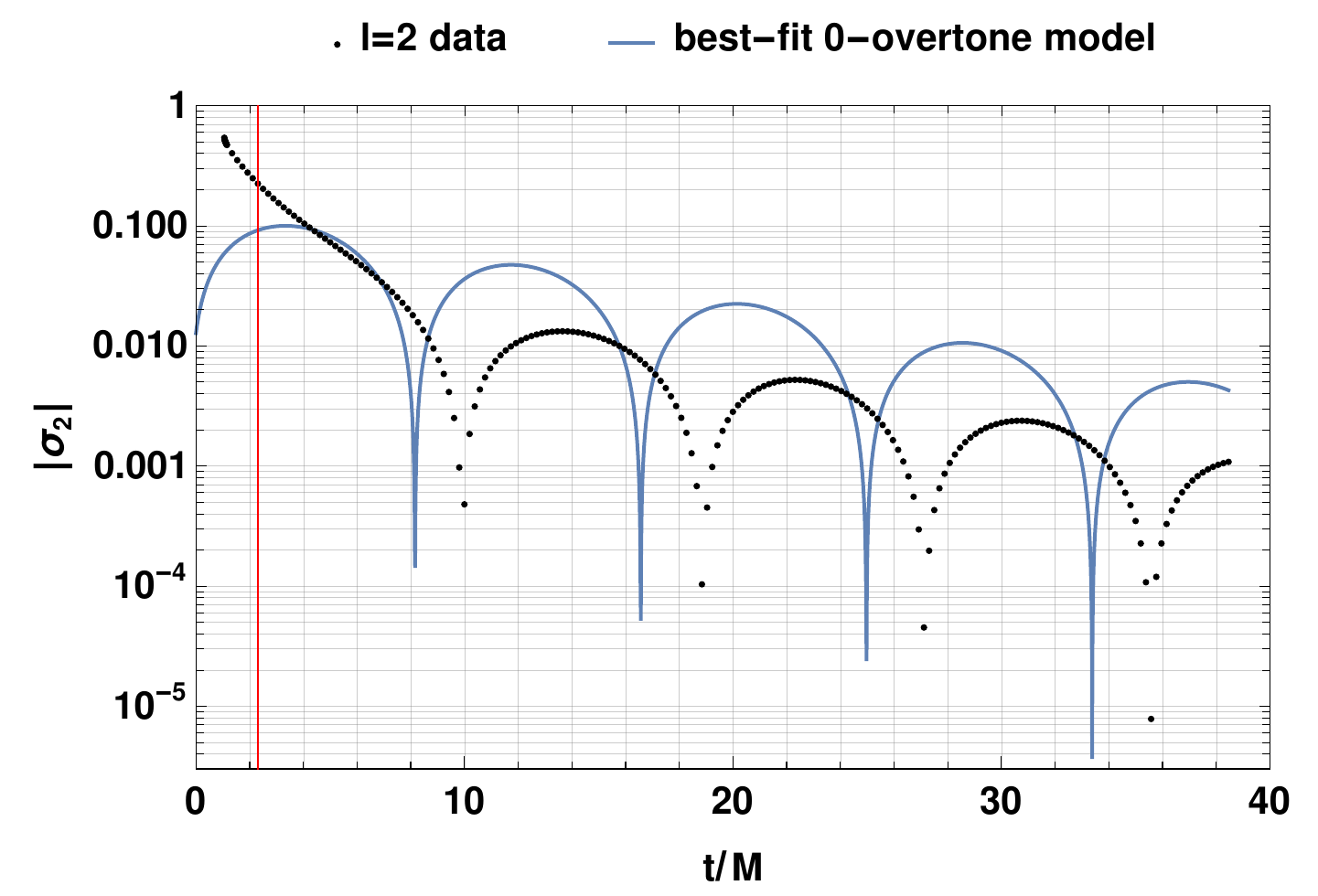}
  \includegraphics[width=\columnwidth]{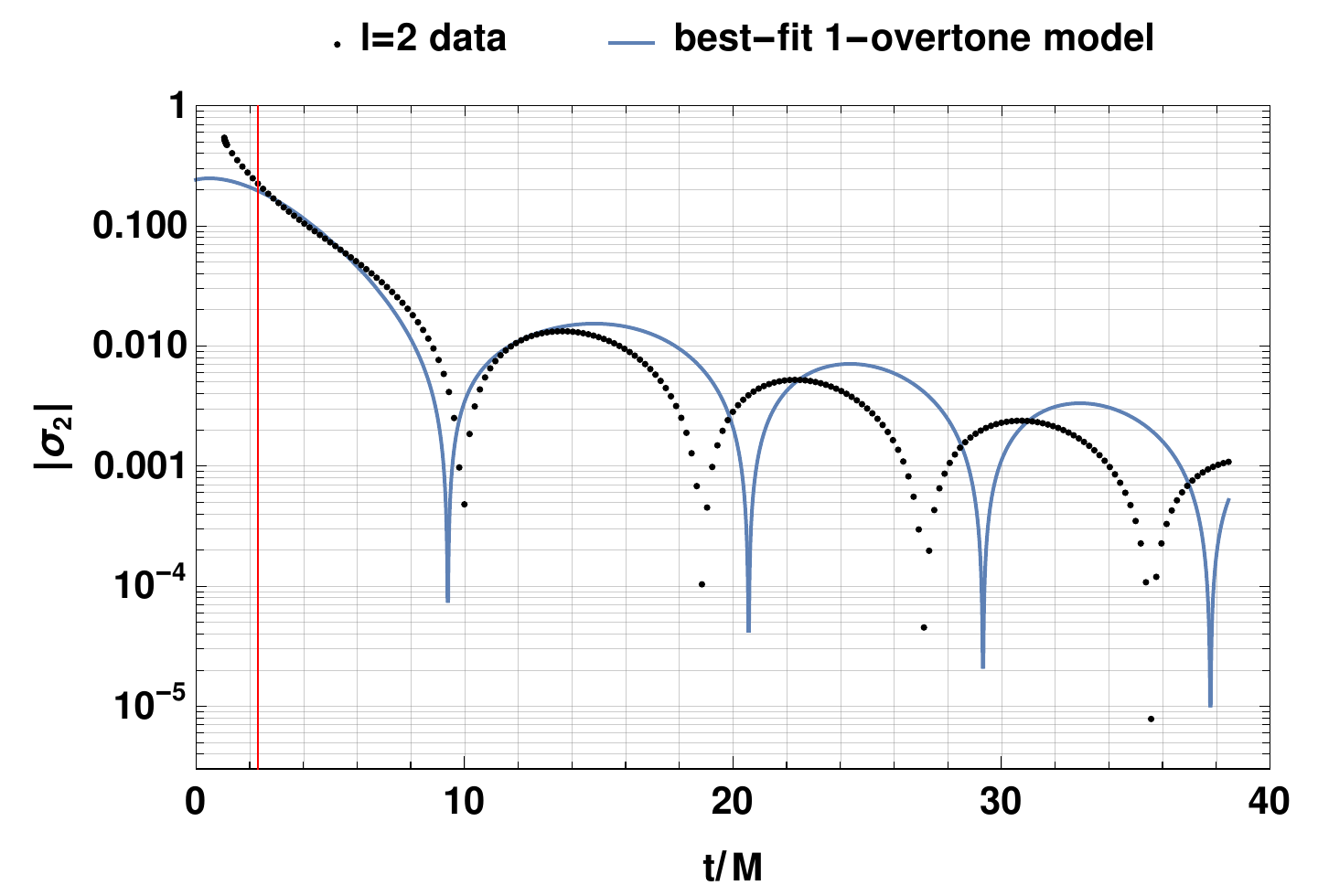}\\
  \includegraphics[width=\columnwidth]{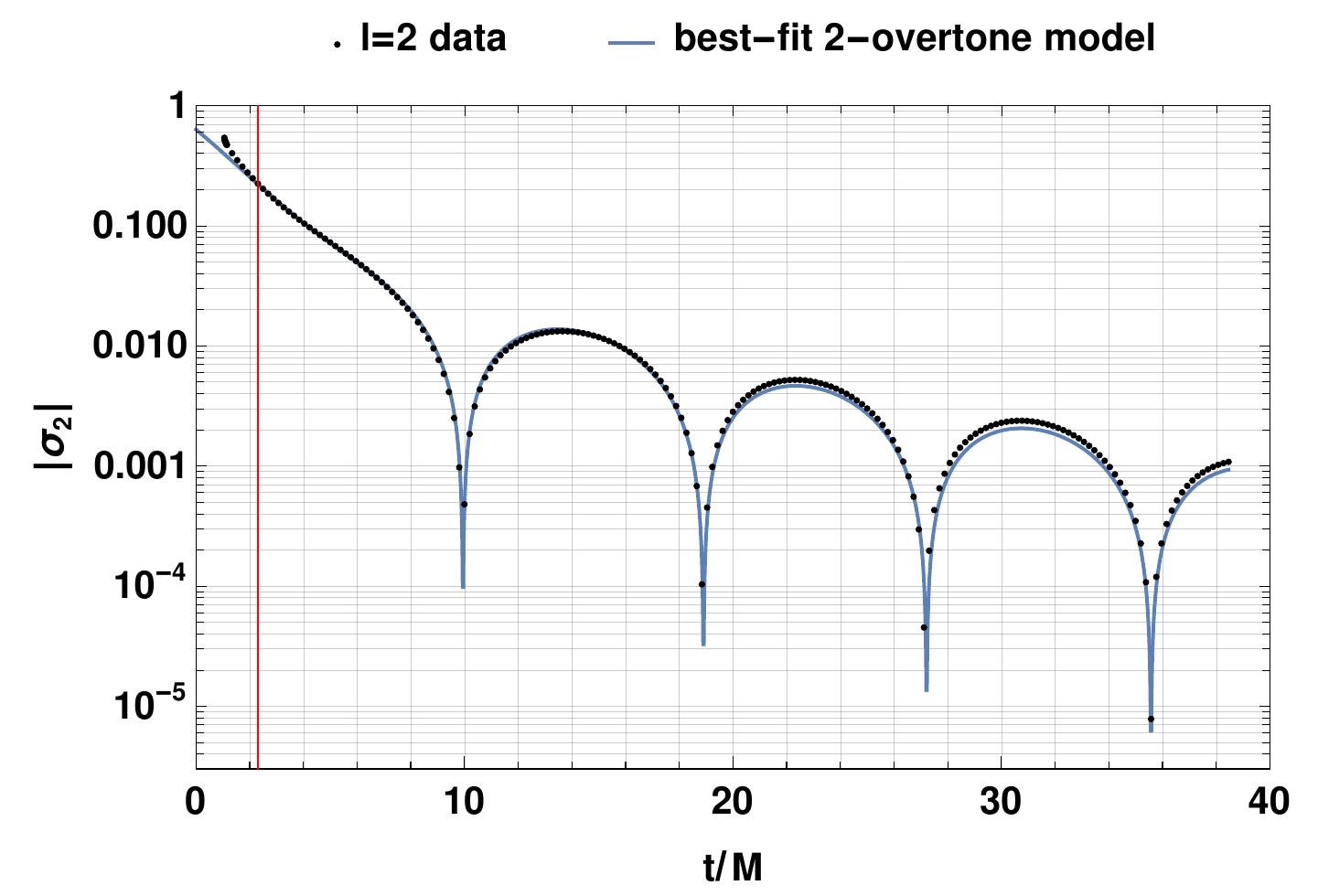}
  \includegraphics[width=\columnwidth]{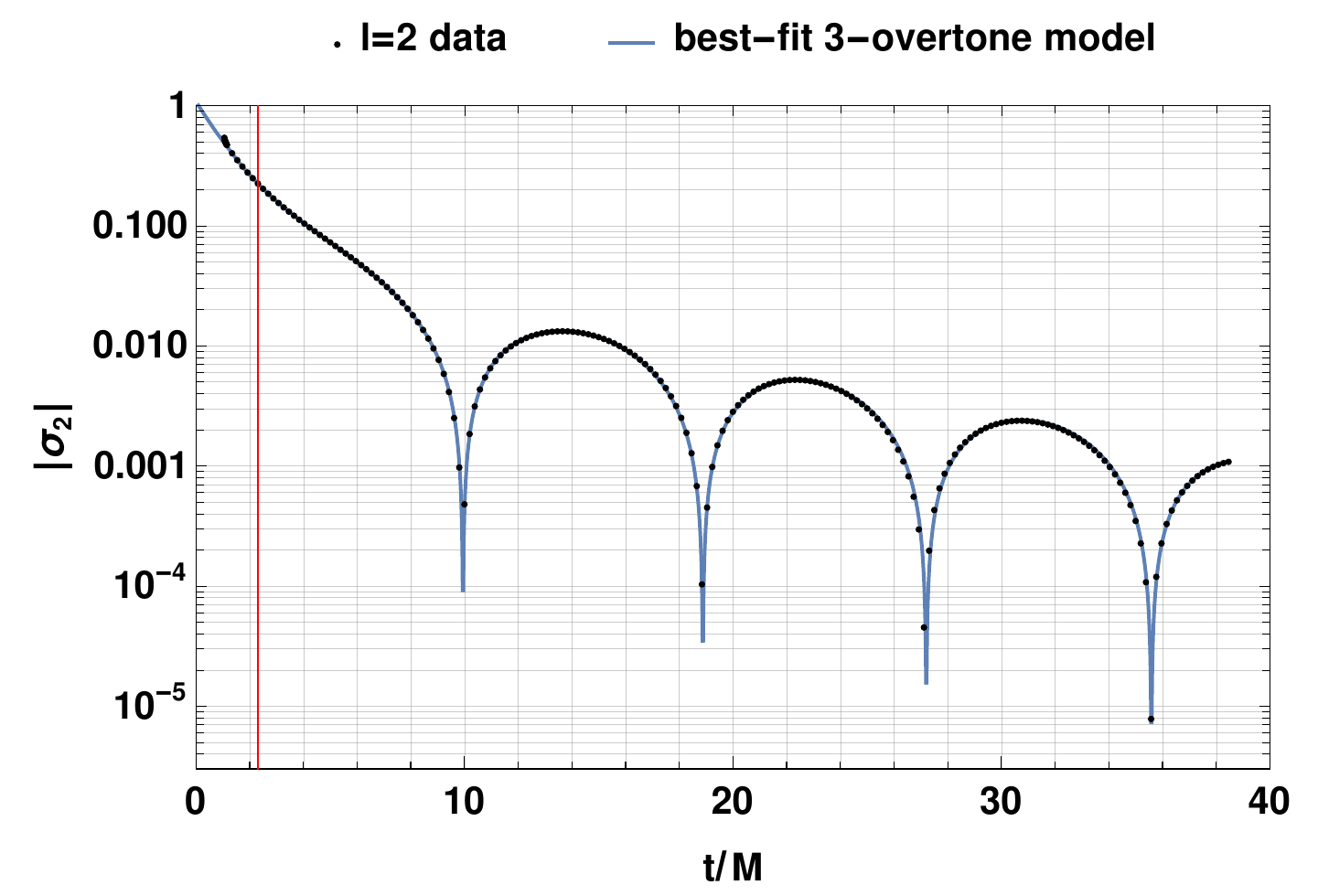}\\
  \includegraphics[width=\columnwidth]{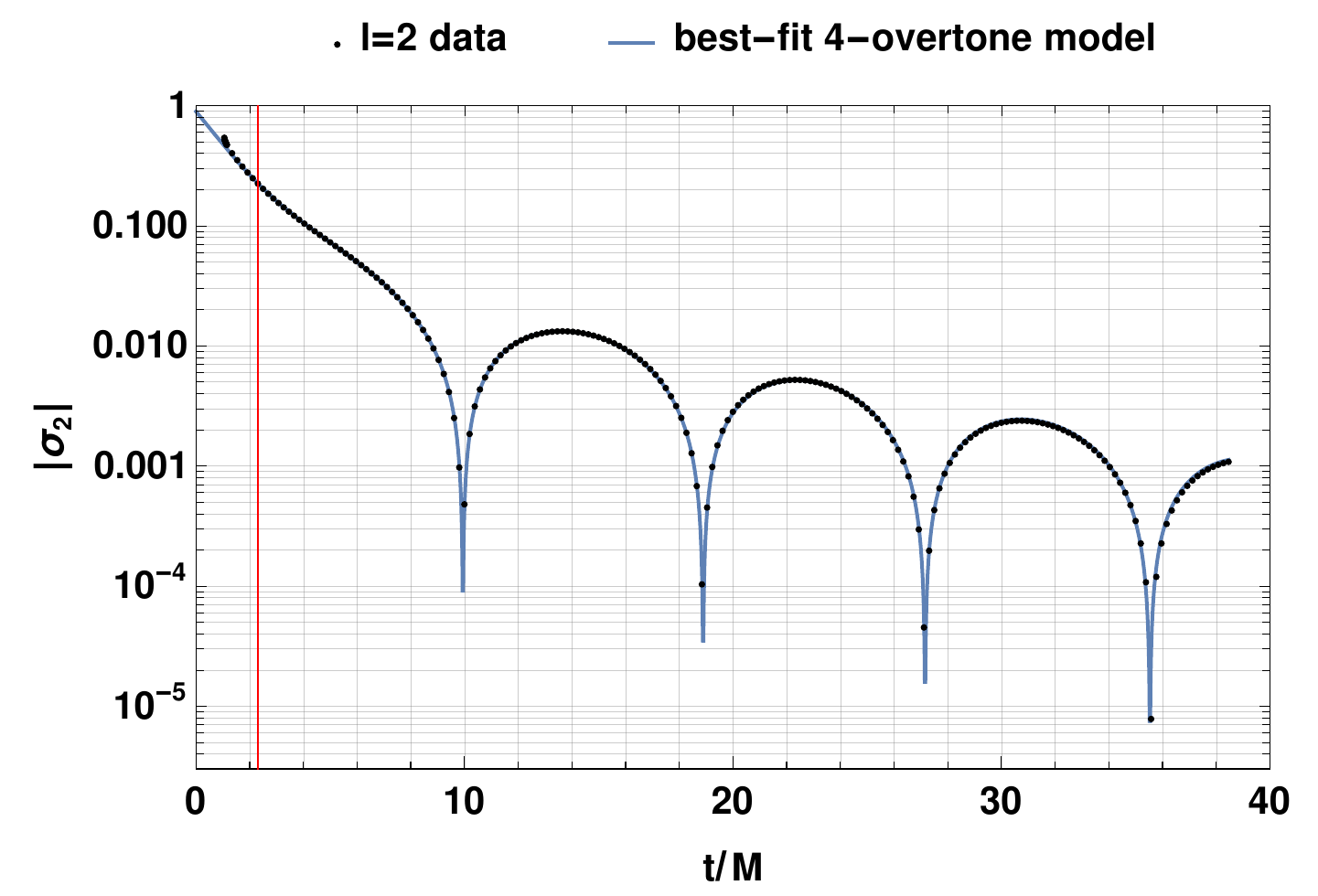}
  \includegraphics[width=\columnwidth]{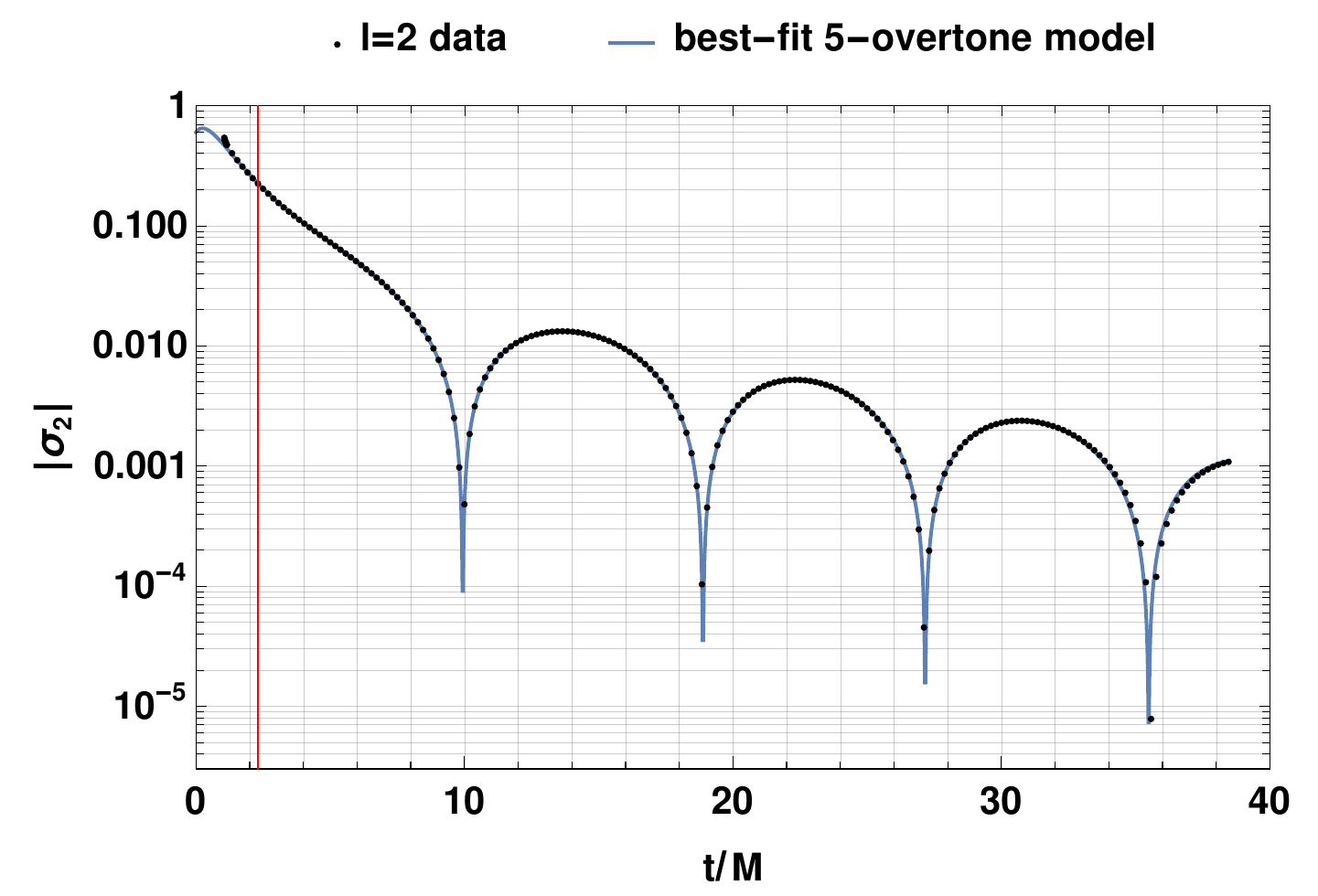}
  \caption{Direct comparison, as a function of the simulation time $t$,
  of the NR $l=2$ shear mode (black dots)
  and the associated best-fit models of the class \eqref{eq:QNM-multitone-model}
  (blue continuous lines) including, from left to right and top to bottom,
  $\nmax=0$ to $\nmax=5$ overtones. The entire NR dataset is shown,
  \emph{i.e.}, from $t = t_{\mathrm{bifurcate}}$ to $t=t_f$.
  All of the fits shown in this figure were obtained
  using the same fit starting time value $t_0 \simeq 2.3 M$.
  The corresponding $\{ t=t_0 \}$ vertical line, indicating which part of the dataset
  (to the right of this line) was actually used to constrain the model,
  is indicated in red on each plot. One can note the good agreement of the model to the data
  both after \emph{and before} this starting time for $\nmax \geq 3$.}
  \label{fig:l2-overtonefits-shear}
\end{figure*}

\begin{figure*}
  \centering    
  \includegraphics[width=\columnwidth]{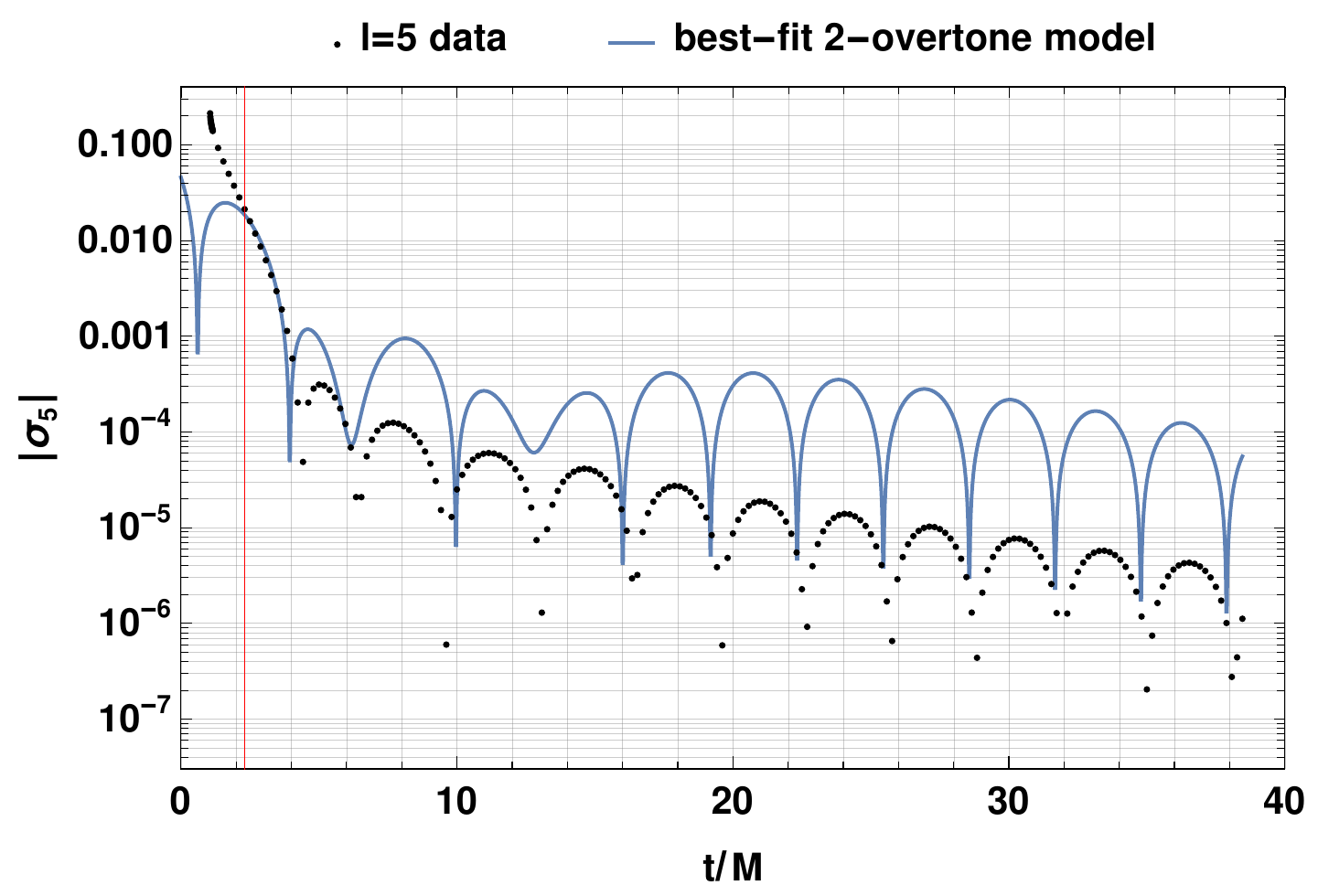}
  \includegraphics[width=\columnwidth]{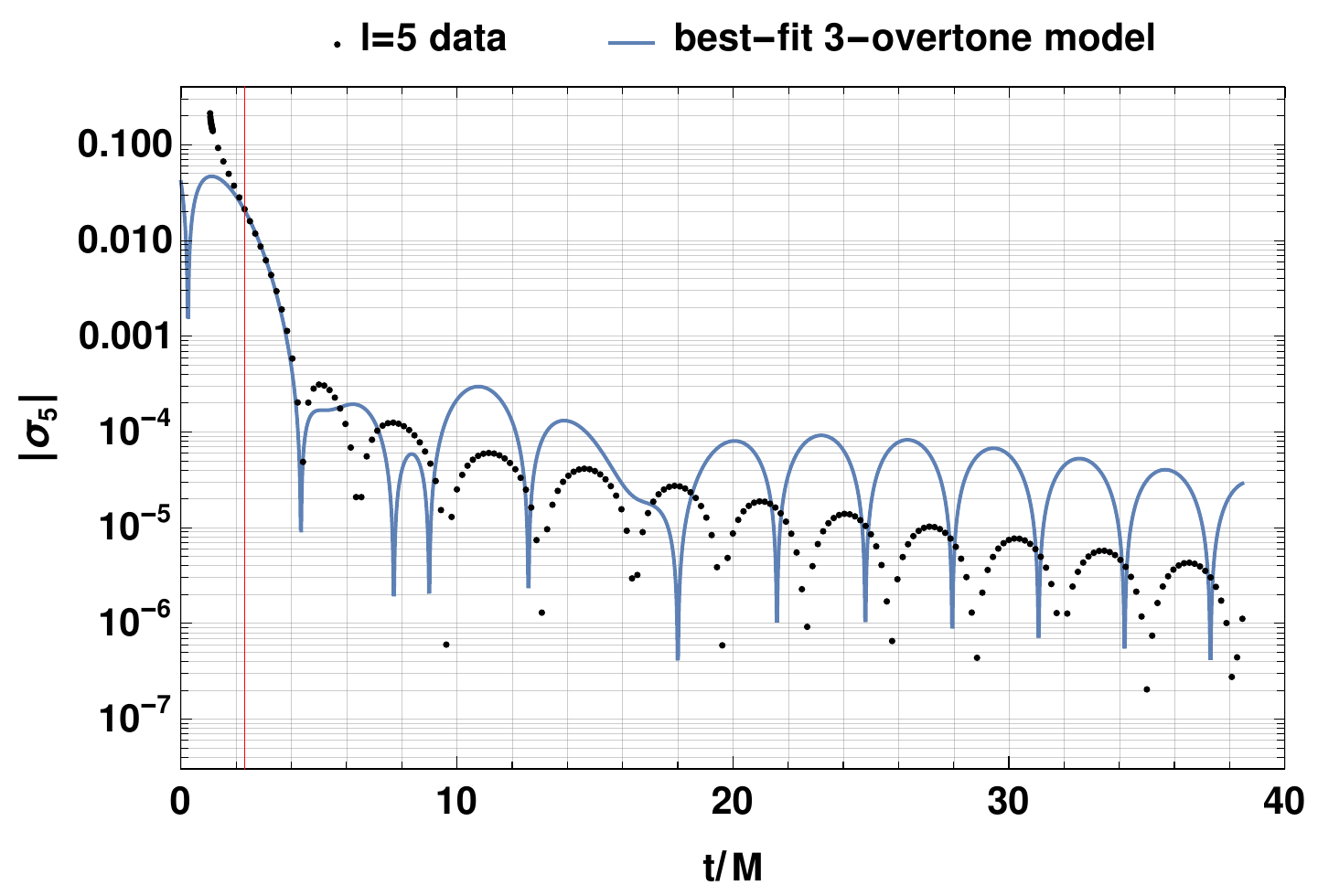}\\
  \includegraphics[width=\columnwidth]{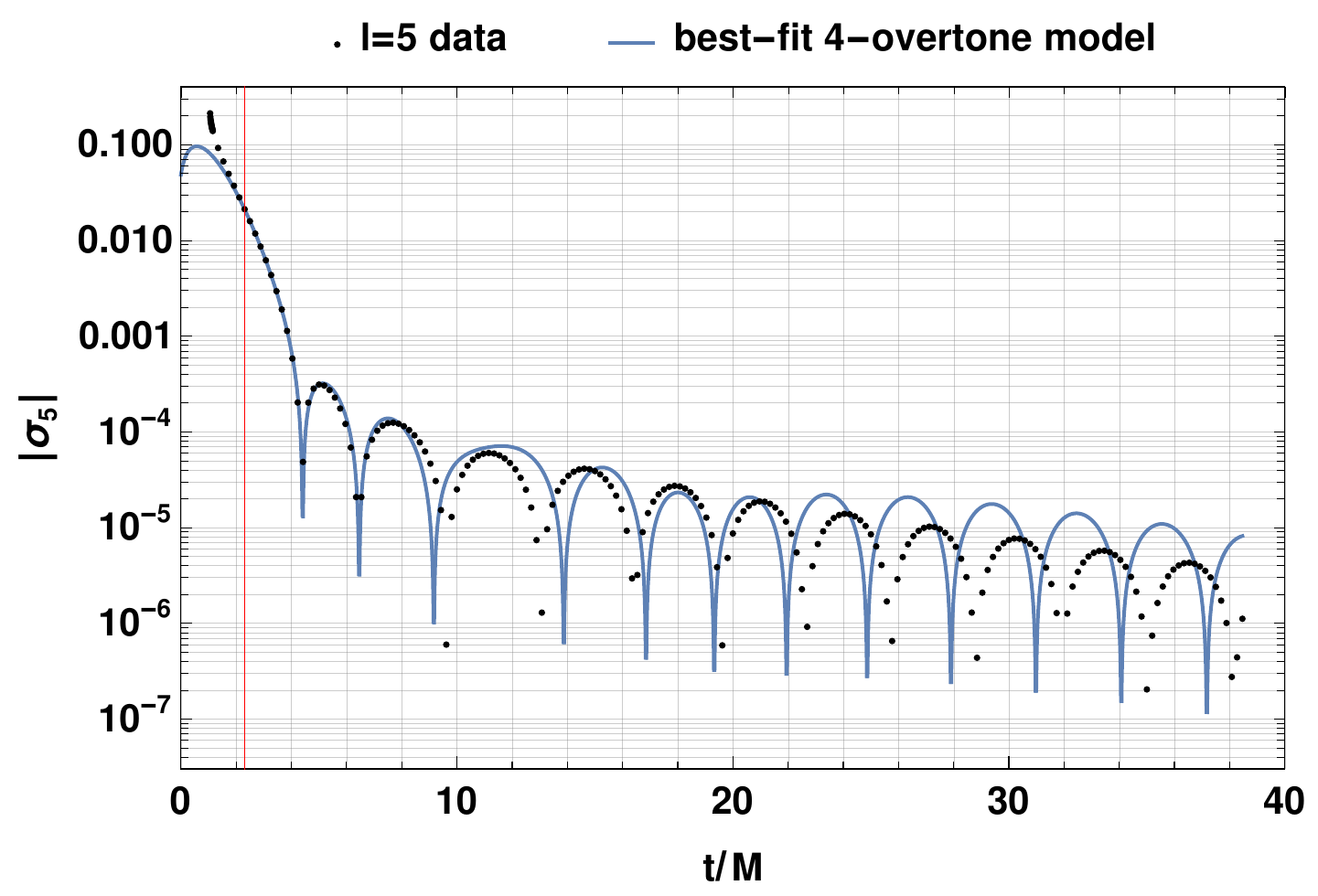}
  \includegraphics[width=\columnwidth]{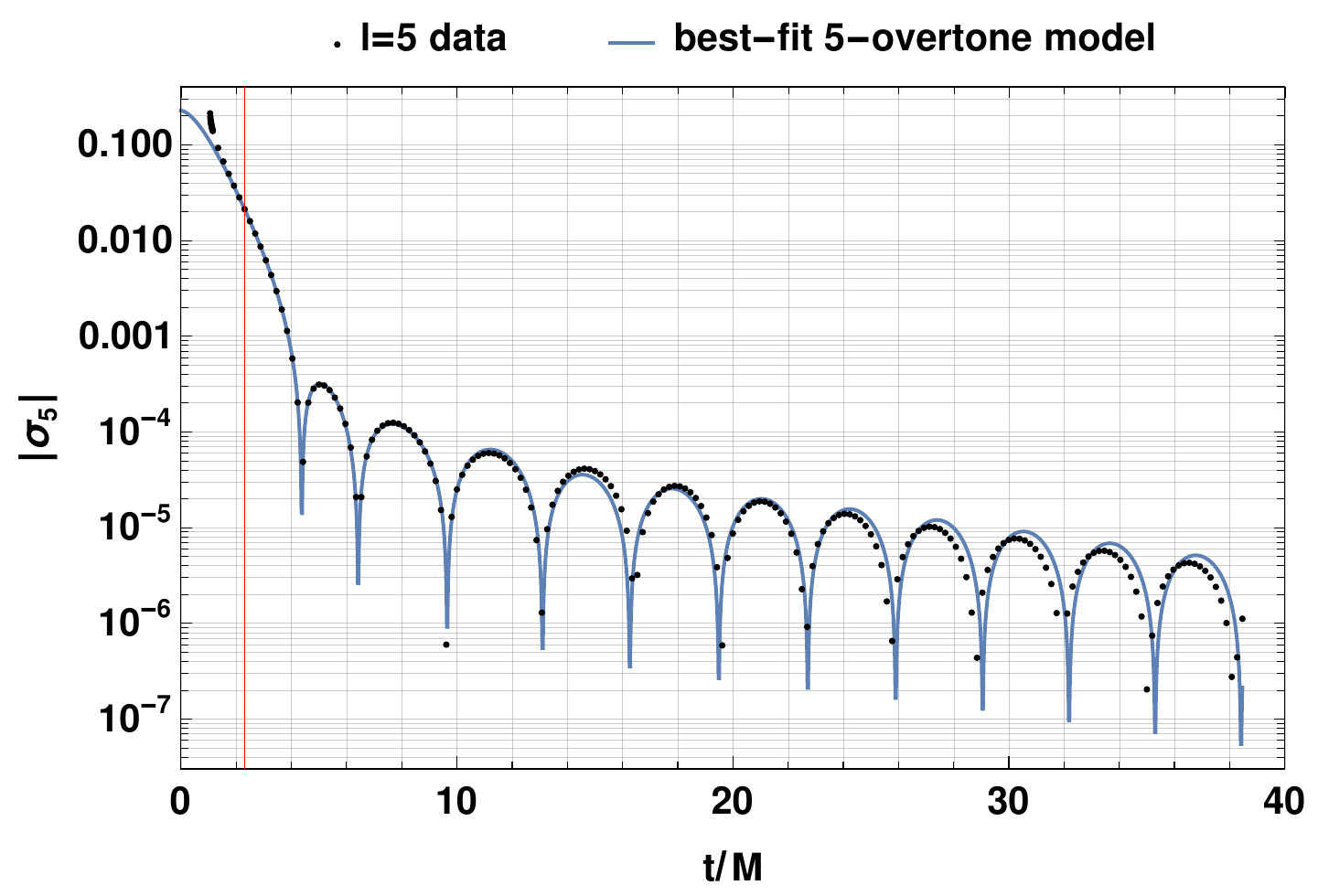}\\
  \includegraphics[width=\columnwidth]{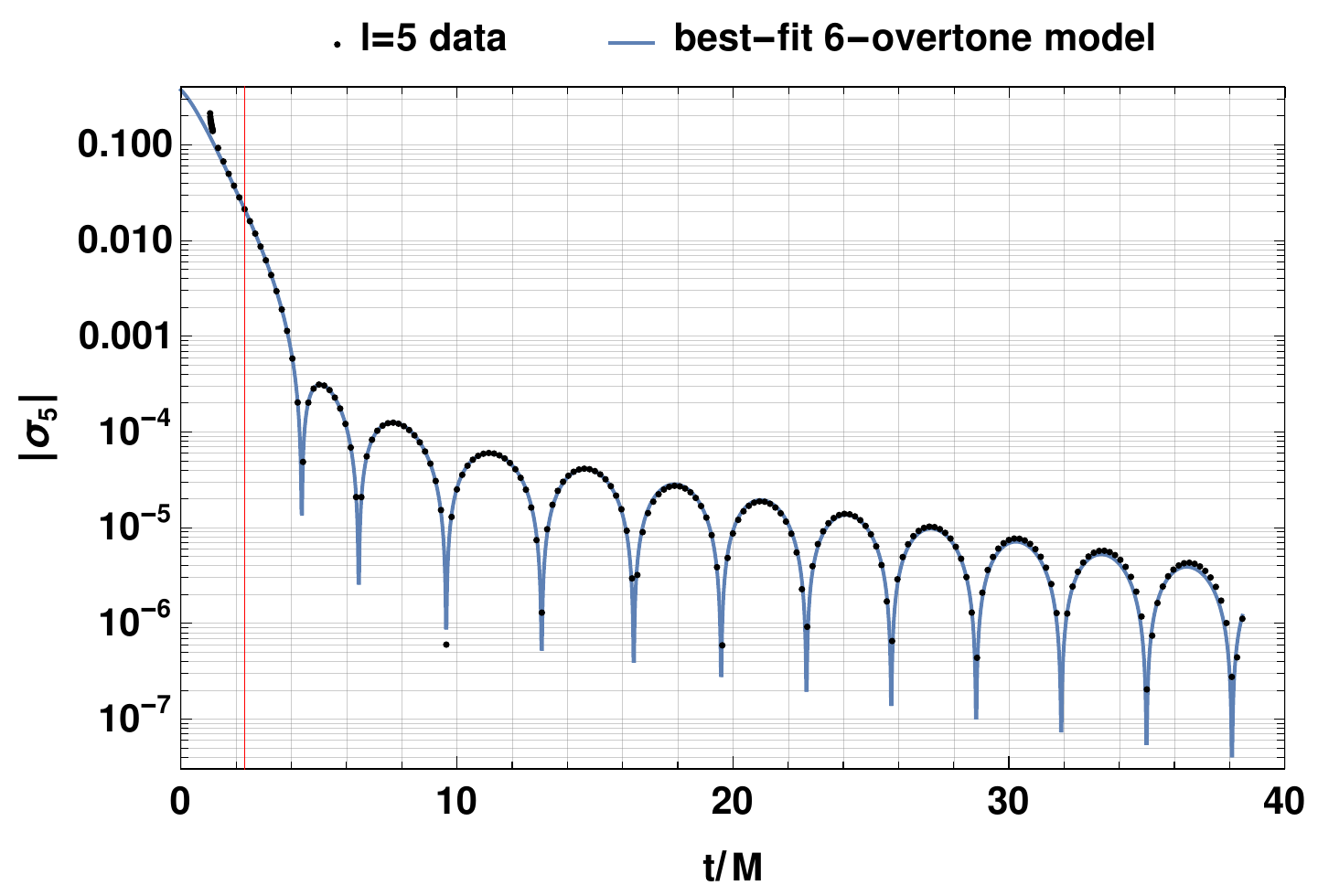}
  \includegraphics[width=\columnwidth]{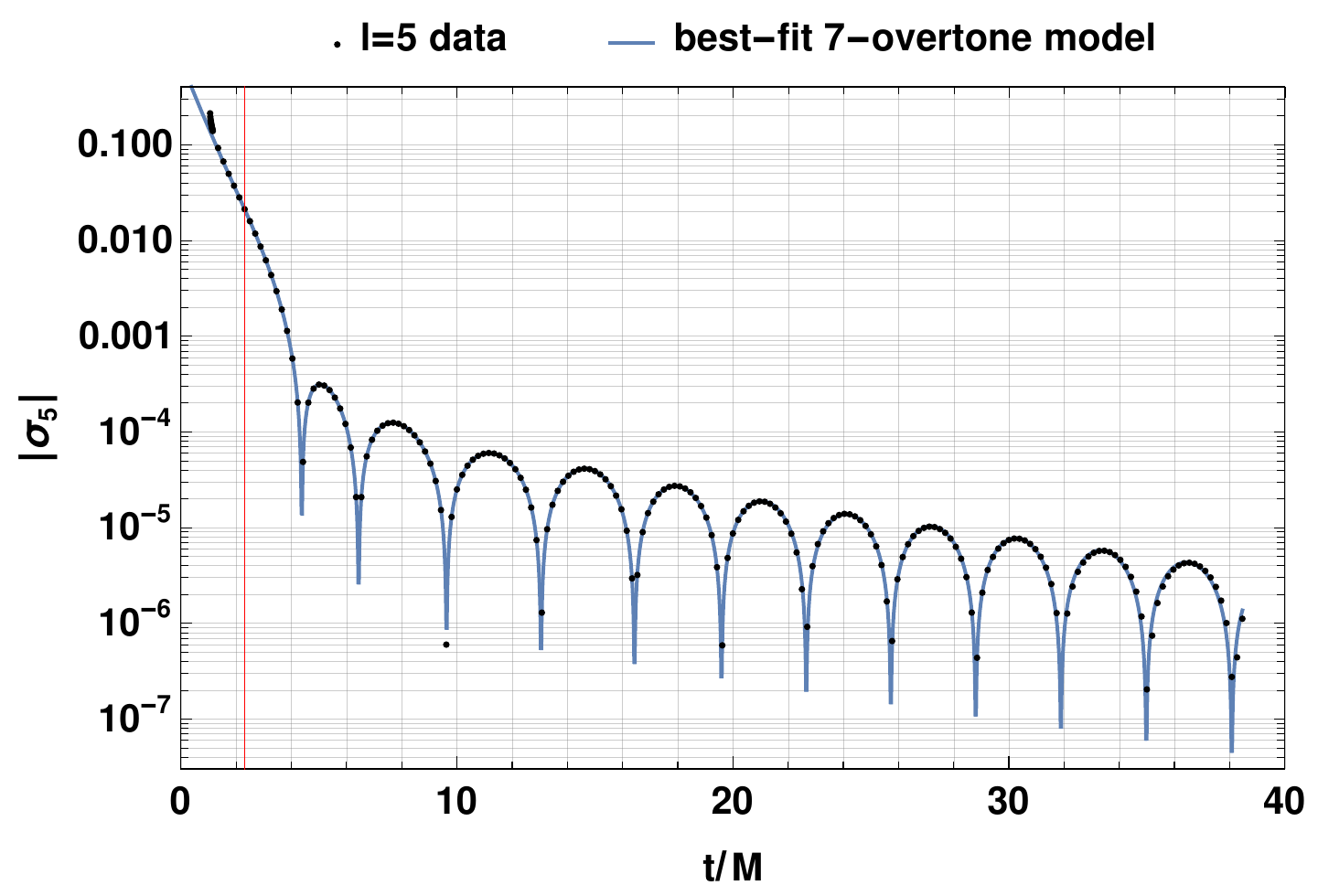}
  \caption{Same as Fig.~\ref{fig:l2-overtonefits-shear} for the $l=5$ shear mode
  and $\nmax=2$ to $\nmax=7$ overtones. The same fit starting time is used.
  A relatively good agreement to the data is obtained both after and before the fit starting time
  at $\nmax =6$, and this is further improved at $\nmax =7$.}
  \label{fig:l5-overtonefits-shear}
\end{figure*}

\begin{figure*}
  \centering    
  \includegraphics[width=\columnwidth]{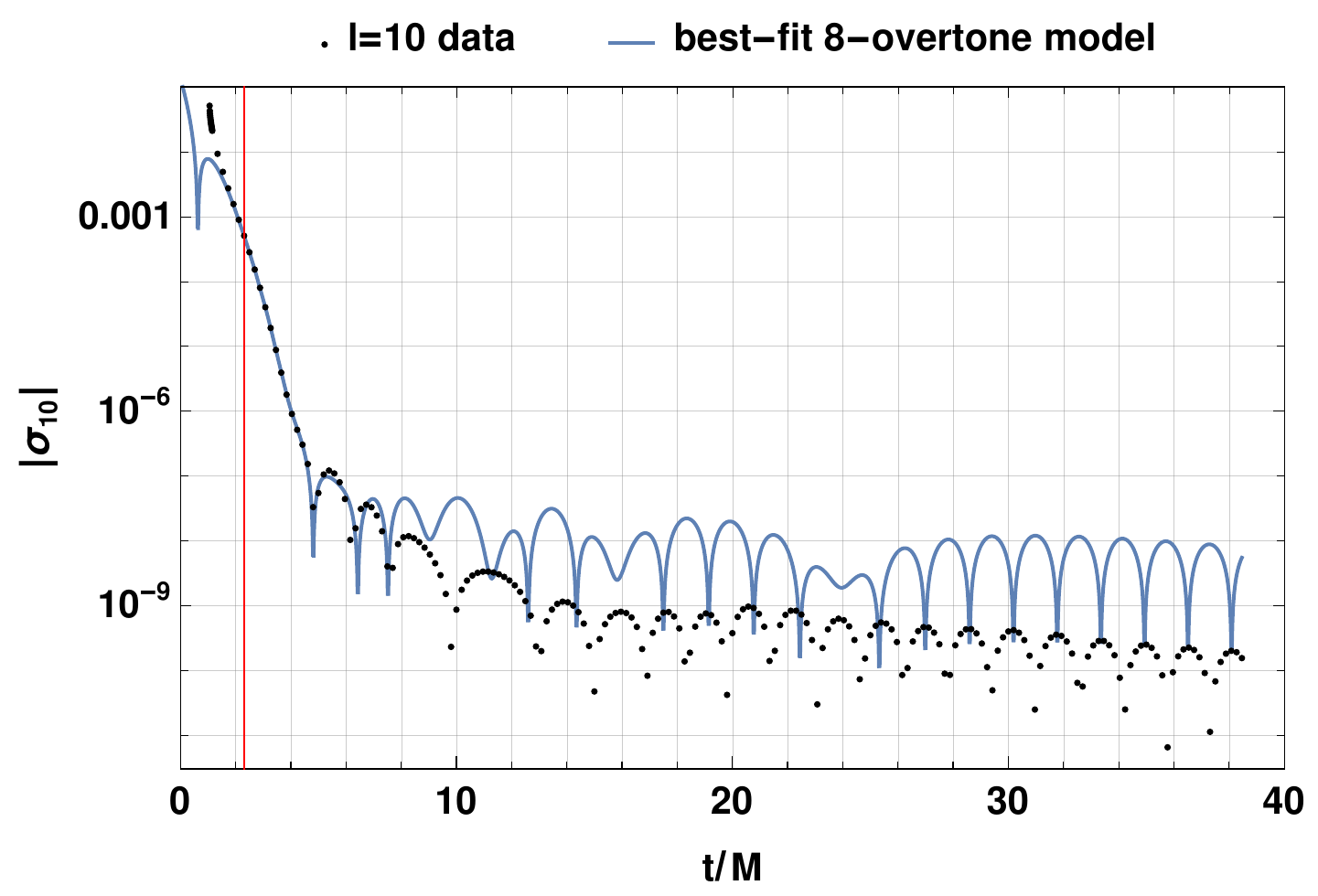}
  \includegraphics[width=\columnwidth]{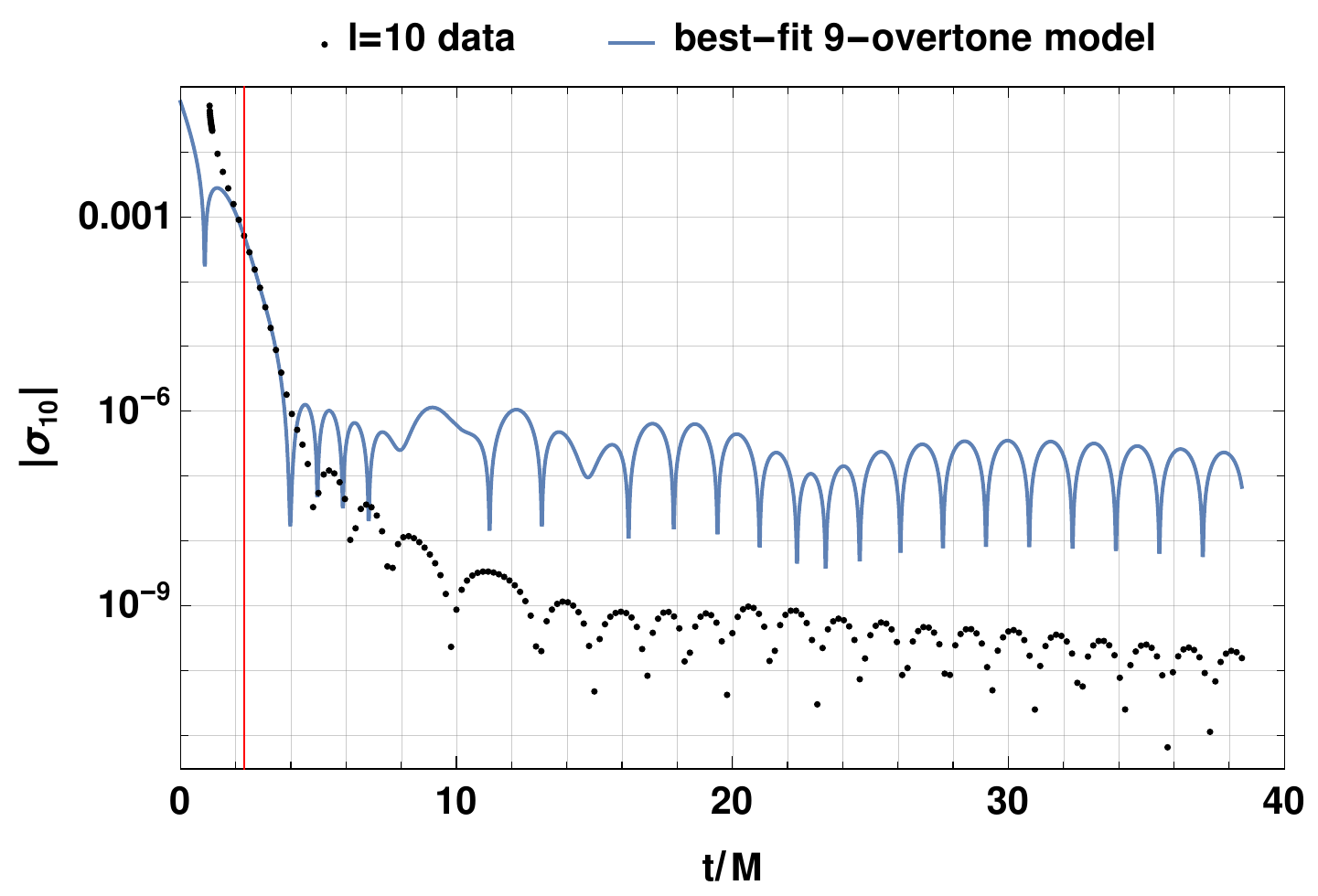}\\
  \includegraphics[width=\columnwidth]{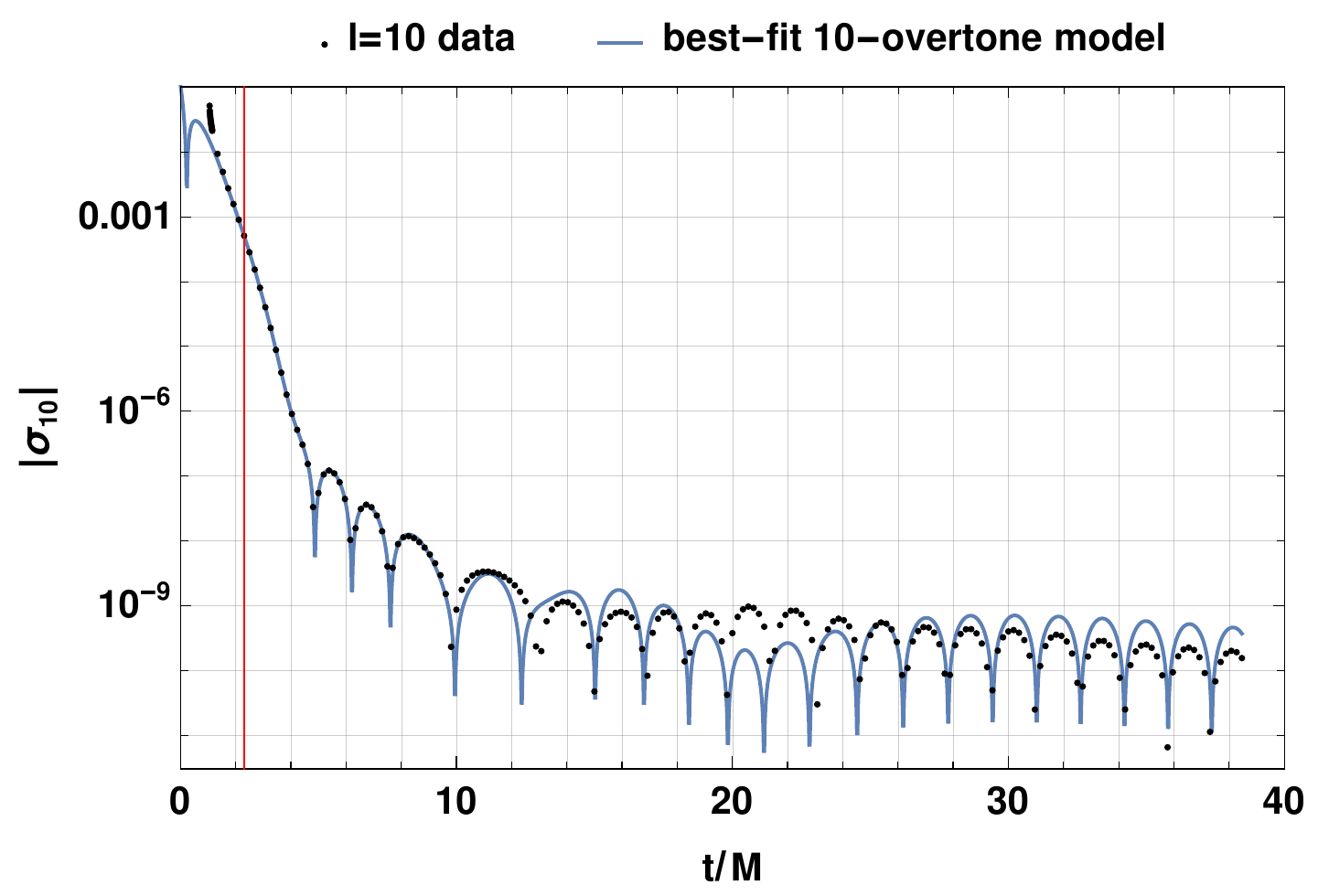}
  \includegraphics[width=\columnwidth]{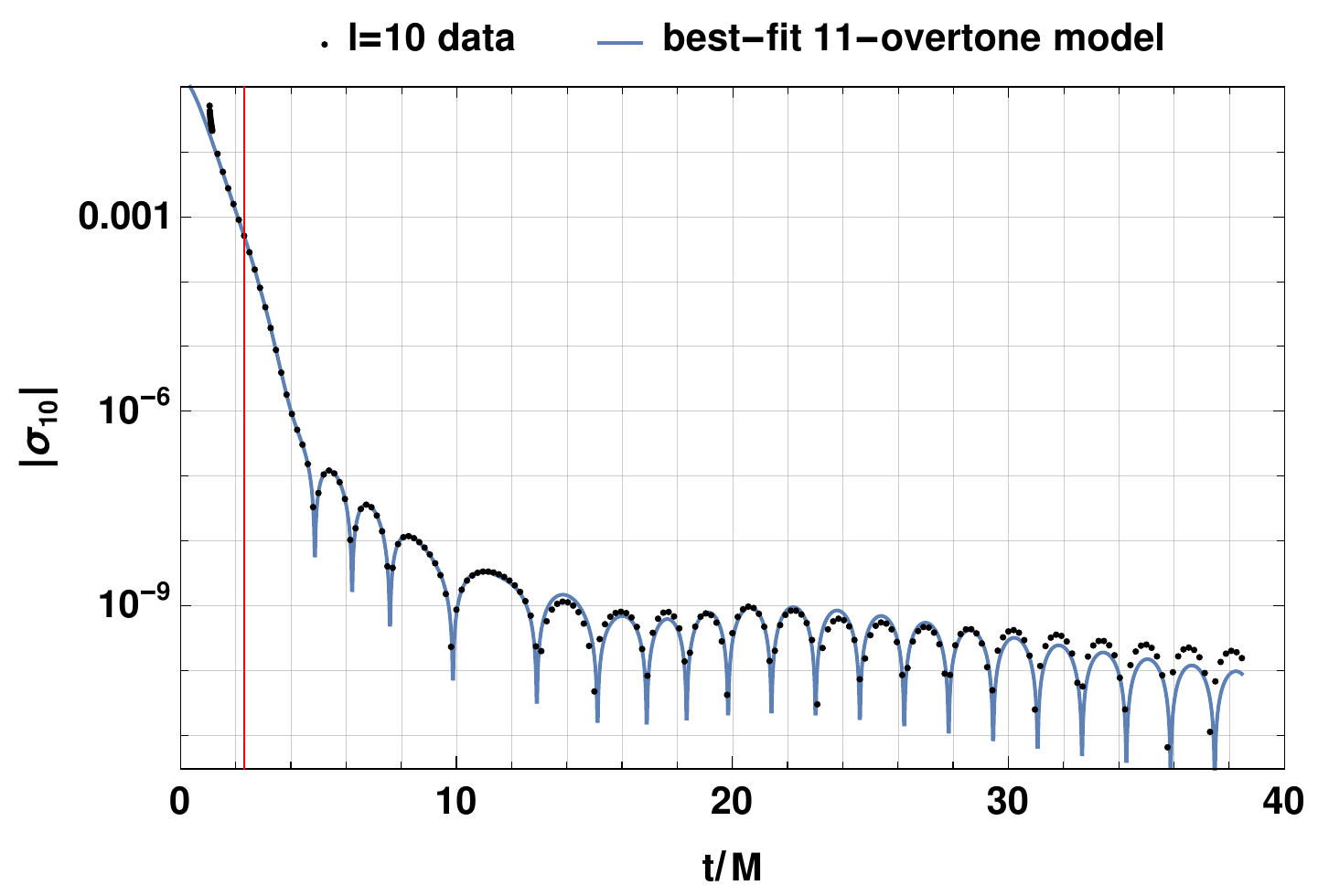}\\
  \includegraphics[width=\columnwidth]{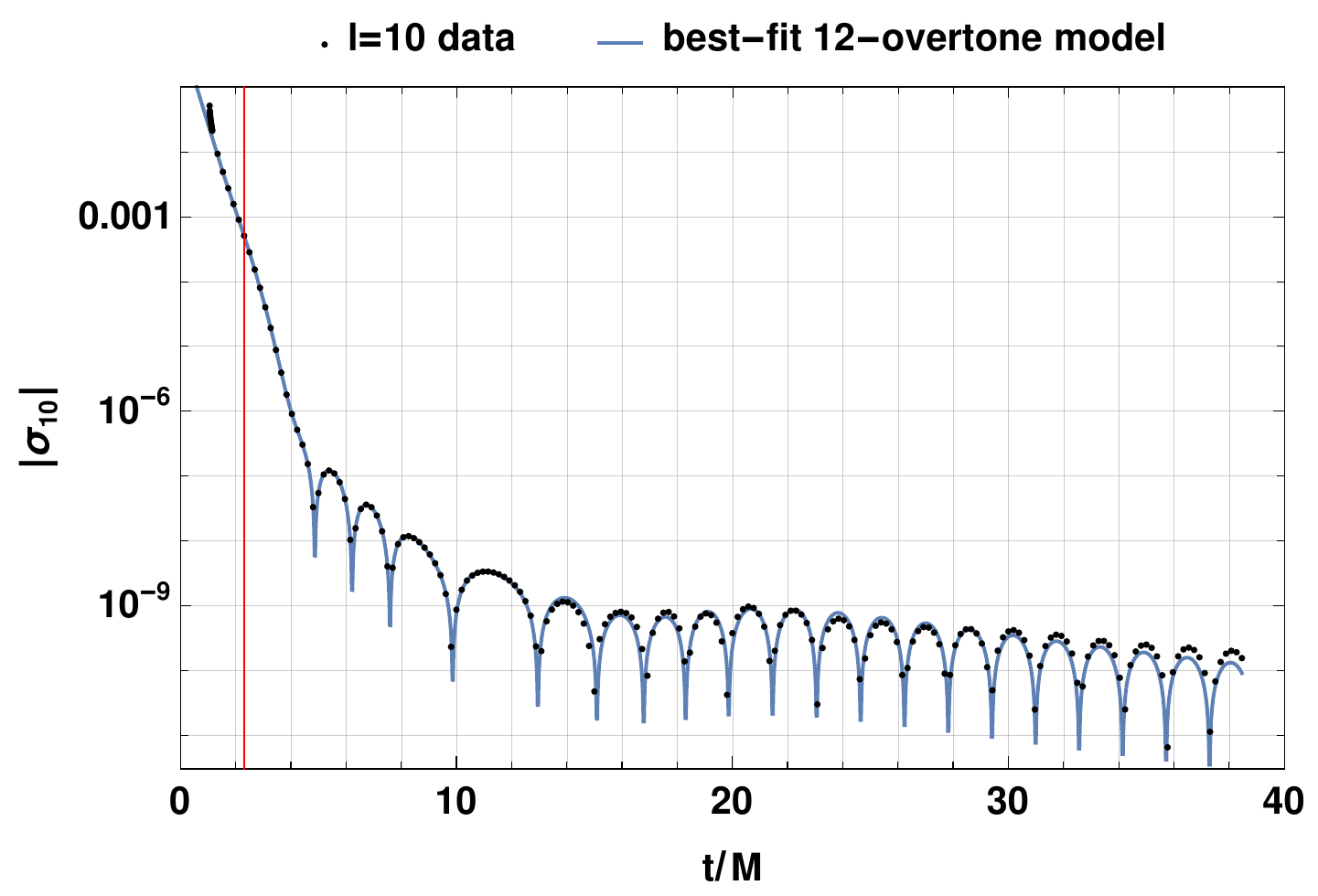}
  \includegraphics[width=\columnwidth]{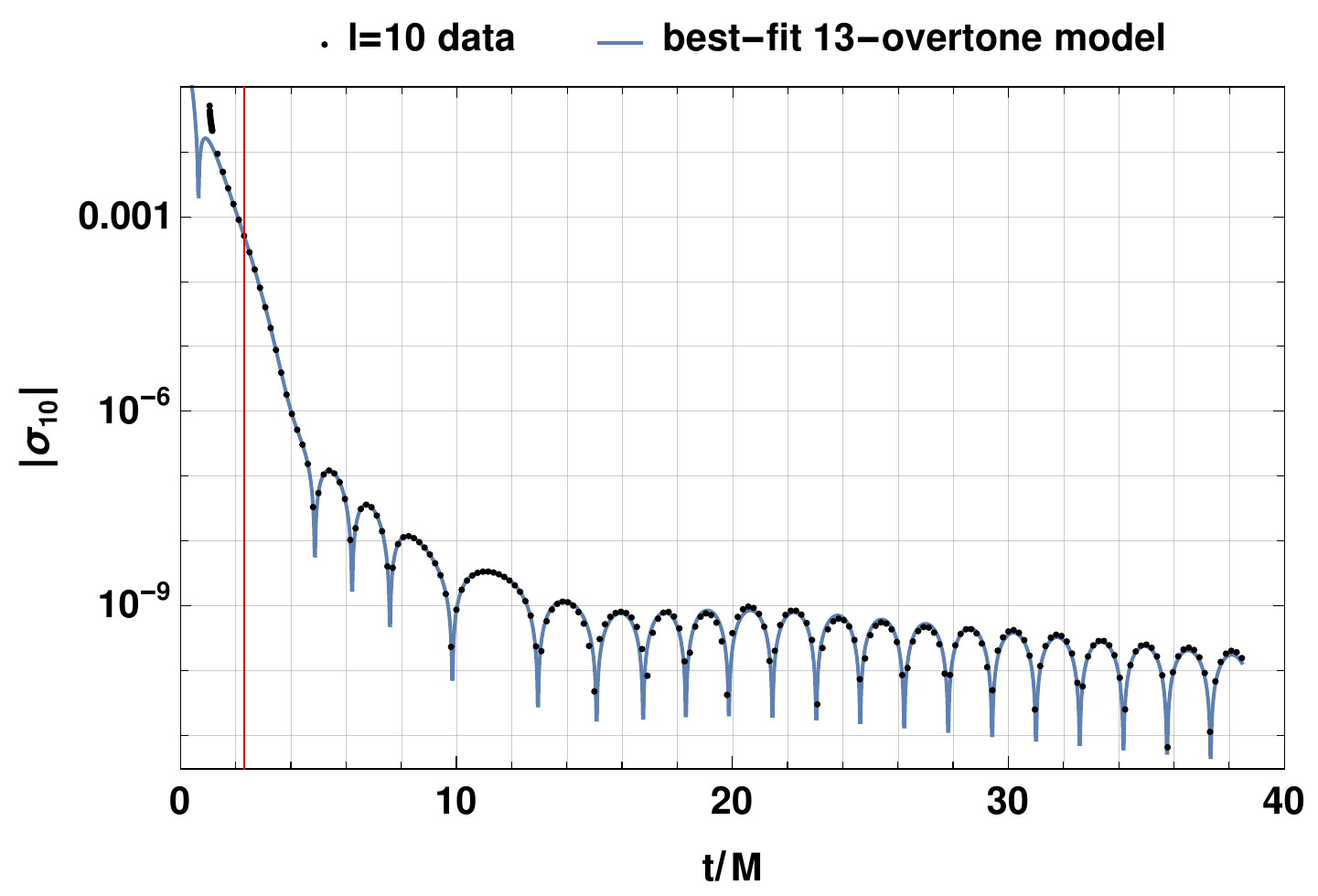}
  \caption{Same as Fig.~\ref{fig:l2-overtonefits-shear} for the $l=10$ shear mode
  and $\nmax=8$ to $13$ overtones. The same fit starting time is used.
  A good qualitative agreement to most of the data, including before the fit starting time,
  is obtained for $\nmax = 11$
  and improves further at late times for $\nmax = 12$ and $13$,
  although an oscillating behavior ---atypically--- reappears in the best-fit model
  at very early times for $\nmax = 13$.}
  \label{fig:l10-overtonefits-shear}
\end{figure*}

For this reason, we also directly examine, for each shear mode $\sigma_l$
with $2 \leq l \leq 12$,
the best-fit multiple-tone QNM models as arising from Eq.~\eqref{eq:QNM-multitone-model},
and we compare them to the corresponding NR data,
as the number of included overtones increases.
A fixed value of $t_0$ shortly after $t_{\mathrm{bifurcate}}$ is used in the fitting process.
We show in Fig.~\ref{fig:l2-overtonefits-shear} on a logarithmic scale
the NR $l=2$ shear mode and the best-fit models for several successive $\nmax$ values.
Two additional examples are similarly presented in Figs.~\ref{fig:l5-overtonefits-shear}
and~\ref{fig:l10-overtonefits-shear},
corresponding to the $l=5$ and $l=10$ modes respectively.
Numbers of overtones lower than those shown
never provide a relevant match to the shear modes
beyond a very short time range past $t_0$.
Conversely, higher numbers of overtones than those shown provide no visible improvement.

For this analysis we have selected a fit starting time $t_0 = (3/1.3) \, M \simeq 2.3 M$.
We thus start fitting the data fully within the early decay regime,
but not immediately at the formation time $t_{\mathrm{bifurcate}} \simeq 1.06M$.
This avoids the times immediately after $t_{\mathrm{bifurcate}}$, up to $t \simeq 1.3 M$,
where an even steeper decrease is observed
due to the infinite slope occurring at the bifurcation with the inner horizon (see Sec.~\ref{subsec:discussion-general} and Fig.~\ref{fig:bifurcation-shear}).
The choice of $t_0$ yet about $1 M$ further beyond this very specific regime
also allows us to evaluate the robustness of the fit results,
by checking for the continued agreement to the data at times preceding $t_0$.

In overall agreement with the mismatch investigations above,
we find that the behavior of each of the shear modes $2 \leq l \leq 12$
is well described, over the broad time range $[t_0,t_f]$ considered,
by a combination of a sufficiently large number of QNM overtones.
For the $l=2$ mode, a combination of two overtones
captures well the damping and oscillation features of the mode at all times,
and three overtones are enough to ensure a very small relative deviation at all times
even including the times $t < t_0$.
The same remarks hold for $\nmax=5$ and $\nmax=6$ overtones for $l=5$,
while a good modeling of the $l=10$ mode requires $\nmax = 11$ to $\nmax = 13$ overtones
(with a larger discrepancy to the data at $t < t_0$ for $\nmax=13$ in this case).
Similar results are found for all modes for adequate values of $\nmax$.

As suggested by the above examples, the number of overtones required
for an accurate representation of the shear modes increases with
$l$, reaching large values of $\nmax > 10$ overtones for $l \geq 10$.
In general, a reliable modeling of the mode $l$ typically requires
$\nmax=l$ or $\nmax = l+1$ overtones, occasionally up to $\nmax = l+3$
(in particular for the ``atypical'' cases $l=7$ and $l=10$).  These
estimates may be unreliable for $l \geq 10$ as the numerical
uncertainty may become larger than the contribution of some of the
highest overtones considered over most of the time range. In the
mismatch analysis above, we pointed out that the improvements of the
mismatch to the $l=11$ mode while adding more overtones beyond
$\nmax=6$ are within our estimate of the numerical error.  From a
similar consideration, for the example of the $l=10$ mode shown here,
the models with $\nmax \geq 10$ overtones may actually be poorly
constrained for this value of $t_0$ due to the uncertainties in the NR
results.

Nevertheless, it is quite remarkable that the late-time damped
oscillations, the intermediate-time regime and the early steep decay
without oscillations of each of the shear modes can be consistently
captured by a sum of QNM tones, with a relatively small number of
overtones for small $l$. As the real frequencies of the first few
modes typically remain close to that of the fundamental mode
and thus close to the frequency of the observed late-time oscillations,
one would in particular expect such a sum of QNMs to feature oscillations and zeros
over the range where the shear modes undergo a steep decay.
Instead, the observed oscillation-free early-time regime
is well reflected by the best-fit QNM model
for large enough $\nmax$, including (in nearly all cases) the domain $t < t_0$
which is not involved in the fitting procedure.

\subsubsection{Multipole moments}
\label{subsubsec:multipoles_overtones}

\begin{figure*}
  \centering    
  \includegraphics[width=\columnwidth]{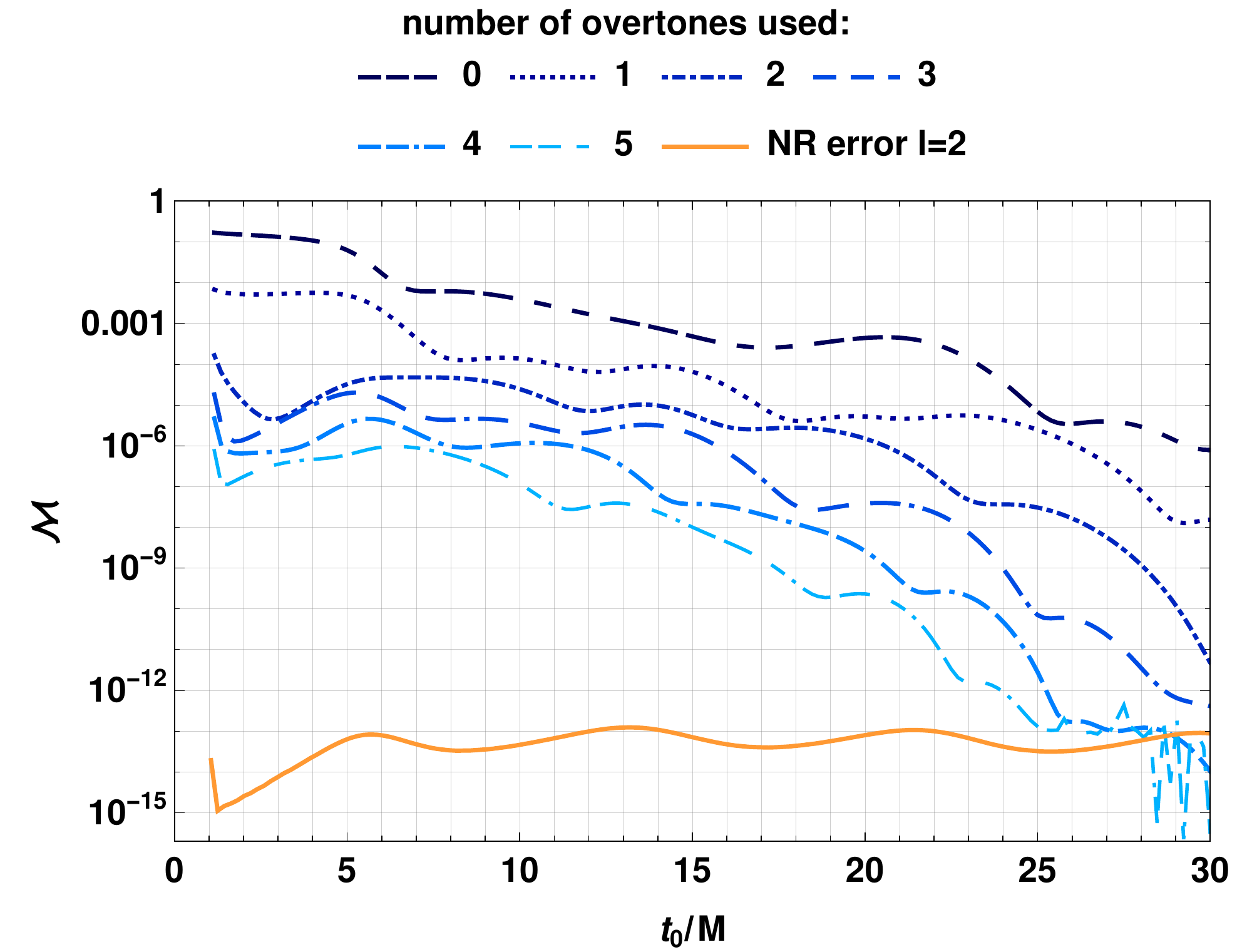}
  \includegraphics[width=\columnwidth]{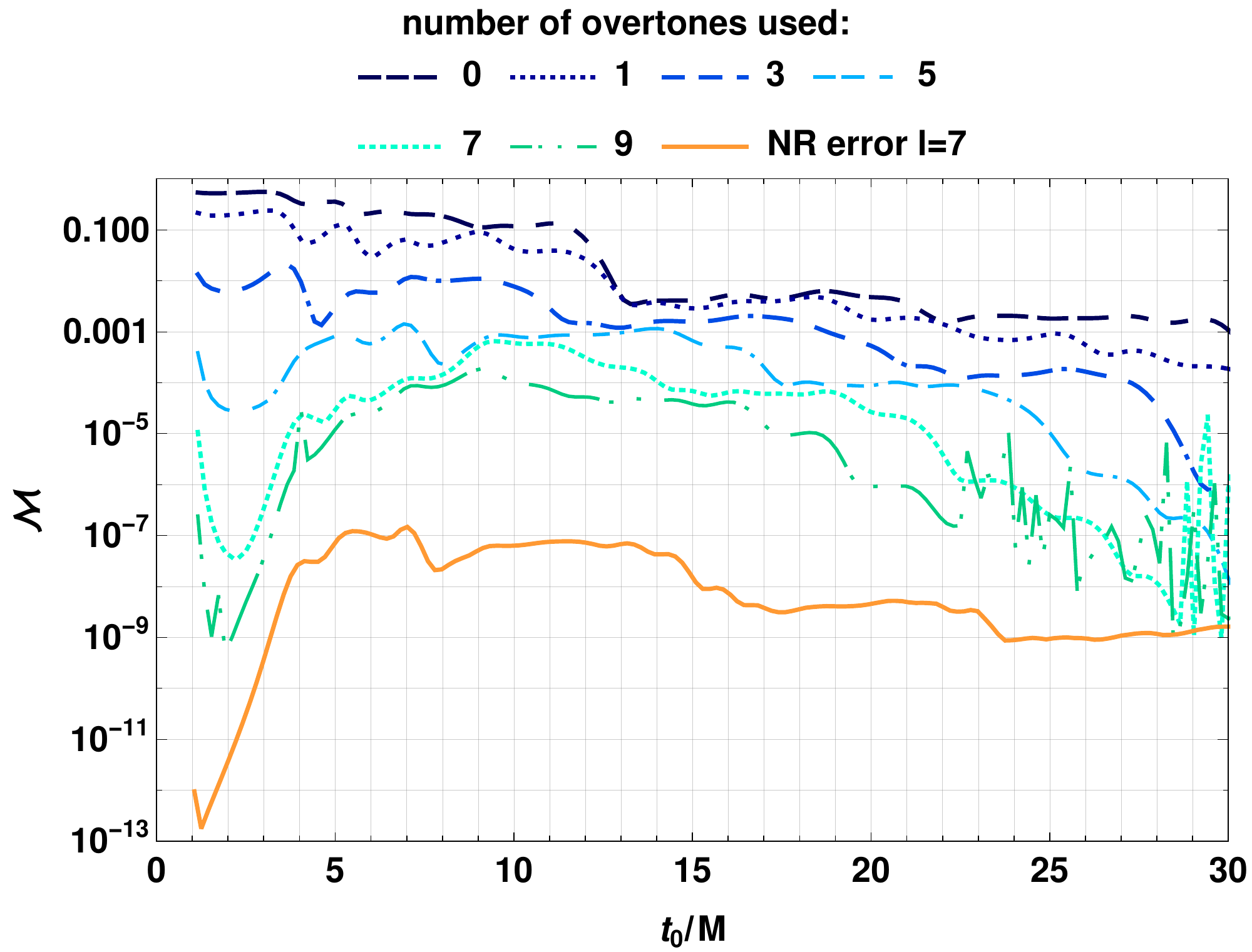}
  \caption{Similar to Fig.~\ref{fig:mismatch-shear} for the $l=2$ (left) and $l=7$ (right)
  mass multipoles as examples. For $l=7$ we again drop the curves
  obtained for (nonzero) even numbers of overtones for readability.
  As for the $l=7$ and $l=11$ shear modes in Fig.~\ref{fig:mismatch-shear},
  the missing curves are in line with the overall mismatch decrease trend
  with increasing $\nmax$  observed in the curves shown and can be extrapolated from them.}
  \label{fig:mismatch-multipoles}
\end{figure*}

We can repeat the above analysis for the mass multipoles for $2 \leq l \leq 12$
(still using the spin-weight-$2$ QNM frequencies).
The results are qualitatively very similar to those obtained for the shear modes.
Fig.~\ref{fig:mismatch-multipoles} shows the mismatch
between each of two example multipoles, $I_2$ (left panel) and $I_7$ (right panel),
and the corresponding best-fit models
of the class described by Eq.~\eqref{eq:QNM-multitone-model},
as $\nmax$ and $t_0$ are varied. The numerical error is estimated
in the same way as for the shear in Sec.~\ref{subsubsec:shear_overtones} above,
and is again shown as an orange continuous line.
We find again a decrease of $\mathcal{M}$
as $t_0$ increases towards $t_f$ at fixed $\nmax$ at least for $\nmax \leq 2$,
and also a decrease in mismatch with increasing number of overtones at fixed $t_0$.
We note that in this case the sharp decrease in mismatch at early $t_0$
with increasing number of overtones, for $l=2$, already occurs at two overtones
(\emph{vs}. three overtones for the shear $l=2$ mode).
We also generally find again larger mismatches for larger $l$,
for a given number of overtones and a given $t_0$.

\begin{figure*}
  \centering    
  \includegraphics[width=\columnwidth]{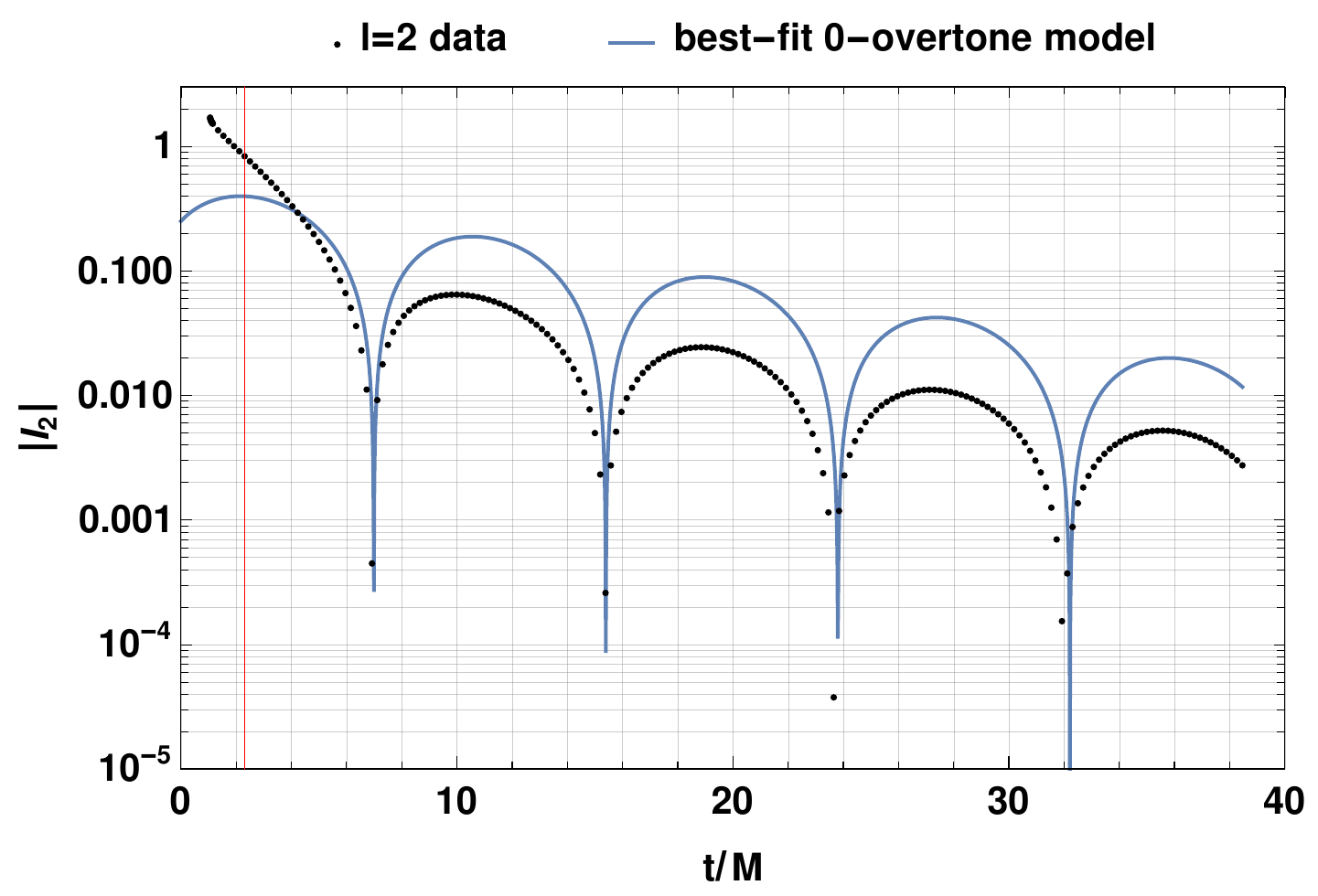}
  \includegraphics[width=\columnwidth]{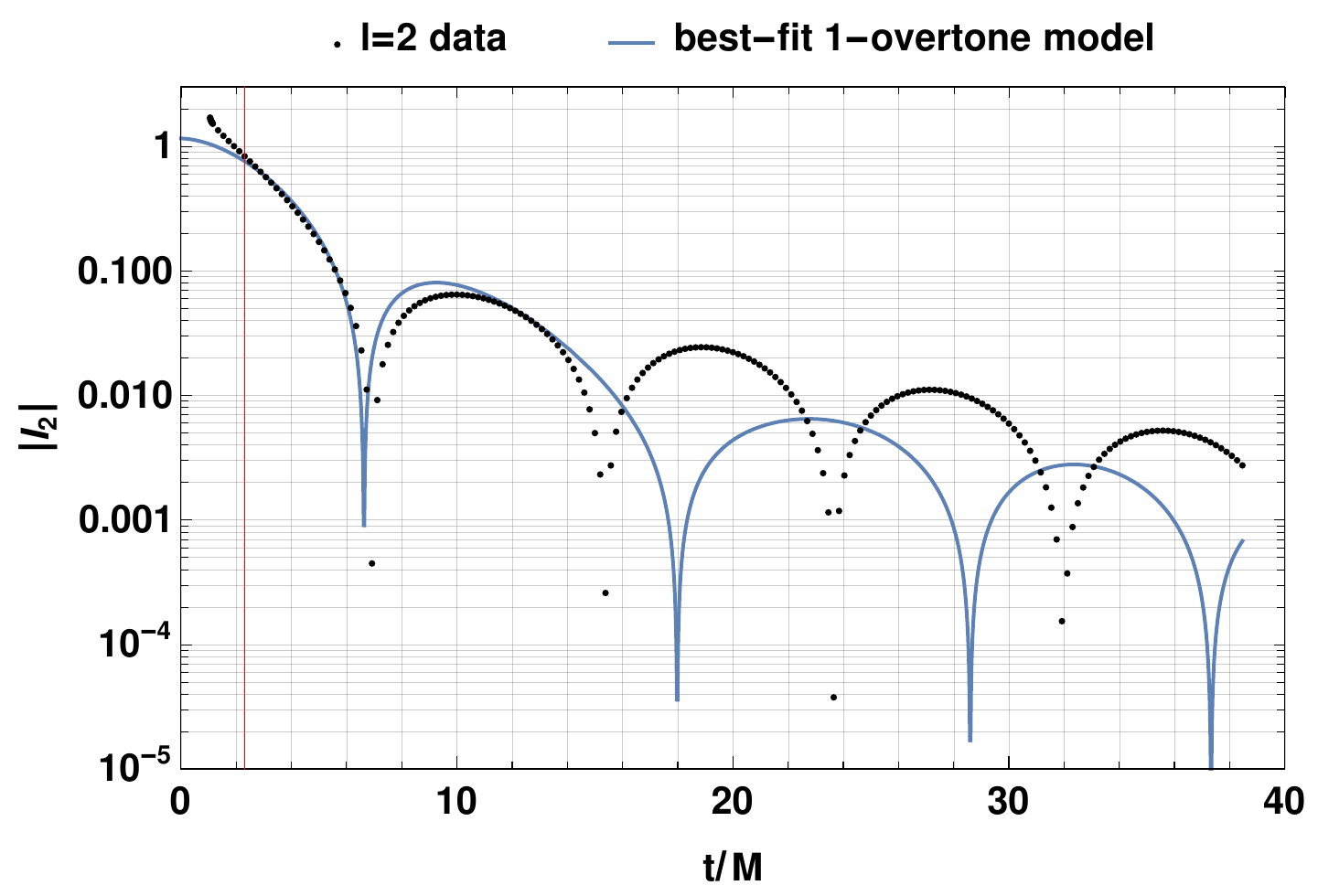}\\
  \includegraphics[width=\columnwidth]{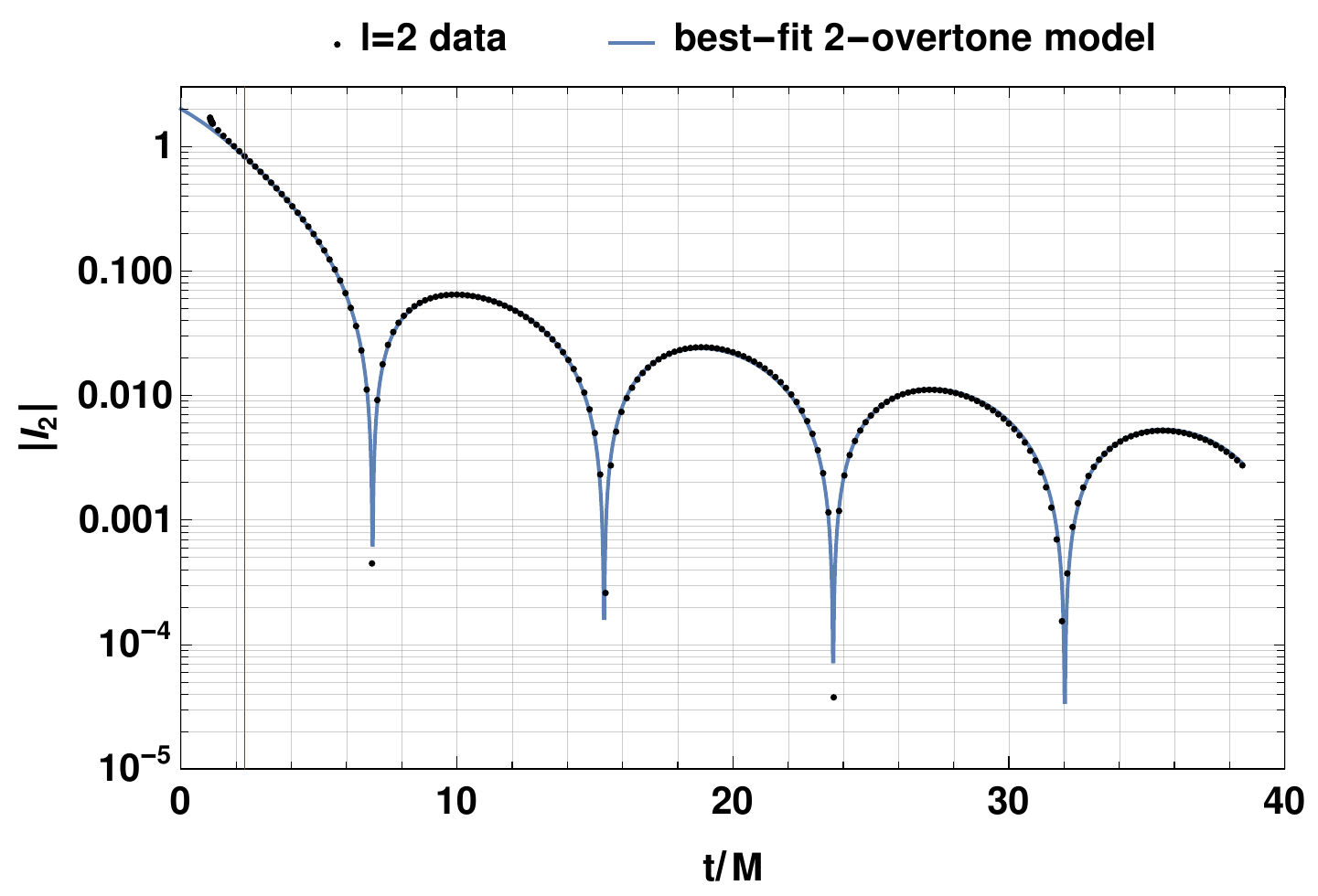}
  \includegraphics[width=\columnwidth]{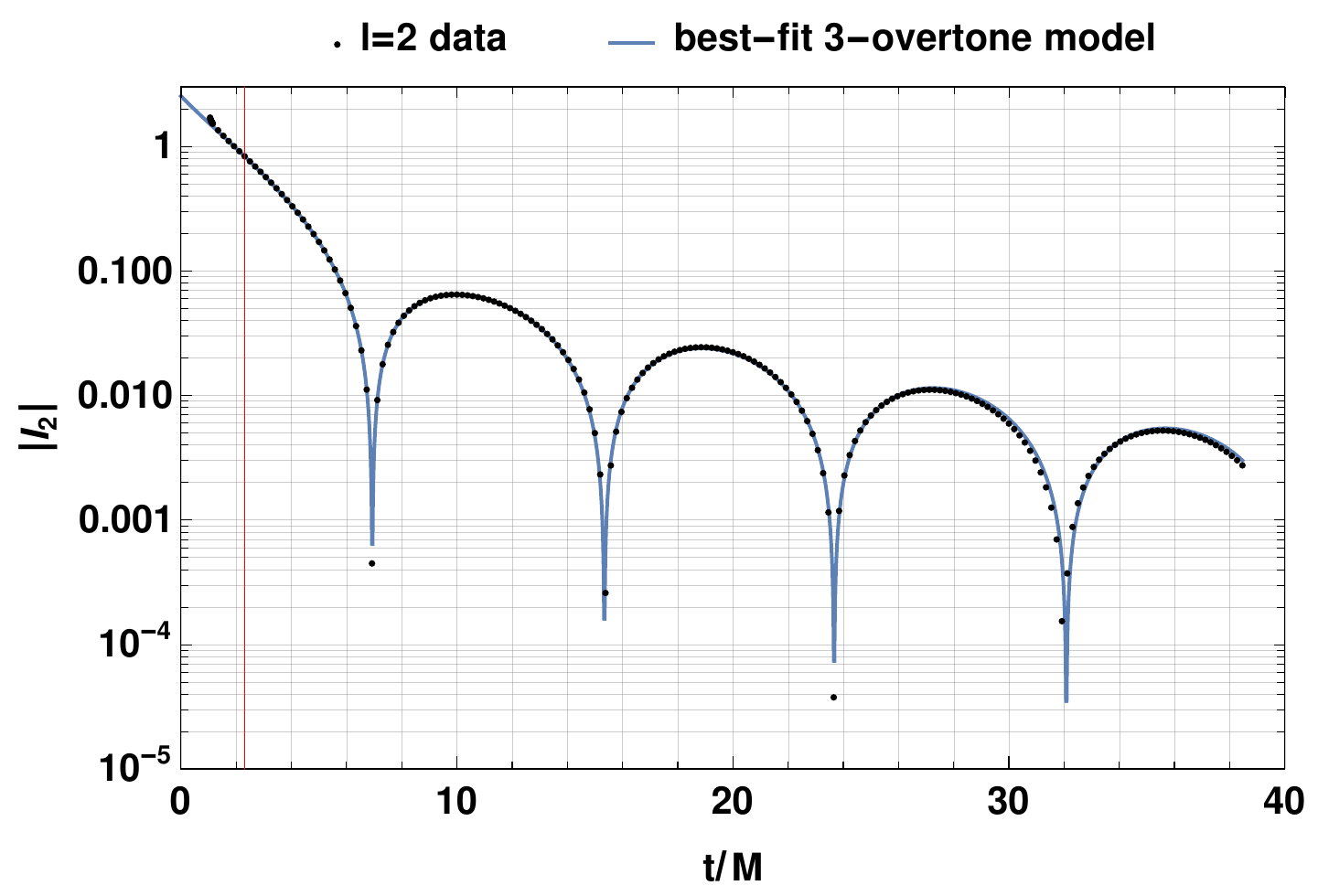}\\
  \caption{Same as Fig.~\ref{fig:l2-overtonefits-shear} for the $l=2$ multipole,
  for $\nmax=0$ to $\nmax=3$ overtones.
  We use again the same fit starting time as for the shear modes, $t_0 \simeq 2.3 M$.
  The model matches well the data both after and before this time at $\nmax = 2$,
  and this improves further (especially at $t < t_0$) at $\nmax = 3$.}
  \label{fig:l2-overtonefits-multipole}
\end{figure*}

\begin{figure*}
  \centering    
  \includegraphics[width=\columnwidth]{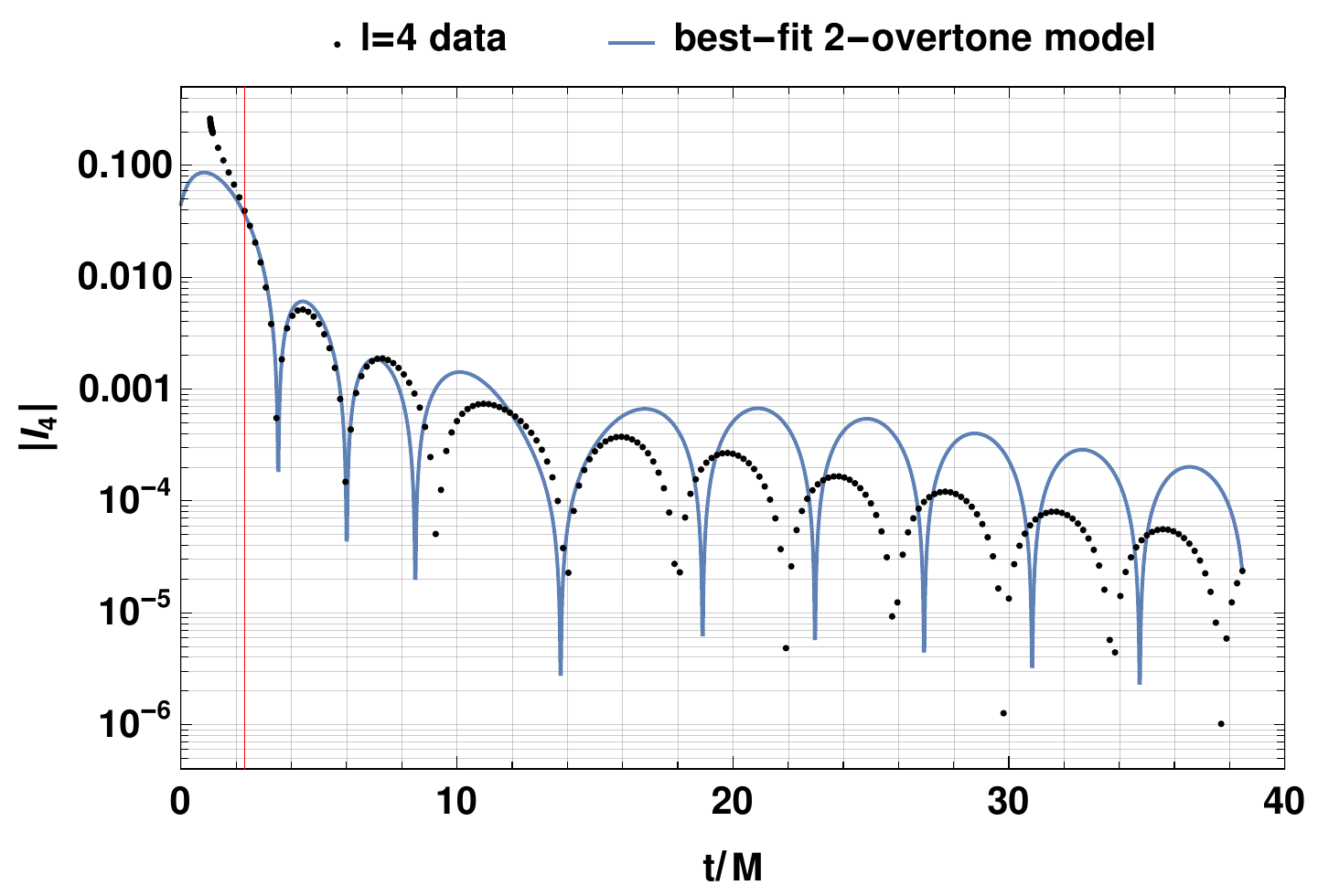}
  \includegraphics[width=\columnwidth]{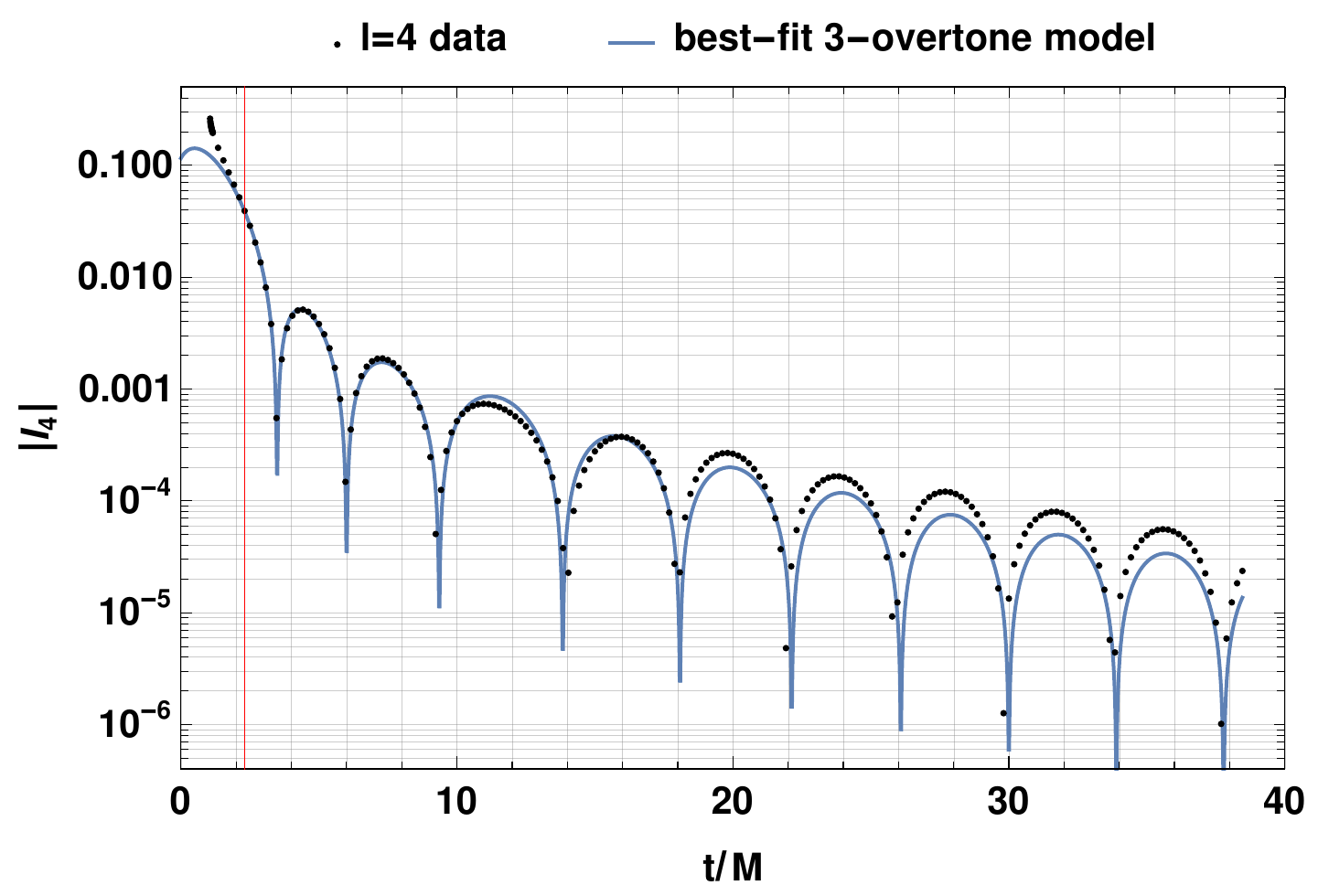}\\
  \includegraphics[width=\columnwidth]{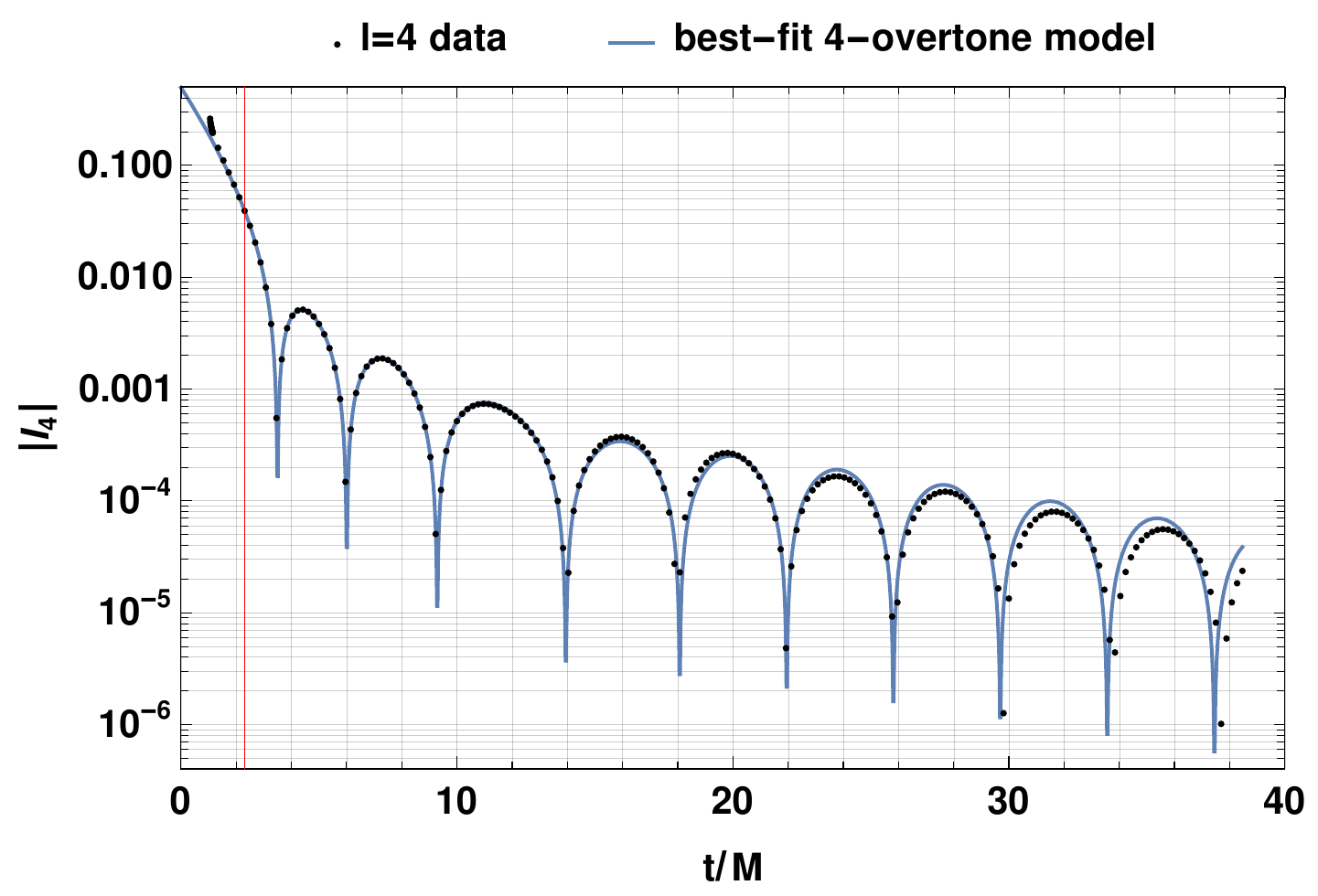}
  \includegraphics[width=\columnwidth]{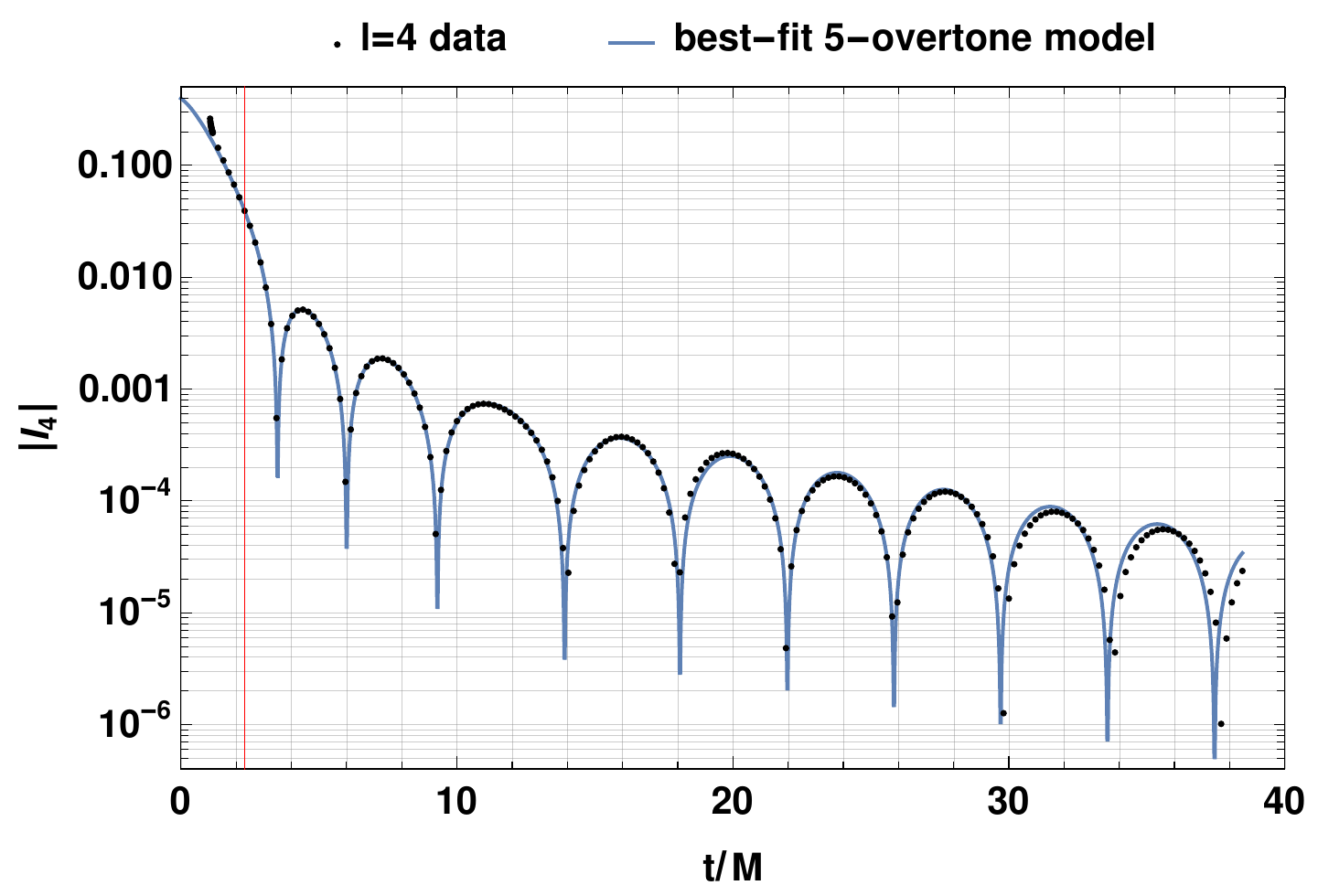}\\
  \includegraphics[width=\columnwidth]{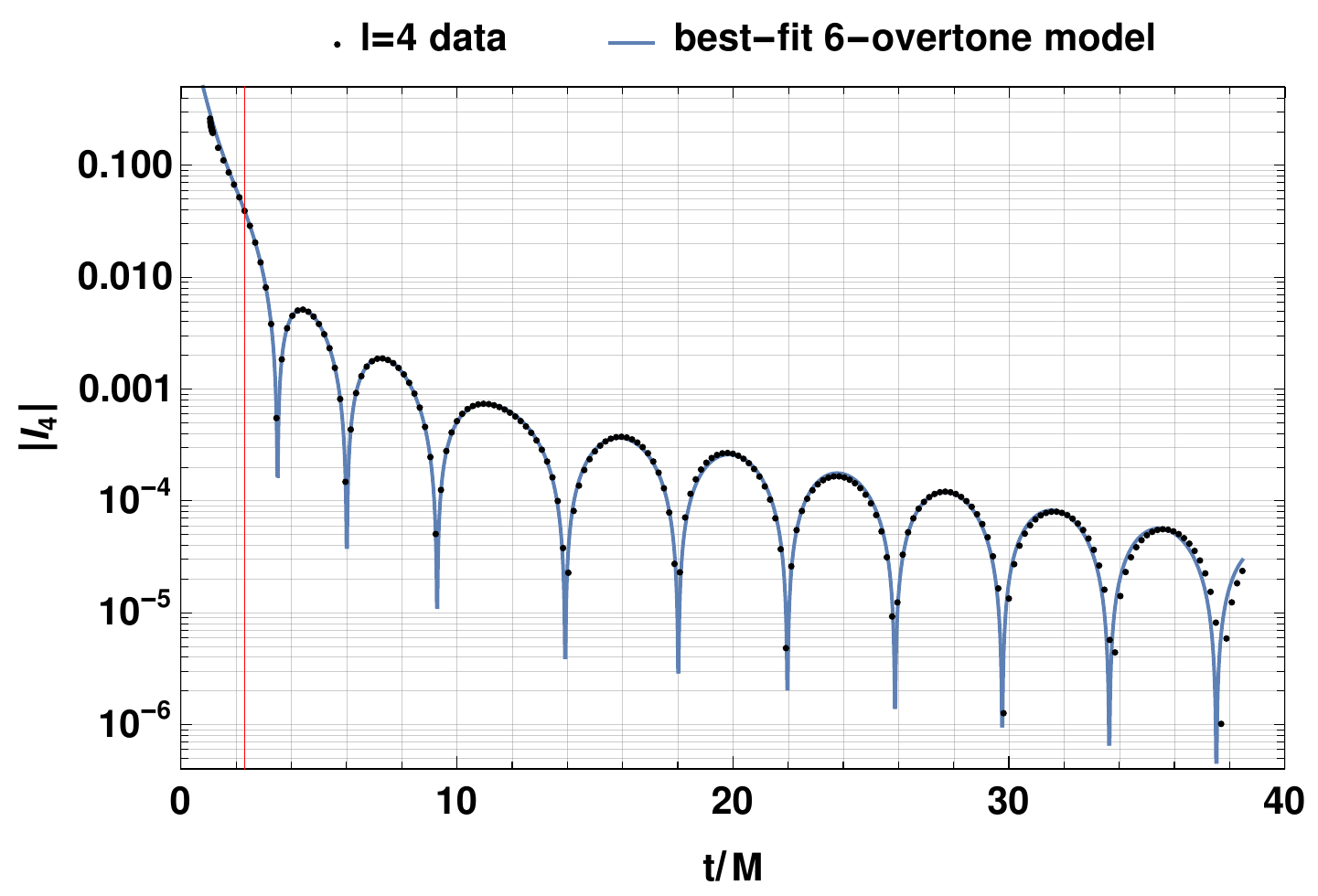}
  \includegraphics[width=\columnwidth]{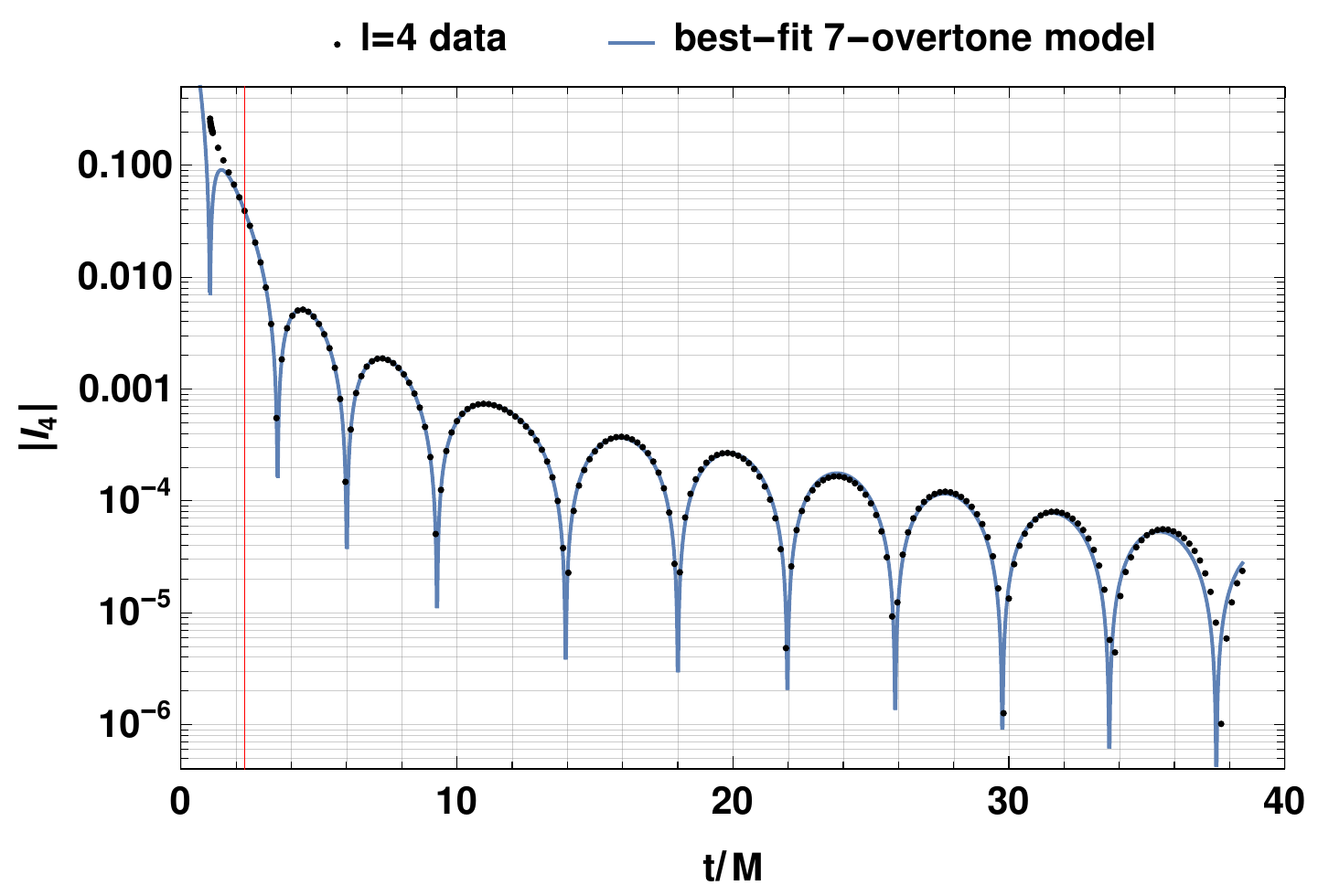}
  \caption{Same as Figs.~\ref{fig:l2-overtonefits-shear} and~\ref{fig:l2-overtonefits-multipole}
    for the $l=4$ multipole,
    for $\nmax=2$ to $\nmax=7$ overtones. The same fit starting time $t_0$
    is used. A relatively good agreement to the data is found after and before this time
    for $\nmax = 4$, improving for increasing $\nmax$,
    with the exception of $\nmax = 7$
    where the best-fit model displays an unusual behavior for $t < t_0$.}
  \label{fig:l4-overtonefits-multipole}
\end{figure*}

We then directly compare the best-fit models
to the NR multipole results for a fixed fit starting time $t_0$ as $\nmax$ increases.
We use again the same value of $t_0 \simeq 2.3 M$
as for the similar direct comparisons made for the shear modes
above in Sec.~\ref{subsubsec:shear_overtones}.
Two examples of such comparisons for the multipoles are shown on a logarithmic scale
in Fig.~\ref{fig:l2-overtonefits-multipole} ($I_2$)
and in Fig.~\ref{fig:l4-overtonefits-multipole} ($I_4$) for the relevant numbers of overtones.
Consistently with the mismatch results for $l=2$, $I_2$ shows a small qualitative difference
with the results obtained for  $\sigma_2$.
Namely, a two-overtone model already ensures
a very small relative deviation to the NR $I_2$ data at all times,
while a comparable accuracy required three overtones for $\sigma_2$.
For $I_4$, on the other hand, a comparable match is not reached under $\nmax=6$,
and the best-fit seven-overtone model atypically features an oscillation
within the range $[t_{\mathrm{bifurcate}}, t_0[$.
We note that this multipole has a lower magnitude at early times
than the surrounding ones $I_2$ to $I_6$ (see Fig.\ref{fig:shear-multipole-common},
bottom panel), which may be related to
this lower fit quality compared to the other modes.

More generally, as for the shear,
good matches to the behavior of each multipole $I_l$ are obtained
over the whole time domain, when including at least $\nmax=l$ or $\nmax=l+1$
overtones (or occasionally slightly more, such as for $l=4$).
In most cases, such a good match also extends back to $t < t_0$.
Note that for the multipoles, the models with $\nmax \geq l$ overtones
may be poorly constrained (due to numerical uncertainty) for $l \geq 9$.


\section{Is the early horizon dynamics really due to overtones?}
\label{sec:overtones-discussion}

\subsection{General considerations}
\label{subsec:discussion-general}

In the previous section we found, rather surprisingly,
that the first few shear modes and multipoles
could be well described by a combination of QNMs including the fundamental mode
and a few overtones, over the whole time interval available
--- or at least ignoring the first $\sim 0.3M$ immediately after horizon formation.
This conclusion also holds for the larger values of $l$ considered
(\emph{i.e.}, at least up to $l=12$),
although larger numbers of overtones are needed as $l$ increases. \revision{In this section, we consider various criteria, described below,
in order to assess the robustness and physical relevance
of such a QNM combination description. We shall see, however,
that a clear answer remains elusive.}

In section~\ref{subsec:singlemode} we confirmed
that \revision{each of} the shear modes and multipoles
\revision{appears} to be fully described asymptotically
by the \revision{corresponding} spin-weight-$2$
\revision{ \emph{fundamental} QNM} alone for $t \rightarrow \infty$.
We however observed deviations from this at intermediate times
(\emph{e.g.}, around $t= t_r \simeq 15.4 M$).
We noted that a detectable residual presence of rapidly decaying QNM overtones
could be expected in this regime,
assuming that the nonlinear deviations to equilibrium
are already negligible at these times~\cite{bhagwat:2017tkm,pook-kolb2020II}.
At earlier times $t \lesssim 8 M$ however, the area of the outer common horizon
is still varying steeply (see Fig.~\ref{fig:bl5-area}),
suggesting a still dynamical regime for the horizon at those times.
Accordingly, one might not expect the QNMs of the final Schwarzschild black hole
to account well for the evolution of the shear flux and geometry of the common horizon
from almost immediately after its formation.

\begin{figure}
  \centering    
  \includegraphics[width=\columnwidth]{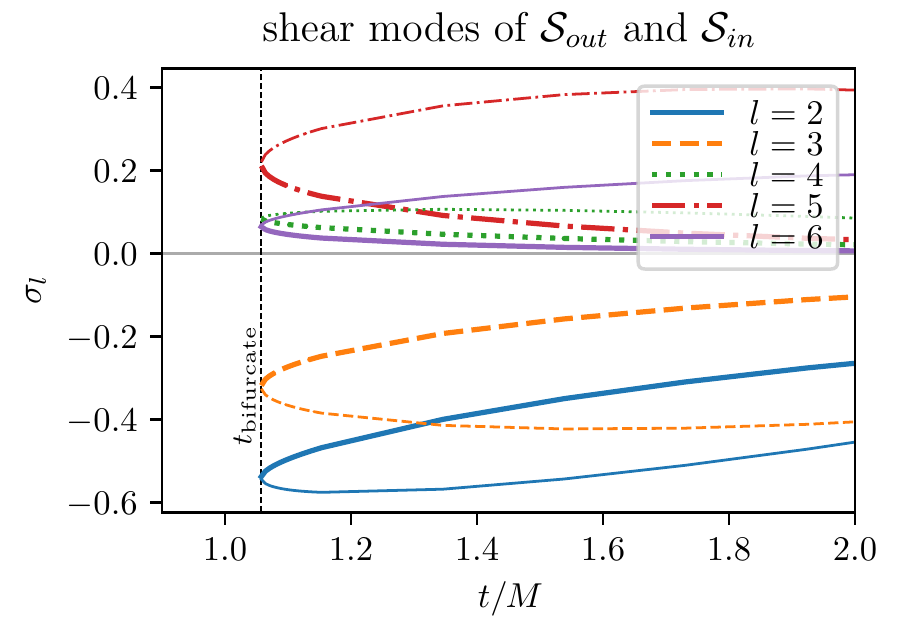}
  \caption{Numerically computed first six shear modes as a function of
  the simulation time $t$ on the outer common horizon $\mathcal{S}_{out}$
  (thicker lines) and on the inner common horizon $\mathcal{S}_{in}$ (thinner lines),
  near the formation/bifurcation time $t=t_{\mathrm{bifurcate}}$
  (highlighted as a vertical dashed line).
  The continuity of these variables across both horizons
  and the resulting vertical tangent at bifurcation are easily visible.
  The higher shear modes, and the multipoles, have the same behavior.}
  \label{fig:bifurcation-shear}
\end{figure}

In particular, at the time $t_{\mathrm{bifurcate}}$ of the common
horizon formation, the observables on this horizon, such as the shear
modes and multipoles, have an infinite slope as a function of our time
coordinate $t$. This is not a numerical artifact but rather a direct
consequence of the bifurcation of the inner and outer common horizons
at their joint formation. This is illustrated in
Fig.~\ref{fig:bifurcation-shear} by considering the shear modes on
both of the common horizons near their formation and bifurcation time
$t_{\mathrm{bifurcate}}$. As a consequence of this infinite slope, no
finite sum of QNMs (or any damped sinusoids) can strictly match qualitatively
the outer horizon shear modes or multipoles as a function of $t$
for $t\simeq t_{\mathrm{bifurcate}}$.
This might however only be due to a coordinate singularity on the horizon.
The simulation time $t$ which we use is a time coordinate adapted to our spacetime slicing,
and such an infinite slope should indeed occur for any choice of slicing by Cauchy surfaces
equipped with an adapted time coordinate. On the other hand,
this coordinate is not suitable around $t \simeq t_{\mathrm{bifurcate}}$
for the description of the smooth $3$-surface formed by the union of both common horizons.
One could imagine using instead, for example, the radius of the horizon
as a more adapted coordinate for this purpose.
This will be discussed elsewhere. In this work we shall be content with using the
simulation time $t$, discarding a short time range of about $0.3M$
after $t_{\mathrm{bifurcate}}$ from our analyses.
This range corresponds to a short faster-than-exponential decrease
that can be observed in the shear modes and multipoles
and that matches the vertical tangent at $t = t_{\mathrm{bifurcate}}$. We have seen that
the shear modes and multipoles can be well described by combinations of QNM
tones at all times past this short regime.

In the present section, we thus aim at investigating
whether the results we presented in section~\ref{sec:results}
should really be interpreted as the physical presence
of initially high-amplitude QNM overtones
determining the entire evolution of the outer common horizon past the first $\sim 0.3 M$.
We must also consider the alternative
--- that these results are simply an artifact of fitting the observables $\sigma_l$ and $I_l$
with damped sinusoids with sufficiently many free parameters.

We note in particular that the modes at large $l$
can only be well matched by a sum of QNM tones
if this sum extends to a large number $\nmax$ of overtones,
leaving a large number of degrees of freedom for fitting the data
(that is, $2 \nmax + 2$ degrees of freedom;
with the minimum $\nmax$ required ranging from $10$ to $14$ for $10 \leq l \leq 12$).
The numerically computed shear modes and multipoles
feature a steeper and steeper early-time
($t_{\mathrm{birfurcate}} + 0.3 M \lesssim t \lesssim t_{\mathrm{birfurcate}} + 3 M$)
damping as $l$ increases. On the other hand, the damping rates of the QNMs
for a given $n$ are independent of $l$ to first approximation, but increase with $n$
(see, \emph{e.g.}, Sec.~3.1 of \cite{kokkotas:1999bd};
an illustration of this can also be seen in Table~\ref{tab:QNMmodes} of the present work, right panel).
It was thus expected from the \emph{observed} behavior of the shear modes and multipoles
that their modeling in terms of QNMs would require higher overtones for larger $l$.
It is not obvious however from a theoretical perspective
that a larger early-time amplitude of high overtones should have been expected
\emph{a priori} for higher shear modes or multipoles.


One may come back to the generalized model of
Eq.~\eqref{eq:rdmodel_ansatz} with multiple tones ($\nmax \geq 1$),
leaving the parameters $\alpha_{l0n}$ and $\beta_{l0n}$ free, and
attempt to check if the best-fit model indeed recovers the QNM
frequency values, corresponding to $\alpha_{l0n} = \beta_{l0n} =
0$. Unfortunately, already for $\nmax=1$ and even more for larger
$\nmax$, the frequency deviation parameters appear to be very hard to
constrain --- even when the rescaling procedure of
Eq.~\eqref{eq:rescaling} is applied.  These parameters typically
feature large fitting uncertainties and overlaps between tones or with
zero-frequency models ($\alpha_{l0n} = -1$).  This is likely due to
the small differences between the QNM real frequencies (used as
reference values) for successive tones, as well as to the rapid decay
of QNM overtones.  Accordingly we cannot really conclude on the actual
presence of QNM overtones from such an analysis.
This does however suggest that constraining
the deviations of the complex frequencies of a combination of damped
sinusoids from the theoretical QNM frequency values can be very
challenging in general, even for zero-noise, low-systematic error
data.

In the following subsections we will thus
\revision{rather probe the robustness of the QNM modelling using several
tests of fit stability and fit comparison. These tests provide more insight
than the approach mentioned hereabove,
even though they still do not allow us to reach a definitive conclusion.}
We will simply focus on the shear for this investigation, and
specifically on the $l=2$ shear mode as an example and as an easy case
that can be modeled with a relatively small number of overtones.

\subsection{Comparing models with different numbers of overtones but an equal number of free parameters}
\label{subsec:discussion-equal-number-of-DoF}

We first consider the relative quality of the fits provided either by a sum of a few QNM tones,
or by another model of the general class of Eq.~\eqref{eq:rdmodel_ansatz}
with less modes but some of the parameters $\alpha_{l0n}$ and $\beta_{l0n}$ left free
rather than being set to zero. We choose the second model
in such a way that both models have the same number of free parameters.
We will consider two such pairs of models, with respectively four and six free parameters.
We compare the models within each pair
in terms of their mismatch to the NR $l=2$ shear mode for the best-fit parameters
(without applying a rescaling such as that of Eq.~\eqref{eq:rescaling}),
as a function of the fit starting time $t_0$.

\begin{figure}
  \centering    
  \includegraphics[width=\columnwidth]{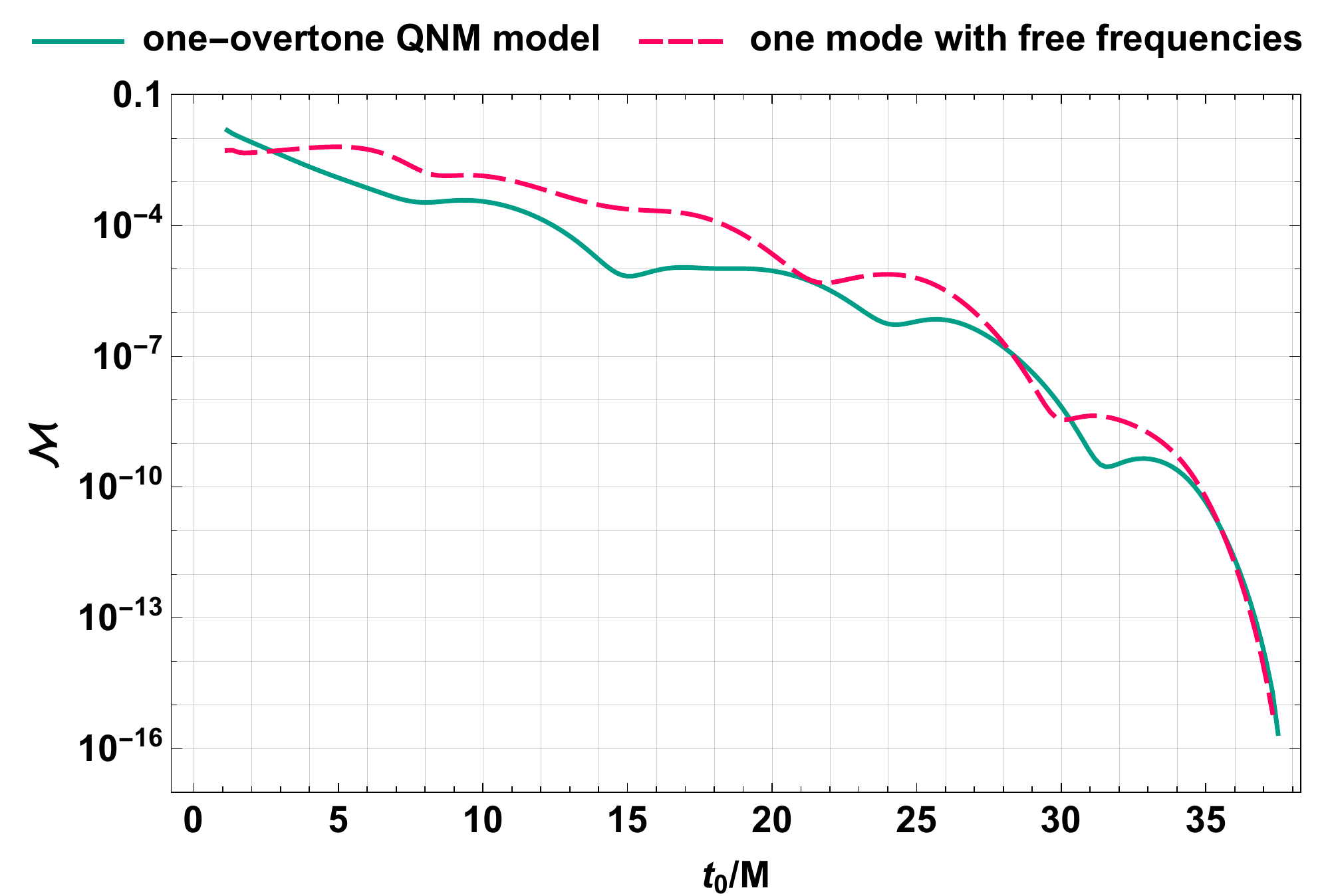}
  \caption{Comparison of the respective mismatches to the NR shear $l=2$ mode
  for two four-parameter models
  (see Sec.~\ref{subsec:discussion-equal-number-of-DoF} for details),
  as a function of the fit starting time $t_0$.}
  \label{fig:4param-models-shear}
\end{figure}

Fig.~\ref{fig:4param-models-shear} shows this mismatch
for the $(\nmax=1)$-overtone model of the class of Eq.~\eqref{eq:QNM-multitone-model}
(green continuous line), where all frequencies are set to the QNM values;
and for the single-mode model with free frequencies
used in section \ref{subsec:singlemode} and given by Eq.~\eqref{eq:onetonemodel}
(red dashed line).
The free parameters are $\{ A_0, \phi_0, A_1, \phi_1 \}$ in the first case
and $\{ A_0, \phi_0, \alpha_0, \beta_0 \}$ in the second case.
These results suggest a small preference at nearly all times
for the one-overtone model with QNM frequencies over a single-damped-sinusoid model
even though the complex frequency of the latter is freely adjusted.
The improvement in mismatch does however occur at most but not all values of $t_0$,
and barely goes beyond $1$ order of magnitude when it occurs.
In particular, for most values of $t_0/M \gtrsim 28$, \emph{i.e.} at very late times,
we obtain similar values of the mismatch $\mathcal{M}$ for both models considered.
This is consistent with Fig.~\ref{fig:single-mode-alpha0-beta0-shear-l2} since, in this regime,
deviations to a fundamental-mode-only model (with QNM complex frequency)
are expected to be mostly negligible.

\begin{figure}
  \centering    
  \includegraphics[width=\columnwidth]{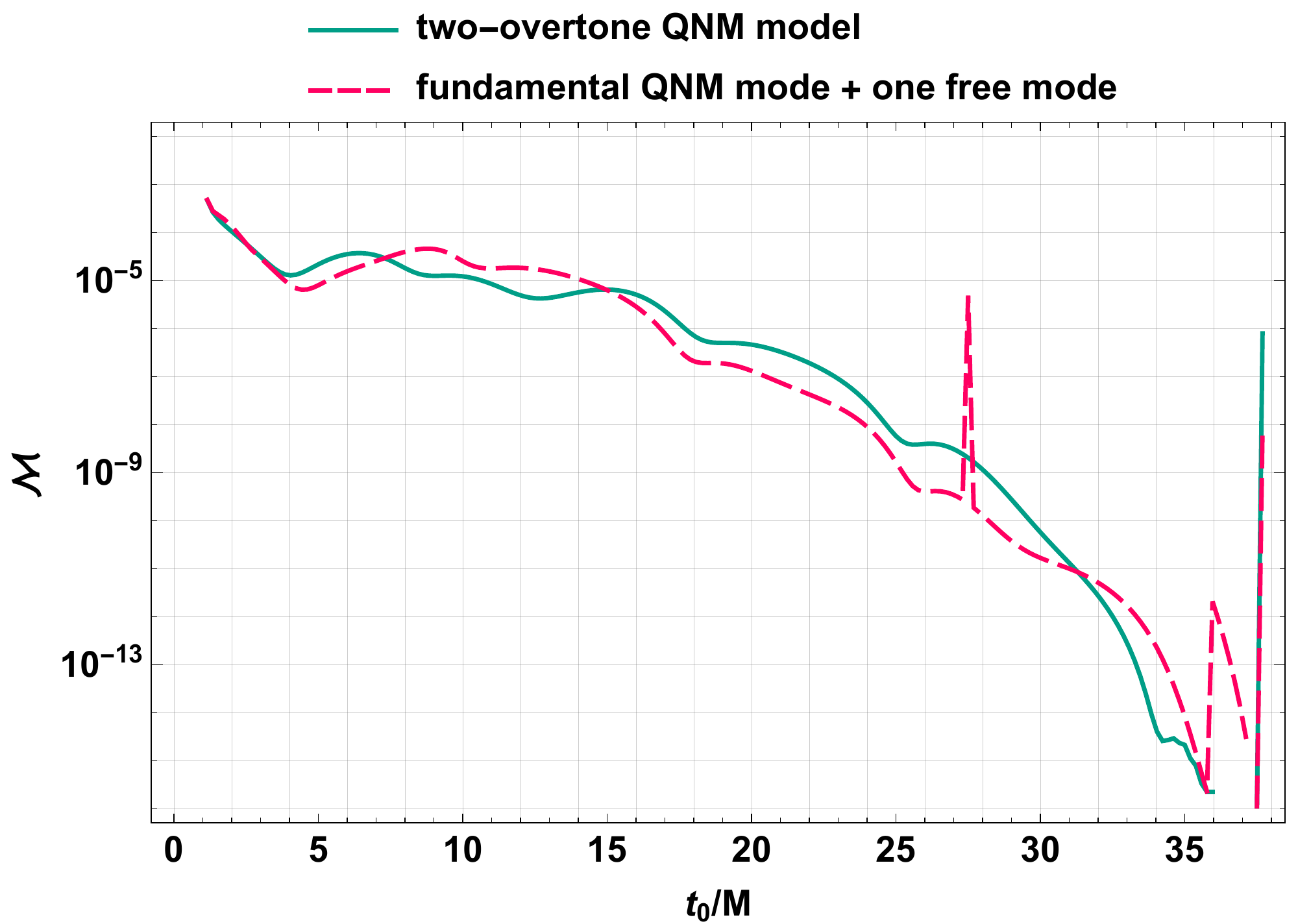}
  \caption{Comparison of the respective mismatches to the NR shear $l=2$ mode
  for two six-parameter models
  (see Sec.~\ref{subsec:discussion-equal-number-of-DoF} for details),
  as a function of the fit starting time $t_0$.}
  \label{fig:6param-models-shear}
\end{figure}

Fig.~\ref{fig:6param-models-shear} shows the similar mismatch
for two six-parameter models.
Both models assume the presence of the fundamental QNM,
which we seem to recover asymptotically at late times.
They both consider an additional contribution,
which takes the form of either the first two QNM overtones,
or of a single damped sinusoid with unconstrained complex frequency.
The first model (green continuous line) thus corresponds to the $(\nmax = 2)$-overtone model
with QNM frequencies of the class of Eq.~\eqref{eq:QNM-multitone-model},
with free parameters $\{ A_0, \phi_0, A_1, \phi_1, A_2, \phi_2 \}$. 
The second model (red dashed line)
corresponds to the general ansatz of Eq.~\eqref{eq:rdmodel_ansatz} for $\nmax=1$
and with $\alpha_{l00}$ and $\beta_{l00}$ set to zero.
The free parameters in this case
are $\{ A_{l00}, \phi_{l00}, A_{l01}, \phi_{l01}, \alpha_{l01}, \beta_{l01} \}$.
No clear preference is found for either model,
both of them alternately having the lowest mismatch for various ranges of $t_0$,
and with very small differences between both mismatch values.

Hence, a \{fundamental QNM + first QNM overtone\} model
is only marginally preferred to a single-damped-sinusoid model,
and assuming the presence of the fundamental QNM,
we cannot conclude about the additional presence of two QNM overtones
\emph{vs.} that of an arbitrary single additional damped sinusoid.
This neither confirms nor rules out the actual presence of QNM overtones,
but hints again quite strongly at the difficulty of confidently determining
a) the presence of overtones
and b) the frequencies of multiple damped sinusoids that may be present in the data.

\subsection{Comparison to a toy model with altered real frequencies}
\label{subsec:discussion-toy-model}

We now investigate how the quality of multiple-tone fits depends
on deviations in the frequencies of the tones with respect to the QNM values.
We here focus on the real frequencies, noting that the imaginary frequencies
(or damping rates) of successive QNM tones $n$ are well separated,
while the corresponding real frequencies vary by smaller amounts for small $n$ values
(Sec.~3.1 of {kokkotas:1999bd}; see also Table~\ref{tab:QNMmodes} hereabove).
For the $(l=2, \, m=0)$ mode considered here,
the real frequencies of the first three QNM overtones $\omega_{20n}, n=1,2,3$, for instance,
are smaller than the fundamental-mode one $\omega_{200}$ by about $7 \%$, $19 \%$
and $33 \%$ respectively (see Table~\ref{tab:QNMmodes}, left panel).

For this purpose, we consider a family of arbitrary toy models
following the general ansatz of Eq.~\eqref{eq:rdmodel_ansatz},
with a variable total number of overtones $\nmax$.
We define this family by setting $\alpha_{l00}$
and all the $\beta_{l0n}$ parameters to zero,
\emph{i.e.}, we keep the fundamental QNM
and we keep all damping rates at the QNM values,
and by setting the other $\alpha_{l0n}$ parameters, $n>0$,
to a specific choice of nonzero values. Our choice here
is to set every $\alpha_{l0n}$ such that the real frequency of each tone in the model
stays equal to the fundamental QNM real frequency:
$\omega_{l0n} (1 + \alpha_{l0n}) = \omega_{l00} \, \forall n$.
We thus end up with the following family of models, parametrized by $\nmax$:
\begin{equation}
 X_l = \sum_{n=0}^{\nmax} A_{l0n} \, \exp \left[ -\frac{t-t_r}{\tau_{l0n}} \right] \,
 \cos \left[ \omega_{l00} \, (t-t_r) + \phi_{l0n} \right] \; .
 \label{eq:toy-model}
\end{equation}
The free parameters of the models are the amplitudes $A_{l0n}$ and the phases $\phi_{l0n}$.
We probe the ability of these artificial models to match the shear $l=2$ mode at all times
as $\nmax$ is varied, in a similar way as was done,
\emph{e.g.}, for Fig.~\ref{fig:l2-overtonefits-shear} in section \ref{subsubsec:shear_overtones}.
That is, we set the same early fit starting time $t_0 \sim 2.3 M$ as for the latter figure
and we directly study the relative deviation of the best-fit model to the NR data
(and to its general behavior) for each $\nmax$,
without using a rescaling such as that of Eq.~\eqref{eq:rescaling} in the fitting process.

\begin{figure*}
  \centering    
  \includegraphics[width=\columnwidth]{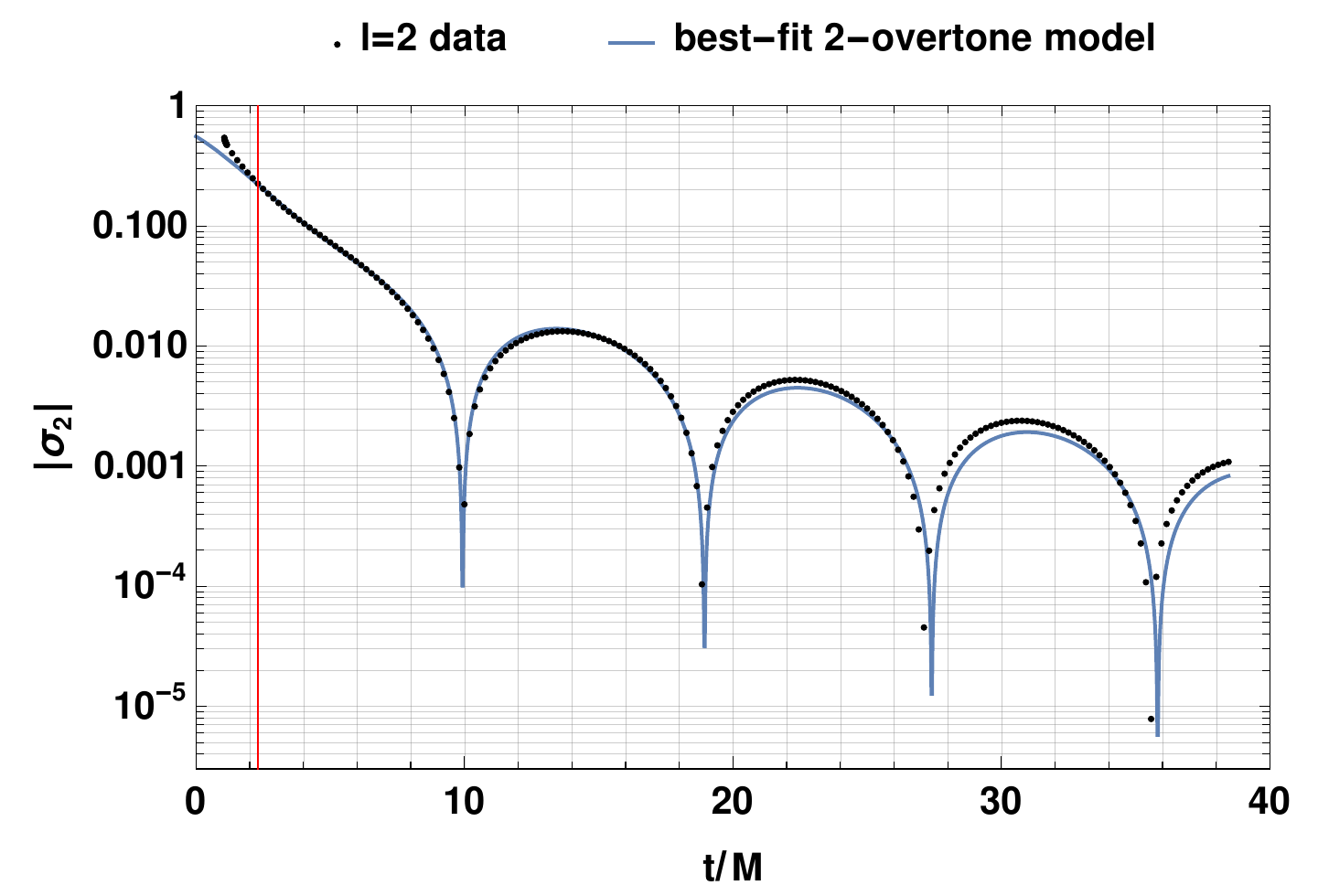}
  \includegraphics[width=\columnwidth]{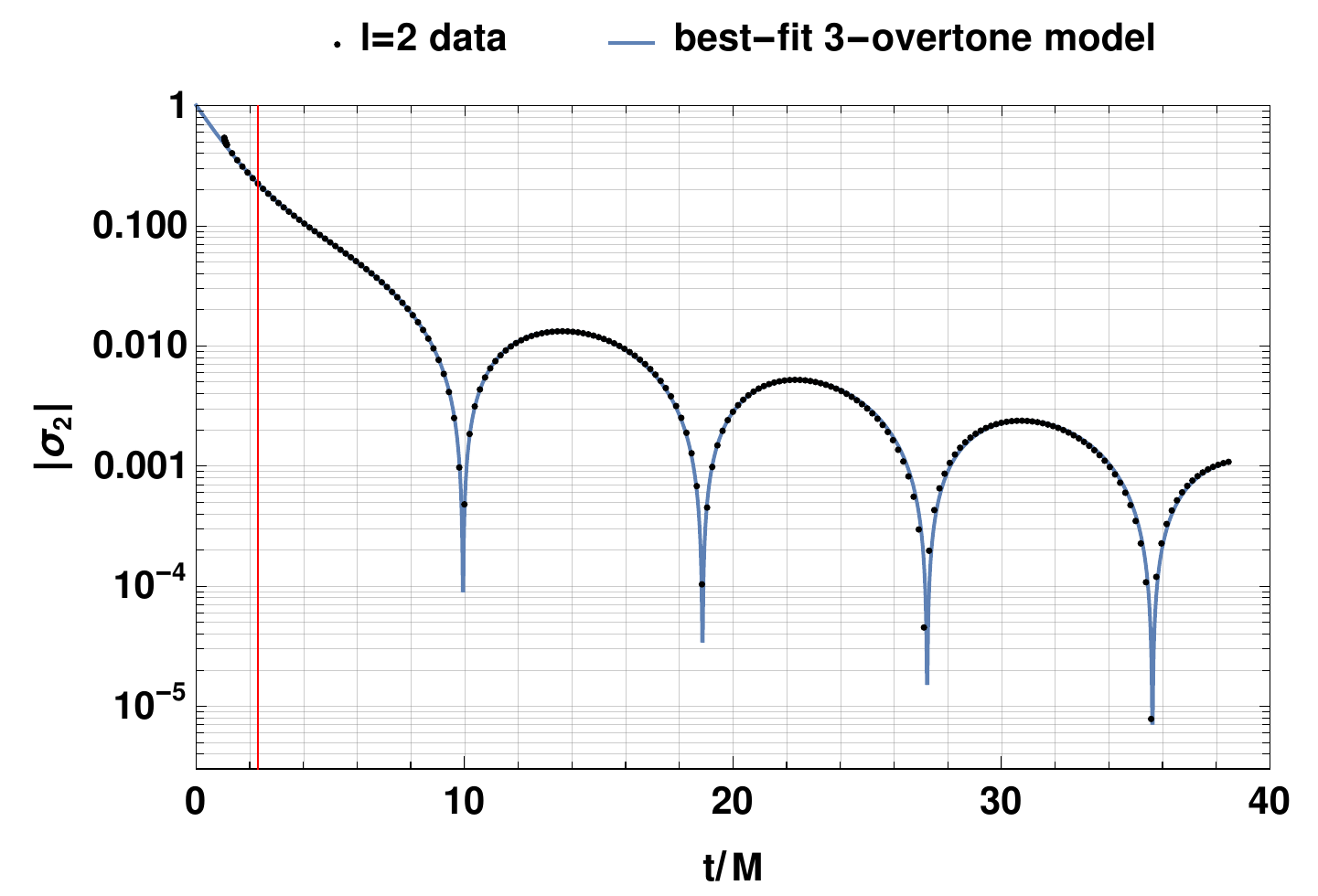}
  \caption{Direct comparison of the NR $l=2$ shear mode (black dots)
  and the associated best-fit toy models
  as introduced in section \ref{subsec:discussion-toy-model}
  (blue continuous lines; see this section for details),
  as a function of the simulation time $t$.
  We show here the results for the most relevant numbers of $\nmax=2$ (left panel)
  and $\nmax=3$ (right panel) additional modes beyond the fundamental QNM.
  The vertical red line on each panel indicates the $\{ t=t_0 \}$ line,
  where the starting time $t_0$ used for the fits is set to a constant value
  given by $t_0/M  = 3/1.3 \simeq 2.3$, as for the similar analyses with a different model
  presented in section \ref{subsec:fit-overtones}.
  For a given $\nmax$, the agreement of the best-fit toy model to the NR data is comparable
  to that obtained with the actual QNM model in Fig.~\ref{fig:l2-overtonefits-shear},
  both after and before $t_0$.}
  \label{fig:l2-modifiedovertonefits-shear}
\end{figure*}

Fig.~\ref{fig:l2-modifiedovertonefits-shear} shows the results
similarly to Fig.~\ref{fig:l2-overtonefits-shear} with the best-fit model for each $\nmax$
as a continuous line and the NR data as dots, as a function of $t$ and on a logarithmic scale.
We show here only the most relevant values of $\nmax=2$ and $\nmax=3$.
$\nmax=0$ (fundamental QNM only, already considered earlier) and $\nmax=1$,
do not provide a good match to the overall behavior of the shear mode,
while values of $\nmax > 3$ show little visible difference to the $\nmax=3$ case.

Interestingly, we find again for this artificial model a rather good match to the data
for $\nmax=2$, and a very good match at all times (including prior to $t_0$
but after $t \simeq t_{\mathrm{bifurcate}} + 0.3 M$) for $\nmax=3$.
The results are qualitatively very similar
to those obtained with the multiple-QNM-tones model of Eq.~\eqref{eq:QNM-multitone-model}
in section~\ref{subsec:fit-overtones}, despite the unphysical real frequency values used here
for the overtones.
Hence, our conclusions of a good modeling of the shear modes or multipoles
by combinations of sufficiently many QNM tones
are not very sensitive to the actual frequencies (at least regarding the real part)
used in the overtones model. We see here that similar conclusions can be reached
with models that do not match the GR QNM values for $n > 0$.

\subsection{Stability of the multiple-QNM fits with the fit time range}
\label{subsec:discussion-amplitudes-varying-t0}

We finally study some aspects of the stability of the best-fit parameters
when fitting the multiple-tone QNM model of Eq.~\eqref{eq:QNM-multitone-model}
to the NR data over the range $[t_0, t_f]$ as $t_0$ is varied.
Such a stability can be seen as a necessary condition
for the consistent presence of a set of QNM overtones in the data.
If these modes are present, then for instance
their amplitudes should be recovered consistently over a range of $t_0$ values
where they are detectable.

We have already mentioned some stability properties of the fits provided by this model
for large enough numbers of overtones in section~\ref{subsec:fit-overtones}.
It is indeed noteworthy that
when selecting a fit starting time $t_0 \simeq 2.3 M \simeq t_{\mathrm{bifurcate}} + 1.2 M$,
in almost all cases
where any given shear mode or multipole is well matched by the model after $t_0$,
the nonoscillating damped regime
extending before $t_0$ to $t \simeq t_{\mathrm{bifurcate}} + 0.3 M$ is also well recovered,
qualitatively and quantitatively. These QNM models
thus consistently match the behavior of the data
even at times where they have not been constrained.
This suggests that the corresponding QNM overtones
are recovered consistently for some range of times around this $t_0$.

Here we turn to the investigation of the best-fit amplitude parameters $A_n$
obtained for each tone $n$
in multiple-tone models of the class of Eq.~\eqref{eq:QNM-multitone-model}
(hence, with all frequencies equal to the QNM values),
still for the example of the shear $l=2$ mode.
These parameters are by definition amplitudes computed at the fixed time $t_r$,
and we check for their constancy as we vary the time $t_0$ at which the fit is started.
The same or a similar test has been used to check for the presence of overtones
in numerical gravitational-wave ringdown models
\emph{e.g.} in \cite{Bhagwat:2019dtm,giesler2019,cook2020}.

As we want to retrieve the amplitudes of the tones,
we here apply the rescaling procedure given by Eq.~\eqref{eq:rescaling}
prior to fitting\footnote{%
More explicitly, the rescaled model reads in this case
$\tilde{h}_x(t) = A_0 \cos [ \omega_{l00} \,\Delta t + \phi_0 ]
+ \Sigma_{n=1}^{\nmax} A_n \cos [ \omega_{l0n} \, \Delta t + \phi_n ]
\exp [- (\tau_{l0n}^{-1} - \tau_{l00}^{-1}) \, \Delta t  ]$, which we fit to the rescaled data
$\tilde{h}_{\mathrm{NR}}(t) = h_{\mathrm{NR}}(t) \exp [ + \tau_{l00}^{-1} \, \Delta t ]$,
with $\Delta t = t - t_r$. The amplitude parameters $A_0$, $A_n$
are by definition the amplitudes of each mode at the fixed time $t_r$,
and they are formally neither affected by this rescaling
nor by changing the fit starting time $t_0$.
The best-fit values found for these parameters, on the other hand, may vary,
\emph{e.g.} if the modes are not well recovered by the fitting procedure
when they have been highly damped,
or if the data contains more than the QNMs included in the model.%
}.
Note that we still expect the amplitudes (at $t_r$) of the overtones
not to be accurately determined for too large $t_0$,
as the overtones are damped much faster than the fundamental mode
and hence still decay in the rescaled data. 

\begin{figure}
  \centering    
  \includegraphics[width=\columnwidth]{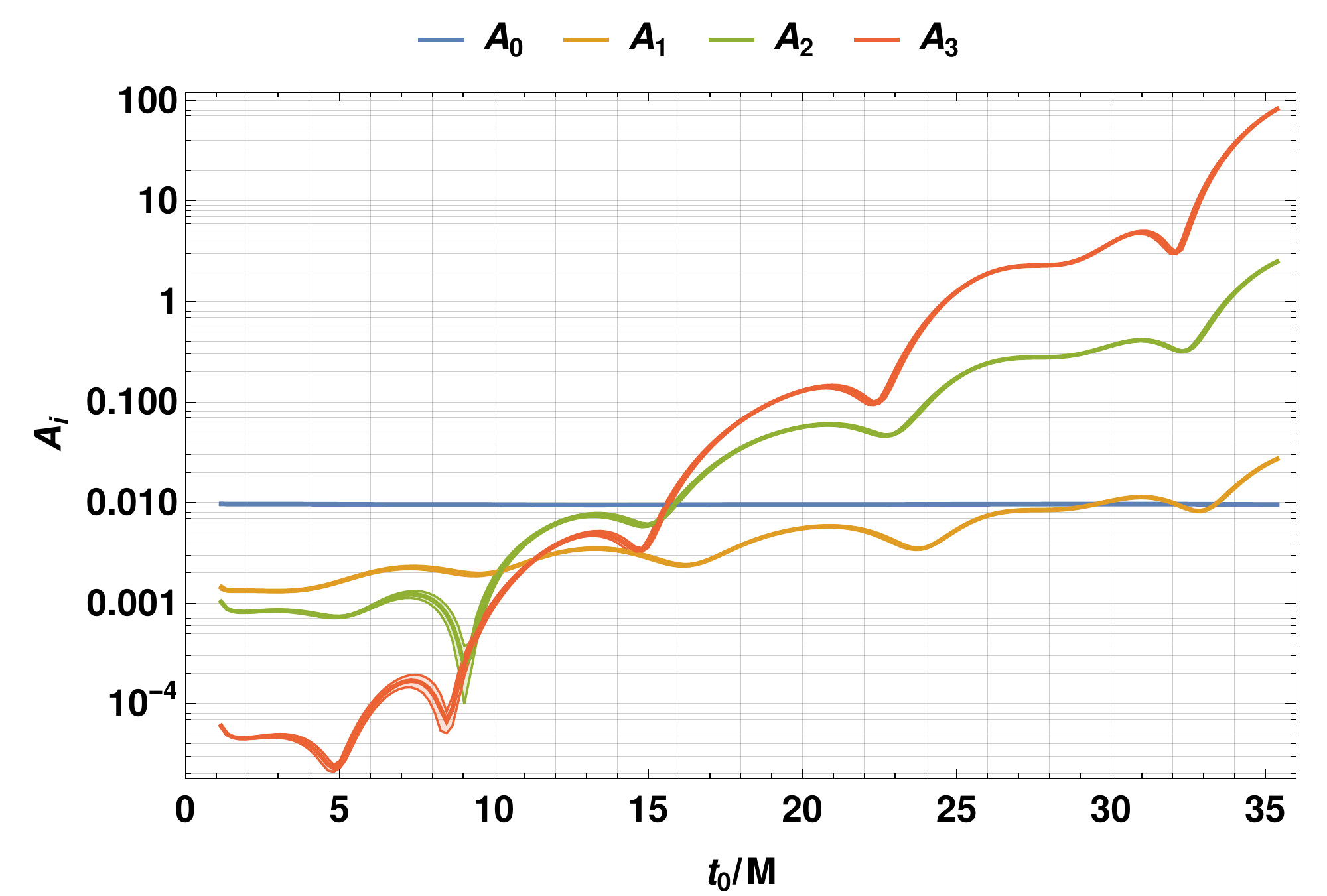}
  \caption{Best-fit amplitudes at $t_r$
  (with $1 \sigma$ uncertainties as given by Eq.~\eqref{eq:fiterrors})
  as a function of the fit starting time $t_0$ for the fundamental mode and overtones
  in the ($\nmax=3$)-overtone QNM model of Eq.~\eqref{eq:QNM-multitone-model},
  for the shear $l=2$ mode.
  The rescaling procedure given by Eq.~\eqref{eq:rescaling} has been used before fitting.}
  \label{fig:3overtone-amplitudes-shear}
\end{figure}

We focus first on the model in the case of $\nmax=3$,
which we found to be the smallest number of overtones
matching very well the behavior of $\sigma_2$ at all times.
Fig.~\ref{fig:3overtone-amplitudes-shear} shows the resulting best-fit amplitude parameters
for the fundamental mode and for the three overtones considered, as functions of $t_0$,
along with their $1 \sigma$ fitting uncertainties (Eq.~\eqref{eq:fiterrors}).

The amplitude parameter $A_0$ of the fundamental mode
is remarkably constant throughout the figure,
providing strong further support for the presence of this mode in the data.
The amplitude parameters $A_n$ ($n > 0$) of the overtones, on the other hand,
are clearly inconsistent between different values of $t_0 \gtrsim 4 M$.
This does not really contradict the presence of overtones in the data
as these amplitudes are expected to be poorly determined beyond early times
once the overtones have decayed. Interestingly however,
all overtones have a stable best-fit amplitude parameter
over the range $1.5 M \lesssim t_0/M \lesssim 4 M$,
which corresponds to the regime of early-time exponential decay.
Each of the amplitudes is thus consistently determined
over multiple values of $t_0$ if this regime is accounted for in the fit.
We note that the ratios of the overtone amplitudes computed at the horizon formation,
$A_n \exp [(t_r - t_{\mathrm{bifurcate}}) / \tau_{l0n} ]$ ($n > 0$),
to the amplitude of the fundamental mode at the same time
$A_0 \exp [(t_r - t_{\mathrm{bifurcate}}) / \tau_{l00} ]$,
as determined here from the stable early-time best-fit values of $A_n$ and $A_0$,
are of the order of $\sim 2$, $\sim 23$ and $\sim 34$
for $n=1$, $n=2$ and $n=3$ respectively.

\begin{figure}
  \centering    
  \includegraphics[width=\columnwidth]{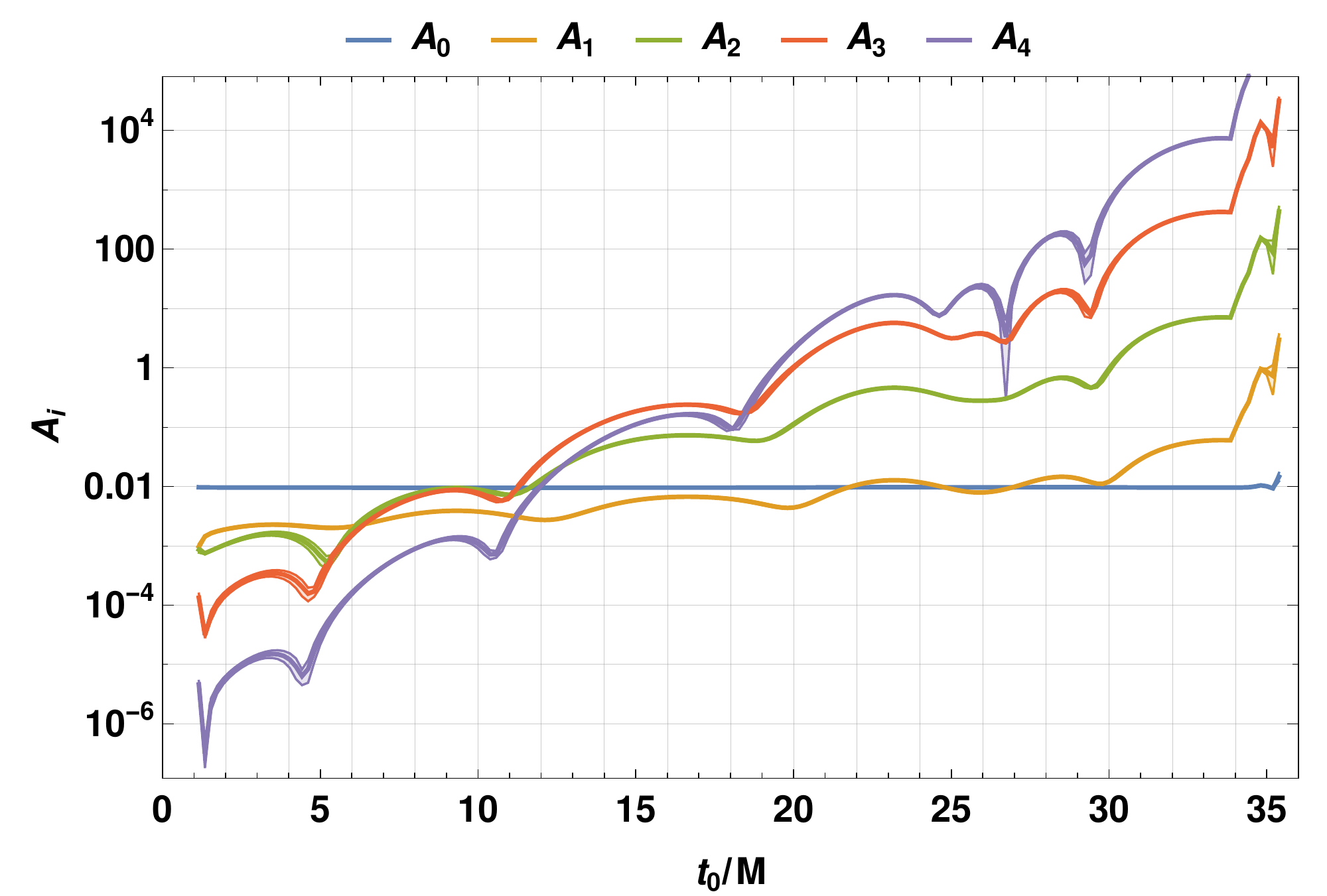}
  \caption{Same as Fig.~\ref{fig:3overtone-amplitudes-shear},
  for the $(\nmax=4$)-overtone model of Eq.~\eqref{eq:QNM-multitone-model}.}
  \label{fig:4overtone-amplitudes-shear}
\end{figure}

We show for comparison in Fig.~\ref{fig:4overtone-amplitudes-shear}
the best-fit amplitude parameters obtained in the same way
with instead $\nmax=4$ overtones included in the model.
The resulting $A_0$ is still constant over the whole time range considered
and $A_1$ is still relatively stable over the same early-time interval as above,
with values roughly consistent with those obtained from the three-overtone model.
The higher overtones ($n \geq 2$) on the other hand do not appear to be stable over any time range.
This may however simply indicate that their amplitudes
cannot be constrained accurately enough even in the exponential damping regime
due to a large number of free parameters and a too quickly decaying fourth overtone.
For $\sigma_2$, $\nmax=3$ seems to be an optimal number of overtones
that models well the data at all times
while still allowing the amplitude of each tone to be correctly constrained.

The considerations of this subsection
--- including the discussion recalled hereabove from section \ref{subsec:fit-overtones} ---
still do not provide any definitive conclusion about the actual presence of QNM overtones
in the shear modes or multipoles,
in particular since the stability of the best-fit model over some time range is a necessary,
but not a sufficient condition for their presence.
Yet these results perhaps represent the most supportive clue that we obtain
in favor of the behavior of the horizon being indeed dominated by QNMs
from shortly after its formation. The results of the previous subsections
would not directly contradict such a statement.
They would rather point towards the difficulty of separating QNM overtones
from any other combination of damped sinusoids
with roughly comparable complex frequencies
(and thus of deciding on the presence of QNM overtones
\emph{vs.} such other decaying modes),
and even of determining how many tones would have non-negligible contributions.

\section{Conclusion and discussion}
\label{sec:conclusion}

We have shown in this paper that the dynamics of the final apparent
horizon in a binary black hole merger can be very well described by
the quasinormal modes of the final black hole,
from shortly after this horizon is formed onward.
We have studied two quantities of interest, namely the shear of the outgoing
normal at the horizon, and the horizon multipole moments; both of
these are well modeled by quasinormal modes provided a large enough
number of overtones is included. We have considered here a high-precision
numerical simulation of a head-on collision of nonspinning
black holes, but we expect these results to qualitatively hold
for other configurations (of higher astrophysical interest) as well.

We have first confirmed that the behavior of each of the shear modes
$\sigma_l$ and of the horizon mass multipole moments $I_l$, for
$2\leq l \leq 12$, is dominated at late times by the corresponding
fundamental quasinormal mode. This is compatible with linear
perturbation theory, which can be expected to hold in this regime and
predicts an asymptotic predominance of the fundamental quasinormal
modes since the associated overtones have shorter damping times. This
result strengthens the conclusions of \cite{pook-kolb2020II}, and
supports the presence of correlations between the emitted
gravitational waves and the dynamics of the final black hole horizon
(\emph{cf.} \cite{Prasad:2020xgr}).  Deviations from a description
only in terms of the fundamental mode are however also evident,
especially at early and intermediate times.  This is accounted for by
also including the higher overtones.  We have shown that the shear and
multipole moments, for essentially the entire time after the common
horizon formation, are well described by superpositions of
quasinormal modes including the overtones.

These results are in qualitative
agreement with studies of the gravitational waveform extracted far
away from the source.  For example, in \cite{giesler2019} it is found
that the dominant $(l = |m|=2)$ harmonic of the gravitational waveform
for a particular quasicircular initial configuration (with mass ratio
$1.22$ and moderate spins aligned with the orbital angular momentum)
is well modeled right up to the peak of the strain by including
up to seven overtones. The ability to detect and separate the successive
overtones in the early stages of the ringdown, before they have
decayed, would improve the prospects for black hole spectroscopy, and for
observational probes of the black hole no-hair theorem.

In the present
work, we have probed another part of spacetime by focusing on the
horizon of the final black hole; we however expect strong correlations
between both dynamics, arising from the same source
\cite{Prasad:2020xgr}.  The simpler geometry in our study, and the
focus on the horizon, allowed for a high numerical precision and for an
investigation of all geometric modes up to $l=12$, rather than just
the dominant $l=2$ mode, for both the shear and the multipole moments.
For all of these modes, we have obtained similar qualitative results.
In particular, in the case of the $l=2$ mode, we have found that
two to three overtones suffice for an accurate modeling
of both variables from shortly after the horizon formation onward.
We have however also noticed the general increase
in the number of overtones
necessary for a good description of the geometric $l$ mode as $l$ increases.

Such results remain surprising because, shortly after the common
horizon is formed, it is highly distorted and cannot be described
as a linear perturbation of a Schwarzschild horizon.  As evidence for this, we
have noted that the area of the horizon increases at a very significant rate in this regime.
The total relative change in area is only of about $6\%$
however, so it could be argued that perturbation theory is still of
some utility.  One can then wonder whether obtaining a description of the
horizon dynamics in terms of ringdown modes in this regime implies
that the horizon is, in some suitable sense still to be understood,
still a small perturbation of a stationary black hole.  We have
accordingly studied, through various possible criteria, whether one
should conclude a) at the linear perturbation spectrum indeed already
driving the horizon dynamics at early times, or b) at the more prosaic
alternative that the quasinormal modes are just a suitable function
basis for the shear and the multipoles, so that there is no deeper
interpretation of these results.

\revision{Despite} this investigation, and given the lack of a calculation from
first principles, \revision{a conclusive answer to this question is still elusive}. 
We have noted that the infinite slope featured by all shear
modes and multipoles at $t_{\mathrm{bifurcate}}$ prevents their formal
description by a finite sum of QNMs at horizon formation, at least in
terms of the $t$ parameter used. Nevertheless, this constraint does
not rule out such a model even at only slightly later times such as
during the observed ``early-time'' decay phase at
$0.3 \lesssim (t-t_{\mathrm{bifurcate}})/M \lesssim 3$. Hypothesis a)
is supported by the stability observed to some extent in the best-fit
amplitudes of each mode when the time $t_0$ at which the fit is
started varies and spans the early-time range quoted above.  This
stability also manifests itself in the continued qualitative agreement
of the model to the data at times prior to $t_0$ that is typically observed
when $t_0$ lies in this range. On the other hand, models with lower
numbers of overtones but some of the overtone frequencies let free to
deviate from the QNM values, did not show a clear preference for the
QNM overtones spectrum. The same was found using an example toy model
with real frequencies artificially set slightly away from the QNM
values.  This is compatible with hypothesis b), but these results do
not rule out the actual predominance of overtones at early times,
hypothesis a), given that a clear preference for \emph{non}-QNM
frequencies was not found either.  This rather hints at the difficulty
of resolving individual modes in a sum of damped sinusoids with
frequencies comparable to that of the QNMs, and of determining how
many such modes can be included and constrained, even with essentially
noise-free data.  We expect these issues ---including the overall
difficulty of firmly ruling out the predominance of nonlinearities
over overtones at early times--- to hold similarly when the ringdown
is analyzed from the emitted gravitational waves, complicating an
overtone-based spectroscopy.  The lack of a clear-cut recovery of the
QNM overtone frequencies, in particular, was indeed also observed for
the dominant $(l=|m|=2)$ gravitational-wave mode during ringdown in a
binary black hole merger simulation in~\cite{Bhagwat:2019dtm}.


Turning now to future directions, there are a few straightforward possible
extensions of the present work.  First, the present analysis may be
completed by a closer look at the more involved behavior of the vector
modes $\xi_l$ (see Fig.~\ref{fig:xi-modes} and the associated brief
discussion in section~\ref{subsubsec:observables}). Within the same
setup as considered here, it would also be natural to try other
parametrizations of ``time'' to circumvent the infinite slope at
$t_{\mathrm{bifurcate}}$ in each of the observables as functions of
$t$.  This could also allow for a consistent joint treatment of both
the inner and outer common horizons, which constitute indeed a single
smooth hypersurface in spacetime.  Second, one can look for a
possible generalization of the results to a wider variety of
configurations, including the astrophysically important quasicircular
orbits and accounting for black hole spin.  Third, a more
fundamental investigation of the mechanisms driving the early-time
dynamics of the outer common horizon could shed more light onto the fast
exponential decays observed at these times for all of the shear modes
and multipoles. We have found here that quasinormal overtones can
indeed combine in such a way as to produce this behavior.
An investigation of the mechanisms driving it
could either provide more insight into why such a combination would take place,
or rule out quasinormal modes as a relevant explanation for this
(possibly still nonlinear) regime altogether.

\begin{acknowledgments}

  We are grateful to Abhay Ashtekar, Ofek Birnholtz, Gregory Cook,
  Jos\'e~Luis~Jaramillo and Neev Khera for valuable discussions and comments,
  and to an anonymous referee for helpful comments on the paper.
  Research at Perimeter Institute is supported in part by the
  Government of Canada through the Department of Innovation, Science
  and Industry Canada and by the Province of Ontario through the
  Ministry of Colleges and Universities.

\end{acknowledgments}
  
\bibliography{./biblio.bib}

\end{document}